\documentclass[twocolumn]{aastex631}

\usepackage{amsmath} 

\begin{document}

\title{Impact of MHD disk wind on early evolutionary stage of protoplanetary disk and dust growth}

\author[0000-0002-9424-1192]{Yoshihiro Kawasaki}
\affiliation{Department of Earth and Planetary Sciences, Faculty of Sciences, Kyushu University, Fukuoka 819-0395, Japan}

\author[0000-0002-0963-0872]{Masahiro N. Machida}
\affiliation{Department of Earth and Planetary Sciences, Faculty of Sciences, Kyushu University, Fukuoka 819-0395, Japan}

\begin{abstract}
  We perform one-dimensional protoplanetary disk evolution calculations to investigate the impact of the magnetohydrodynamic (MHD) disk wind on disk evolution and dust particle growth.
  To examine the effect of the MHD disk wind, we compare calculations with and without it. 
  In disk evolution calculations, episodic accretion events (or outbursts) occur repeatedly, as reported in previous studies, regardless of the presence of the MHD disk wind.
  However, the time interval between outbursts is shorter in cases with the MHD disk wind than in those without it.
  For dust particle growth, during the infall phase, there is no significant difference between cases with and without the MHD disk wind, and dust particles grow to approximately 1-10\,cm.
  Inside the $\mathrm{H_{2}O}$ snowline, the maximum dust particle size is limited by the collisional fragmentation of dust particles.
  Outside the snowline, the maximum dust particle size is primarily determined by radial drift.
  After the infall phase, when the MHD disk wind is considered, the disk temperature decreases noticeably, and the snowline migrates inward. 
  As a result, the dust particles can grow beyond 10\,cm.
  Therefore, we find that the MHD disk wind plays a crucial role in  dust growth and planet formation after the infall phase.
\end{abstract}


\section{Introduction} 
\label{sec:intro}

Clarifying planet formation processes has been a longstanding issue in the fields of astronomy and planetary science.
In protoplanetary disks, the collisional growth of submicron dust particles is the process that initiates planet formation \citep{2014prpl.conf..339T}.
The processes of dust particle growth are affected by the gas density, temperature, and the composition of dust particles.
Therefore, the density, temperature, and chemical structure of the disk gas are crucial for understanding the growth and motion of dust particles.

Atacama Large Millimeter/Submillimeter Array (ALMA) observations have revealed various structures such as spirals and rings within Class II and subsequent evolved disks \citep{2018ApJ...869L..41A}.  
These structures suggest the growth and accumulation of dust particles within the disks and the possible presence of planets. 
Observations targeting young Class 0/I disks have been conducted in recent years \citep{2023ApJ...951....8O}.
Compared to Class II disks, young Class 0/I disks with observable structures are rarely detected. 
However, \citet{2020ApJ...902..141S} detected sub-structures in Class 0/I disks, though their origin is not clear.
Observational results also indicate that dust particles grow in Class 0/I disks \citep{2023ApJ...954..110O, 2023arXiv231202504S,2023arXiv231213573L}. 
Therefore, understanding dust particle growth in Class 0/I disks and determining when planet formation begins after disk formation are crucial questions.

The disk formation and evolution processes are studied using multidimensional numerical simulations.
Non-ideal magnetohydrodynamic simulations with molecular cloud cores as initial conditions have demonstrated the formation processes for protostars and disks \citep{2010ApJ...724.1006M,2011MNRAS.413.2767M,2020ApJ...896..158T,2021MNRAS.502.4911X,2021MNRAS.508.2142X}.
These simulations showed that disks become gravitationally unstable, leading to the development of spiral arms and angular momentum transport due to gravitational torque \citep{2017ApJ...835L..11T}. 
Additionally, outflows, which transport mass and angular momentum from the central region of the cores, are also reproduced in these simulations \citep{2013ApJ...763....6T,2019ApJ...876..149M}.

Recently, multi-dimensional numerical simulations, including dust particles, have been performed \citep{2019A&A...626A..96L, 2020A&A...641A.112L,2021ApJ...913..148T,2021ApJ...920L..35T,2022MNRAS.515.6073K,2023MNRAS.519.3595K,2023A&A...670A..61M}.
These simulations have shown differences in spatial distribution between gas and dust \citep{2020A&A...641A.112L}, and the impact of the growth of dust particles on the disk evolution and outflow activity \citep{2023PASJ...75..835T}.
Multidimensional numerical simulations are a highly useful tool in the study of the formation and evolution of stars and disks.
However, the calculation cost of these simulations is high and it is difficult to perform long-term calculations including detailed dust particle growth and disk temperature evolution.

To study long-term disk evolution, radially one-dimensional calculations are often used \citep{1994ApJ...421..640N, 2005A&A...442..703H, 2010ApJ...713.1143Z, 2013ApJ...764..141B}.
During the disk formation stage, the infall rate onto the disk is calculated by considering the conservation of angular momentum of the infalling gas within the molecular cloud core \citep{1981Icar...48..353C}.
\citet{2013ApJ...770...71T} performed one-dimensional disk calculations employing an analytical model for accretion onto the disk.
They demonstrated that their one-dimensional calculation is in good agreement with the results of three-dimensional hydrodynamic simulations \citep{2010ApJ...724.1006M,2015ApJ...801...81Z, 2016ApJ...818..152B}. 
Thus, one-dimensional calculations are a useful tool for investigating long-term disk evolution.
However, it should be noted that many of the one-dimensional calculations did not include the magnetohydrodynamic (MHD) disk wind (outflow), in contrast to the three-dimensional simulations.

MHD disk winds are considered important in disk evolution as they remove mass and angular momentum from the disk. 
We note that two types of mass loss mechanisms from the disk, namely photoevaporative winds and MHD (or magnetized) disk winds, have been considered.
Photoevaporative winds are associated with angular momentum transport in a viscously evolving disk, while MHD disk winds play a crucial role in both angular momentum removal and mass accretion by directly driving disk material outward.
The latter is considered more significant during the early infall phase.
Compared to photoevaporative winds, MHD disk winds exert a stronger influence on the inner regions of the disk, significantly affecting its evolution \citep{2023MNRAS.524.3948A}.
Since we focus on disk evolution within approximately 100\,au before significant disk mass dissipation occurs, photoevaporative winds are not expected to be particularly dominant on either spatial or temporal scales.
Incorporating photoevaporative winds requires longer-term calculations, which we leave for future studies.
For this reason, we consider only MHD disk winds in this study.

Since the 2010s, one-dimensional calculations for disk evolution considering MHD disk winds have been performed \citep{2016A&A...596A..74S, 2016ApJ...821...80B,2022MNRAS.512.2290T}. 
\citet{2018ApJ...865..102T} calculated the disk evolution of gas and dust including the magnetized disk wind.
Their results showed that the amount of disk gas decreases from the inner disk region due to the MHD disk wind, leading to the accumulation of dust in regions with pressure maxima and the formation of ring structures.
While their study demonstrated the importance of the MHD disk wind in the evolution of gas and dust, it has several limitations.
For instance, their calculations did not consider dust growth, the MHD disk wind is not driven during infall onto the disk (or infall phase), removal of angular momentum from the disk by the MHD disk wind is not considered, and the disk temperature profile is fixed. 
These limitations are crucial for considering more realistic disk evolution. 

Many studies on dust growth in disks assume evolved Class II isolated disks \citep{2008A&A...480..859B}, 
but calculations considering dust growth in disks assuming Class 0 disks have been performed recently \citep{2017ApJ...838..151T,2023ApJ...946...94X}.
Some studies have investigated the evolution of dust particles in an evolving disk considering mass accretion from the envelope onto the disk \citep{2010A&A...513A..79B,2018ApJ...868..118H},
though these studies did not consider the MHD disk wind.
Hence, the evolution of dust particle size and spatial distribution in an evolving disk considering the MHD disk wind remains unclear.
In this study, we perform one-dimensional disk evolution calculations, incorporating the effects of self-gravity, thermal evolution by solving the energy equation, and the MHD disk wind to investigate their effects on disk structure and dust growth.

In addition to adequately investigating the properties of the disk and dust,
thermodynamics and the thermal evolution of the disk are also important in terms of gravitational instability.
Whether gravitational instability occurs depends on the detailed thermodynamics, particularly the cooling rate of the disk \citep[e.g.,][]{2024arXiv241012042X}.

In the following, we use ``disk wind'' instead of ``MHD disk wind'' to avoid redundancy. The term "MHD disk wind" is used only when emphasizing magnetohydrodynamic effects.
This paper is structured as follows. 
We describe the basic equations and numerical settings for calculating the disk evolution in Section 2. 
The calculation results are presented in Section 3. 
The implications for planet formation processes are discussed in Section 4. 
A summary is presented in Section 5.

\section{Methods}
\label{sec:method}

\subsection{Disk evolution and basic equations}
\label{sec:method:surface_density_evolution}

In this section, we describe the equations for gas and dust surface density evolution.
The formulas used are based on \citet{2018ApJ...865..102T}, but have been extended in this study.
While \citet{2018ApJ...865..102T} treated dust as a single fluid, we treat dust as a multi-fluid, considering each size separately.
The evolution of the surface density of gas and dust is described by the following equations,
\begin{equation}
	\frac{\partial \Sigma_{\mathrm{g}}}{\partial t} + \frac{1}{r} \frac{\partial}{\partial r} \left( r\Sigma_{\mathrm{g}} v_{\mathrm{g}, r} \right)
  = \dot{\Sigma}_{\mathrm{g, inf}} - \dot{\Sigma}_{\mathrm{g, wind}}
  \label{eq:gas_surface_evolution}
\end{equation}
\begin{align}
	& \frac{\partial \Sigma_{\mathrm{d}}\left(m\right)}{\partial t} + \frac{1}{r} \frac{\partial}{\partial r} \left[ r\Sigma_{\mathrm{d}}\left(m\right) v_{\mathrm{d}, r}\left(m\right) \right] \notag \\ 
	& = \frac{1}{r} \frac{\partial}{\partial r} \left[ r D_{\mathrm{d}} \left( m \right) \Sigma_{\mathrm{g}} \frac{\partial}{\partial r} \left( \frac{\Sigma_{\mathrm{d}}\left( m\right) }{\Sigma_{\mathrm{g}}} \right) \right] \notag \\
  & + \dot{\Sigma}_{\mathrm{d, inf}}\left( m \right) - \dot{\Sigma}_{\mathrm{d, wind}}\left( m \right) + \left(\frac{\partial \Sigma_{\mathrm{d}}\left(m\right)}{\partial t}\right)_{\rm{coag/frag}},
  \label{eq:dust_surface_evolution}
\end{align}
where $\Sigma_{\rm g}$ is the gas surface density, $\Sigma_{\mathrm{d}}\left(m\right)$ is the dust surface density with mass $m$, $v_{\mathrm{g}, r}$ is the gas radial velocity, $v_{\mathrm{d}, r}\left( m \right)$ is the dust radial velocity, and
$\dot{\Sigma}_{\mathrm{g, inf}}$, $\dot{\Sigma}_{\mathrm{d, inf}}$, $\dot{\Sigma}_{\mathrm{g, wind}}$, and $\dot{\Sigma}_{\mathrm{d, wind}}$ represent the mass supply from the infalling envelope onto the disk and the mass loss by the disk wind from the disk of gas and dust, respectively (for details, see \S\ref{sec:method:infall_and_disk_wind_model}).
The first term of the right-hand side of equation (\ref{eq:dust_surface_evolution}) is the turbulent diffusion of dust in a radial direction and $D_{\mathrm{d}}$ is the dust turbulent diffusion coefficient.
The last term of the right-hand side of equation (\ref{eq:dust_surface_evolution}) represents the change rate for dust mass $m$ due to coagulation or fragmentation caused by dust particle collisions (for details, see \S\ref{sec:method:dust_growth}).

The gas and dust velocity are obtained by solving the equations of motion in the radial and azimuthal directions.
The radial and azimuthal velocities of the dust $v_{\mathrm{d}, r}$ and $v_{\mathrm{d}, \phi}$, respectively, are given by
\begin{equation}
	v_{\mathrm{d}, r} \left( m \right) =  \frac{(1 + A \mathrm{St}^{2})v_{\mathrm{g}, r} + 2\mathrm{St} \delta v_{\mathrm{g}, \phi} - \mathrm{St}^{2} \frac{r\dot{M}_{r, \rm{tot}}}{M_{r}} }
		{1 + \left(1 + A\right)\mathrm{St}^{2}},
  \label{eq:dust_radial_velocity}
\end{equation}
\begin{equation}
	\delta v_{\mathrm{d}, \phi} \left( m \right)
	= \frac{ -\mathrm{St}v_{\mathrm{g}, r} + 2\delta v_{\mathrm{g}, \phi} - \mathrm{St} \frac{r\dot{M}_{r, \rm{tot}}}{M_{r}} }
		{2 \left\{ 1 + \left(1 + A\right) \mathrm{St}^{2} \right\} },
  \label{eq:dust_azimuthal_velocity}
\end{equation}
where the dust azimuthal velocity is described by the deviation from the mean azimuthal flow, $\delta v_{\mathrm{d}, \phi} \left( m \right) = v_{\mathrm{d}, \phi} \left(m\right) - r\Omega $.
The same description is used for the gas azimuthal velocity, $\delta v_{\mathrm{g}, \phi} = v_{\mathrm{g}, \phi} - r\Omega$.
We estimate the angular velocity $\Omega = \left( GM_{r} / r^{3} \right)^{1/2}$ using the enclosed mass $M_{r}$ within radius $r$, $M_{r} = M_{\ast} + \int \Sigma_{\rm g} 2\pi r \mathrm{d}r$, where $M_{\ast}$ is the central stellar mass.
The factor $A = 2\pi r^{2}\Sigma_{\mathrm{g}} / M_{r}$ represents the effect of the self-gravity of the disk.
$\dot{M}_{r, \mathrm{tot}}$ represents the net disk mass change rate due to the mass supply from the envelope and the mass loss by the disk wind, and is described by 
\begin{equation}
  \dot{M}_{r, \mathrm{tot}} = \int_{0}^{r} 2\pi r\left( \dot{\Sigma}_{\mathrm{g, inf}} - \dot{\Sigma}_{\mathrm{g, wind}} \right) \mathrm{d}r.
  \label{eq:enclosed_mdot}
\end{equation}

The Stokes number $\mathrm{St}$ is a dimensionless stopping time and quantifies the degree of coupling between gas and dust.
In this study, each dust particle is assumed to be compact and spherical, while actual dust particles may be distorted or have porosity. 
Thus, a dust mass with size $a$ is described as $m = (4\pi/3)\rho_{\mathrm{di}}a^{3}$, where $\rho_{\mathrm{di}}$ is the internal (material) density of the dust particles.
It is noted that dust porosity is important for both dust growth and dust motion in disks \citep{,2007A&A...461..215O, 2012ApJ...752..106O,2013A&A...557L...4K}.
$\mathrm{St}$ is given by 
\begin{equation}
  \mathrm{St} = \left\{
      \begin{aligned}
        & \frac{\pi \rho_{\mathrm{di}} a}{2 \Sigma_{\mathrm{g}}} \ \left( a < \frac{9}{4}\lambda_{\mathrm{mfp}} \right) \\ 
        & \frac{2\pi \rho_{\mathrm{di}}a^{2}}{9\lambda_{\mathrm{mfp}}\Sigma_{\mathrm{g}}} \ \left( a > \frac{9}{4}\lambda_{\mathrm{mfp}} \right)
      \end{aligned}
  \right.,
\end{equation}
where $\rho_{\mathrm{g}}$ is the gas mass density, $v_{\mathrm{th}} = \sqrt{8/\pi}c_{s}$ is the thermal velocity,  $c_{s}$ is the sound speed, and $\lambda_{\mathrm{mfp}}$ is the gas mean free path.
The gas mean free path is described as $\lambda_{\mathrm{mfp}} = m_{\mathrm{g}}/(\sigma_{\mathrm{mol}}\rho_{\mathrm{g}})$, where $m_{\mathrm{g}}$ is the gas-particle mass and $\sigma_{\mathrm{mol}} = 2\times 10^{15} \ \mathrm{cm^{2}}$ is the gas molecular cross section.
Additionally, the turbulent diffusion coefficient is given by \citep{2007Icar..192..588Y}
\begin{equation}
  D_{\mathrm{d}} = \frac{\alpha_{\mathrm{turb}} c_{s} h_{\rm g}}{1 + \mathrm{St}^{2}},
\end{equation}
where $h_{\rm g} = c_{s} / \Omega$ is the gas scale height and $\alpha_{\mathrm{turb}}$ is the strength of the turbulence.

The radial and azimuthal velocities of the gas $v_{\mathrm{g}, r}$ and $\delta v_{\mathrm{g}, \phi}$, respectively, are given by
\begin{equation}
	v_{\mathrm{g}, r} = \frac{ \frac{2r}{j}N(1 + \lambda_{2}) + 2\lambda_{1} \left(1 + A \right) r \Omega\eta}{\left( 1 + \lambda_{2} \right)^{2} + \lambda_{1}^{2}\left(1 + A\right)}
		- \frac{r \dot{M}_{r, \rm{tot}}}{M_{r}},
\end{equation}
\begin{equation}
	\delta v_{\mathrm{g}, \phi} = 
    \frac{ \frac{2r}{j} N \lambda_{1} -2\left(1 + \lambda_{2}\right)r \Omega\eta }
		{2 \left\{ (1 + \lambda_{2})^{2} + \lambda_{1}^{2} \left(1 + A\right) \right\} },
\end{equation}
where $j = r^{2}\Omega$ is the specific angular momentum in a disk and $\eta$ represents the effect of the pressure gradient force, which we estimate as
\begin{equation}
	\eta = -\frac{1}{2} \left( \frac{c_{s}}{r\Omega} \right)^{2} \frac{\partial \ln p}{\partial \ln r},
\end{equation}
with $p$ the gas pressure at the disk midplane.
$\lambda_{1}$ and $\lambda_{2}$ are quantities related to the dust-to-gas back-reaction and are given by
\begin{equation}
	\lambda_{1} = \int \frac{\mathrm{St}\left( m \right)}{1 + \left(1 + A\right)\mathrm{St}\left( m \right)^{2}} \epsilon\left( m \right) \mathrm{d}m,
\end{equation}
\begin{equation}
	\lambda_{2} = \int \frac{1}{1 + \left(1 + A\right)\mathrm{St}\left( m \right)^{2}} \epsilon\left( m \right) \mathrm{d}m,
\end{equation}
where $\epsilon \left( m \right) = \Sigma_{\mathrm{d}} \left( m \right) / \Sigma_{\mathrm{g}}$ represents the dust-to-gas mass ratio for dust mass $m$.

The specific torque acting on the disk $N$ is described by
\begin{equation}
  N = - \frac{1}{r\Sigma_{\mathrm{g}}} \frac{\partial}{\partial r} \left( \frac{3}{2} r^{2} \Sigma_{\mathrm{g}} \alpha_{\rm SS} c_{s}^{2} \right) - \frac{3}{4}\alpha_{\rm DW}c_{s}^{2}.
  \label{eq:disk_torque}
\end{equation}
The first term of equation~(\ref{eq:disk_torque}) corresponds to the viscous torque.
The Shakura--Sunyaev alpha parameter $\alpha_{\rm SS}$ \citep{1973A&A....24..337S} represents the efficiency of angular momentum transport.
We consider both self-gravitational instability (GI) and magnetorotational instability (MRI) as sources of effective viscosity, expressed as $\alpha_{\rm SS} = \alpha_{\mathrm{GI}} + \alpha_{\mathrm{MRI}}$,
where $\alpha_{\mathrm{GI}}$ and $\alpha_{\mathrm{MRI}}$ are GI and MRI viscous parameters, respectively.
In this study, we estimate $\alpha_{\mathrm{GI}}$ as a function of the Toomre Q parameter \citep{2010ApJ...713.1143Z}, $\alpha_{\mathrm{GI}} = \exp\left(-Q^{4}\right)$,
where $Q$ is described by $Q = c_{s}\Omega / (\pi G \Sigma_{\mathrm{g}})$.
This GI viscosity $\alpha_{\mathrm{GI}}$ successfully models disk evolution due to angular momentum transport by spiral arms in a gravitationally unstable disk, as shown by multidimensional numerical simulations \citep{2013ApJ...770...71T}.

We estimate $\alpha_{\rm MRI}$ as a function of the midplane temperature \citep{2016ApJ...827..144F, 2019ApJ...871...10U},
\begin{equation}
  \alpha_{\mathrm{MRI}} = \frac{\alpha_{\mathrm{active}} - \alpha_{\mathrm{dead}}}{2} \left[ 1 - \tanh\left( \frac{T_{\mathrm{MRI}} -T }{50 \ \mathrm{K}} \right) \right]
    + \alpha_{\mathrm{dead}},
\end{equation}
where $T$ is the gas temperature in the disk and $T_{\mathrm{MRI}} = 1000 \ \mathrm{K}$ is a critical temperature above which MRI becomes active \citep{2015ApJ...811..156D}, $\alpha_{\mathrm{active}} = 10^{-2}$ is the MRI strength (or viscous parameter) in the MRI active region, and $\alpha_{\mathrm{dead}} = 10^{-4}$ is that in the MRI inactive region, i.e., the dead zone.

These MRI related values are fixed in this study.
In this study, it is assumed that gravitational instability contributes only to angular momentum transport. 
In other words, gravitational instability does not contribute to the turbulence strength $\alpha_{\mathrm{turb}}$; only MRI is considered to contribute, $\alpha_{\mathrm{turb}} = \alpha_{\mathrm{MRI}}$.
This turbulent strength contributes to the turbulent diffusion of dust particles and the collisional velocity between dust particles.

The second term of the right-hand side in equation~(\ref{eq:disk_torque}) is the wind torque term.
The dimensionless parameter $\alpha_{\rm DW}$ represents the wind torque strength (see \citealt{2022MNRAS.512.2290T}).
In this study, for simplicity, we adopt a constant value of $\alpha_{\rm DW} = 10^{-2}$ in time and space
\footnote{
$\alpha_{\rm DW}$ can be roughly related to the magnetic field strength using equation (7) of \citet{2025MNRAS.536L..13W}.
Based on our calculation results presented in §3, the magnetic field strength at a radius of 10\,au is estimated to be approximately 0.2\,G.
This value is reasonably consistent with recent core collapse simulations \citep[e.g.,][]{2016A&A...587A..32M,2020MNRAS.494..827M}.
}.
As seen in Fig.1 of \citet{2023MNRAS.523.2630W}, $\alpha_{\rm DW}$ induces an additional radial motion in the disk midplane region, distinct from the viscosity-driven diffusion caused by turbulence $\alpha_{\rm MRI}$, in terms of angular momentum transport.
Furthermore, the disk winds contribute to disk accretion, rather than merely transferring angular momentum as in the case of gravitational instability.

\subsection{Energy equation}
\label{sec:method:disk_temperature}

To obtain the disk midplane temperature,  we solve the energy equation
\begin{eqnarray}
 \frac{\partial E}{\partial t} + \frac{1}{r}\frac{\partial}{\partial r} \left(  E  v_{\mathrm{g}, r} \right)
  &=& - \frac{P}{r} \frac{\partial (r v_{\mathrm{g}, r})}{\partial r} 
   \nonumber \\
  +  &Q_{\mathrm{vis}}& + Q_{\mathrm{shock}} + Q_{\mathrm{irr}} - \Lambda_{\mathrm{rad}},
  \label{eq:energy_equation}
\end{eqnarray}
where $E$ is the vertically integrated thermal energy per unit area,
\begin{equation}
  E = \frac{P}{\gamma - 1} = \frac{k_{B} T}{\mu m_{\mathrm{H}} \left(\gamma - 1\right)} \Sigma_{\rm g},
\end{equation}
in which $\gamma$ is the specific heat ratio, $k_{B}$ is the Boltzmann constant, $\mu$ is the mean molecular weight, and $m_{\mathrm{H}}$ is the hydrogen mass. 
Although the properties of the gas vary with temperature, we adopt constant values of $\gamma = 7/5$ and $\mu = 2.34$.
The vertically integrated pressure $P$ is described by $P = \Sigma_{\rm g} c_{s}^{2}$.
The heating ($Q_{\mathrm{vis}}$, $Q_{\mathrm{shock}}$, and  $Q_{\mathrm{irr}}$) and cooling ( $\Lambda_{\mathrm{rad}}$) terms are described in the following paragraph. 

The heating and cooling rates adopted are based on \citet{1994ApJ...421..640N}.
The viscous heating is described by
\begin{equation}
  Q_{\mathrm{vis}} = - \frac{3}{2}\Sigma_{\rm g} \alpha_{\rm SS} c_{s}^{2} r \frac{\partial \Omega}{\partial r}.
\end{equation}
The shock heating due to infalling gas is described by
\begin{equation}
  Q_{\mathrm{shock}} = \frac{4\tau + 1}{3\tau^{2} +1} \frac{1}{2} \dot{\Sigma}_{\mathrm{inf}} \left( r\Omega \right)^{2},
\end{equation}
where $\tau$ is the optical depth, which is calculated as $\tau = \kappa_{R} \Sigma_{\rm g}/2$ with the Rosseland mean opacity $\kappa_{R}$ taken from \citet{2012ApJ...746..110Z}.
The irradiation heating is given by
\begin{equation}
  Q_{\mathrm{irr}} = \frac{8\tau}{3\tau^{2} + 1} \sigma_{\mathrm{SB}} \left(T_{\mathrm{irr}}^{4} + T_{\mathrm{amb}}^{4} \right),
\end{equation}
where $T_{\mathrm{amb}} = 10 \ \mathrm{K}$ is the temperature of ambient gas in a collapsing cloud core. 
The irradiation flux $\sigma_{\mathrm{SB}} T_{\mathrm{irr}}^{4}$ is due to the central star and is represented as \citep{1991ApJ...375..740R}
\begin{equation}
  \sigma_{\mathrm{SB}}T_{\mathrm{irr}}^{4} = \frac{L_{\star} + L_{\rm acc}}{4\pi r^{2}} \left[ \frac{2}{3\pi}\frac{R_{\star}}{r} 
    + \frac{1}{2}\frac{h_{\rm g}}{r}\left( \frac{\mathrm{d} \ln h_{\rm g}}{\mathrm{d} \ln r} - 1\right) \right],
\end{equation}
where $\sigma_{\mathrm{SB}}$ is the Stefan--Boltzmann constant, the term ($\mathrm{d}\ln h_{\rm g}/\mathrm{d} \ln r -1$) is the shielding factor,
$L_{\star}$ is the stellar luminosity, and $L_{\mathrm{acc}}$ is the accretion luminosity.
To avoid numerical instabilities, we set the shielding factor to be constant $\mathrm{d}\ln h_{\mathrm{g}} /\mathrm{d} \ln r = 9/7$ \citep{2005A&A...442..703H}.
We calculate $L_{\star}$ by the mass--luminosity relation proposed by \citet{2013ApJ...764..141B},
\begin{equation}
  \log_{10}\left( \frac{L_{\star}}{L_{\sun}} \right) = 0.2 + 1.74 \log_{10}\left( \frac{M_{\star}}{M_{\sun}} \right).
\end{equation}
The accretion luminosity $L_{\mathrm{acc}}$ is estimated as 
\begin{equation}
  L_{\rm acc} = \frac{GM_{\star}\dot{M}}{2 R_{\star}},
\end{equation}
where we assume a central star radius of $R_{\star} = 2 R_{\sun}$.

The radiative cooling rate is described by
\begin{equation}
  \Lambda_{\mathrm{rad}} =  \frac{8\tau}{3\tau^{2} + 1} \sigma_{\mathrm{SB}} T^{4}.
  \label{eq:lambda_rad}
\end{equation}

\subsection{Infall and disk wind models}
\label{sec:method:infall_and_disk_wind_model}

To obtain the rate of increase of the disk surface density due to mass infall from the cloud core $\dot{\Sigma}_{\mathrm{inf}}$, we use a model proposed by \citet{2013ApJ...770...71T}.
As the initial state of the core, we consider a critical Bonner--Ebert (B.E.) density profile \citep{1955ZA.....37..217E, 1956MNRAS.116..351B}.
We then increase the density by a factor $f$ so that the gravitational force is stronger than the pressure gradient force \citep{2006ApJ...645.1227M}.
For the B.E. core, the time for the spherical shell initially at position $R_{\mathrm{ini}}$ to reach the central region is given by
\begin{equation}
  t_{\mathrm{inf}} = \sqrt{\frac{R_{\mathrm{ini}}^{3}}{2G M(R_{\mathrm{ini}})}} \int_{0}^{1} \frac{\mathrm{d}x}{\sqrt{f^{-1}\ln x + x^{-1} -1}},
\end{equation}
where $R_{\rm ini}$ is the initial cloud shell radius and $M(R_{\mathrm{ini}})$ is the mass within a radius $R_{\mathrm{ini}}$.
Using $t_{\mathrm{inf}}$, the mass accretion rate onto the disk and star from the cloud core is described by
\begin{equation}
  \dot{M}_{\mathrm{inf}}\left(t\right) = 4\pi R_{\mathrm{ini}}^{2} \rho_{\mathrm{ini}}\left( R_{\mathrm{ini}} \right)  \frac{\mathrm{d}R_{\rm ini}}{\mathrm{d}t_{\rm inf}},
  \label{eq:mdot_inf}
\end{equation}
where $\rho_{\mathrm{ini}}\left(R_{\mathrm{ini}} \right)$ is the initial density of the B.E. core at a radius $R_{\mathrm{ini}}$.
We assume that the core rigidly rotates with angular velocity $\Omega_{\mathrm{core}}$ and the specific angular momentum of the infalling gas during collapse is conserved.
Assuming that the cloud gas accretes at the centrifugal radius, the mass infall rate at each radius $r$ is described by \citep{2013ApJ...770...71T,2018ApJ...865..102T}
\begin{equation}
  \dot{\Sigma}_{\mathrm{inf}} = \frac{1}{2\pi r} \frac{\dot{M}_{\mathrm{inf}}}{2\Omega_{\mathrm{core}} R_{\mathrm{ini}}^{2}}
    \left(1 - \frac{j}{\Omega_{\mathrm{core}}R_{\mathrm{ini}}^{2}} \right)^{-1/2} \frac{\partial j}{\partial r},
\end{equation}
We regard $\dot{\Sigma}_{\mathrm{inf}}$ as the sum of the mass increase rate for the disk of gas and dust.
In this study, we assume that the dust-gas-mass ratio in the molecular cloud core is $f_{\mathrm{dg}} = 0.01$ and the infalling dust particle size is $a_{0} = 0.1 \ \mathrm{\mu \ m}$
\footnote{
As the dust-to-gas mass ratio, we use two different notations: $\epsilon \left( m \right)$ and $f_{\rm dg}$. 
$\epsilon \left( m \right)$ represents the dust-to-gas mass ratio in the disk and is time-dependent. 
In contrast, $f_{\rm dg}$ denotes the dust-to-gas mass ratio of the infalling envelope or the initial cloud core and remains constant over time. 
}. 
Then, the mass infall rates into the disk of gas and dust are described as, respectively,
\begin{equation}
  \dot{\Sigma}_{\mathrm{g, inf}} = (1 - f_{\mathrm{dg}}) \dot{\Sigma}_{\mathrm{inf}},
\end{equation}
\begin{equation}
  \dot{\Sigma}_{\mathrm{d, inf}}(m) =
  \begin{cases}
    f_{\mathrm{dg}} \dot{\Sigma}_{\mathrm{inf}}, & (m = m_0) \\
    0, & \text{(otherwise)}
  \end{cases}
\end{equation}
where $m_{0}$ is the dust mass with radius $a_{0}$.

For the mass loss rate due to the disk wind, we use a model adopted in \citet{2022MNRAS.512.2290T}: 
\begin{equation}
  \dot{\Sigma}_{\mathrm{g, wind}} = \frac{3\alpha_{\rm DW}\Sigma_{\rm g} c_{s}^{2}}{4 j \left(\lambda - 1\right)},
  \label{eq:sigma_dot_wind}
\end{equation}
where $\lambda$ is the parameter for the magnetic lever arm. 
This formula is obtained from the angular momentum conversation of the disk wind.
Although some studies have suggested possible sizes of dust particles blown away by the MHD disk wind \citep{2016ApJ...821....3M,2021ApJ...920L..35T},
the extent to which dust particles of a certain size are carried away by the disk wind is not well understood. 
In this study, it is assumed that dust particles with a Stokes number less than $10^{-4}$ are removed by the disk wind \citep{2022MNRAS.515.6073K}. 
The mass loss rate for dust particles with mass $m$ due to the disk wind is assumed to be given by
\begin{equation}
  \dot{\Sigma}_{\mathrm{d, wind}} \left( m \right) = \frac{3\alpha_{\rm DW}\Sigma_{\mathrm{d}}\left(m\right) c_{s}^{2}}{4 j \left(\lambda - 1\right)}.
\end{equation}

As described above, we consider both mass gain by the disk and mass loss from the disk. 
Thus, we need to determine what mass gain or mass loss occurs at each radius of the disk.
The mass gain is attributed to gas inflow from the infalling envelope, while the mass loss is attributed to the disk wind.

The wind driving condition depends on the strength of the magnetic field and the surrounding gas density \citep{2020MNRAS.499.4490M}.
In this study, we compare the ram pressure of the infalling gas with that of the outflowing gas.
The velocity of the gas infalling to the central region is estimated to be $v_{\mathrm{inf}}\sim \sqrt{GM_{r} /r}$.
On the other hand, assuming that the energy source of the wind drive is gravitational energy,  the velocity of the wind is also described by $v_{\mathrm{wind}}\sim \sqrt{GM_{r} /r}$ \citep[e.g.,][]{1997ApJ...474..362K}.
Since $v_{\rm inf}$ and $v_{\rm wind}$ are similar in magnitude, the densities of the infalling and outflowing gas determine whether the ram pressure of the infalling or outflowing gas is greater. 
Thus, by comparing the infall rate $\dot{\Sigma}_{\rm inf}$ and the mass loss rate $\dot{\Sigma}_{\rm wind}$, we can determine whether  mass infall into the disk or mass loss from the disk occurs.
When $\dot{\Sigma}_{\mathrm{inf}} > \dot{\Sigma}_{\mathrm{wind}}$, the infall occurs without driving wind and we set $\dot{\Sigma}_{\mathrm{wind}} = 0$, and vice versa.

\subsection{Evolution of dust particle size}
\label{sec:method:dust_growth}

The size of dust particles changes due to coagulation and fragmentation caused by the collision of dust particles.
The time evolution of the surface density of dust with mass $m$ within the disk due to coagulation and fragmentation is described by the following equation \citep{1916ZPhy...17..557S},
\begin{align}
    & \left( \frac{\partial \Sigma_{\rm d}\left( m \right)}{\partial t} \right)_{\mathrm{coag/frag}} \notag \\ 
    & = \frac{1}{2} \int_{0}^{m} m \tilde{K}\left(m-m^{\prime}, m\right) N_{\rm d}\left(m-m^{\prime}\right) N_{\rm d} \left( m^{\prime} \right) \mathrm{d}m^{\prime} \notag \\ 
	  & - \int_{0}^{\infty} \tilde{K}\left( m , m^{\prime}\right) N_{\rm d} \left(m\right) N_{\rm d} \left(m^{\prime}\right) \mathrm{d}m^{\prime}  \notag \\
    & + \frac{1}{2}\int \int_{0}^{\infty} m \tilde{F} \left(m_{1}, m_{2}\right) N_{\rm d} \left(m_{1}\right) N_{\rm d} \left(m_{1}\right) \notag \\
    & \times \varphi_{\rm f}\left(m ; m_{1}, m_{2}\right) \mathrm{d}m_{1}\mathrm{d}m_{2} \notag \\ 
    & - \int_{0}^{\infty} m \tilde{F} \left(m,m_{1}\right) N_{\rm d} \left(m\right) N_{\rm d}\left(m_{1}\right) \mathrm{d}m_{1},
    \label{eq:coag-frag-equation}
\end{align}
where $N_{\rm d}(m) = \Sigma_{\rm d}/m$ is the vertically integrated number density of dust mass $m$, $\tilde{K}$ and $\tilde{F}$ are the vertical integrated coagulation and fragmentation kernels, and $\varphi_{f} \left(m; m_{1}, m_{2}\right)$ is the distribution function for fragments after a collision between $m_{1}$ and $m_{2}$ dust particles.
The kernels $\tilde{K}$ and $\tilde{F}$ are given by \citep{2022ApJ...935...35S}
\begin{equation}
	\tilde{K}\left( m, m^{\prime} \right) = \frac{K\left(m, m^{\prime}\right)}{\sqrt{2\pi \left[ h_{d}\left(m\right)^{2} + h_{d}\left( m^{\prime} \right)^{2} \right]}},
\end{equation}
\begin{equation}
	\tilde{F}\left( m, m^{\prime} \right) = \frac{F\left(m, m^{\prime}\right)}{\sqrt{2\pi \left[ h_{d}\left(m\right)^{2} + h_{d}\left( m^{\prime} \right)^{2} \right]}},
\end{equation}
where $K$ and $F$ are the coagulation and fragmentation kernels and $h_{\rm d} \left(m\right)$ is the dust scale height for dust mass $m$, described as \citep{2007Icar..192..588Y}
\begin{equation}
  h_{\mathrm{d}}\left(m\right) = h_{\mathrm{g}} \left( 1 + \frac{\mathrm{St}}{\alpha_{\rm turb}} \frac{1 + 2\mathrm{St}}{1 + \mathrm{St}} \right)^{-1/2}.
\end{equation}

To calculate the kernels, we assume that the relative velocities between dust particles follows the Maxwell--Boltzmann distribution \citep{2012A&A...540A..73W, 2022ApJ...935...35S},
\begin{equation}
  P \left(\Delta v ; v_{\rm rms}\right) = \sqrt{\frac{54}{\pi}} \frac{\Delta v^{2}}{v^{3}_{\mathrm{rms}}}\exp\left(-\frac{3 \Delta v^{2}}{2 v^{2}_{\mathrm{rms}}}\right),
\end{equation}
where $\Delta v$ is the relative velocity of the dust and $v_{\mathrm{rms}}$ is the rms velocity.
Using this velocity distribution, the coagulation and fragmentation kernels can then be expressed as follows,
\begin{align}
  \label{eq:coagulation_kernel}
  &K\left(m_{1}, m_{2}\right) \notag \\
  &= \sigma_{\rm coll} \sqrt{\frac{8\pi}{3}} v_{\mathrm{rel}, 12} \left[ 1 - \left( 1 + \frac{3 v_{\mathrm{frag}}^{2}}{2 v^{2}_{\mathrm{rel}, 12}} \right) \exp \left(-\frac{3 v_{\mathrm{frag}}^{2}}{2 v^{2}_{\mathrm{rel},12}}\right) \right],
\end{align}
\begin{align}
  \label{eq:fragmentation_kernel}
  &F\left(m_{1}, m_{2}\right) \notag \\
  & = \sigma_{\rm coll, 12} \sqrt{\frac{8\pi}{3}} v_{\mathrm{rel}, 12} \left( 1 + \frac{3 v_{\mathrm{frag}}^{2}}{2 v^{2}_{\mathrm{rel}, 12}} \right) \exp \left(-\frac{3 v_{\mathrm{frag}}^{2}}{2 v^{2}_{\mathrm{rel},12}}\right),
\end{align}
where $\sigma_{\rm coll, 12} = \pi \left(a_{1} + a_{2} \right)^{2}$ is the collisional cross section and $v_{\mathrm{frag}}$ is the lower velocity limit for fragmentation (see below).

The relative velocity among dust particles $v_{\mathrm{rel, 12}}$ is an important factor in determining the collision rate and the outcome of dust collisions. 
As the origin of the relative velocity between dust particles, we consider turbulence, thermal motion (or Brownian motion), radial drift, azimuthal drift, and vertical settling.
The relative velocities induced by the radial and azimuthal drift are calculated using (\ref{eq:dust_radial_velocity}) and (\ref{eq:dust_azimuthal_velocity}).
The relative velocities induced by turbulence, Brownian motion, and vertical settling are calculated using the prescription described in 
\citet{2007A&A...461..215O}, \citet{2021ApJ...907...80O} and \citet{2021ApJ...922...16K}.

In this study, we adopt the fragmentation model used in \citet{2022MNRAS.515.2072K}.
The fragmentation model is constructed based on numerical dust collision calculations \citep{2013A&A...559A..62W}.
The distribution function for fragments $\varphi_{\rm f}$ is the same as the formula in \citet{2022MNRAS.515.2072K}.

In the fragmentation model, the velocity $v_{\rm col, crit}$ required for most of the dust particles to fragment after the collision is a key physical quantity.
Even at velocities below this threshold, partial fragmentation of dust particles occurs \citep{2009ApJ...702.1490W, 2013A&A...559A..62W}.
Therefore, in this study, we set the lower limit velocity for fragmentation as $v_{\mathrm{frag}} = 0.2\, v_{\mathrm{col, crit}}$.
The value of $v_{\mathrm{col, crit}}$ depends on the composition and structure of the dust particles.
The typical value of $v_{\mathrm{col, crit}}$ for silicate dust particles has been estimated to be in the range $1$--$10\,\mathrm{m\,s^{-1}}$ through laboratory experiments and numerical simulations.
For water-ice dust particles, the typical value of $v_{\mathrm{col, crit}}$ is considered to be higher than that for silicate dust particles. 
In this study, we assume that in regions with temperatures higher than 160 K, the composition of dust particles is silicate, while in regions with temperatures lower than 160 K, the composition of dust particles is water ice.
Although $v_{\mathrm{col, crit}}$ may depend on the mass ratio for colliding dust, for simplicity, we assume that it does not depend on the mass ratio and use the following values,
\begin{equation}
	v_{\mathrm{col,crit}} = \left\{ 
		\begin{aligned}
			& 1\,\mathrm{m\,s^{-1}} \ \ \  \left(T > 160\,\mathrm{K,} \  \mathrm{silicate} \right) \\
			& 10\,\mathrm{m\,s^{-1}} \ \left( T < 160\,\mathrm{K,} \ \mathrm{H_{2}O \ ice} \right).
		\end{aligned}
	\right.
\end{equation}

\subsection{Dust evaporation and condensation}
\label{sec:method:dust_evaporation_condensation}

When the temperature becomes high, dust particles can completely evaporate. 
In this study, vapor components evaporated from dust particles are treated separately from the disk gas. 
We assume that the dust particles evaporate when the disk temperature is higher than $1500 \ \mathrm{K}$.
On the other hand, when the disk temperature decreases to below 1500 K, vapor condenses into dust particles.
The size of the condensed dust particles is assumed to be $a_{0} = 0.1\,\mathrm{\mu\,m}$.
The surface density $\Sigma_{\mathrm{v}}$ of vapor follows the equation
\begin{equation}
	\frac{\partial \Sigma_{\mathrm{v}}}{\partial t} + \frac{1}{r} \frac{\partial}{\partial r} \left[ r\Sigma_{\mathrm{v}} v_{\mathrm{v}, r} \right]
	= \frac{1}{r} \frac{\partial}{\partial r} \left[ r D_{\mathrm{v}} \Sigma_{\mathrm{g}} \frac{\partial}{\partial r} \left( \frac{\Sigma_{\mathrm{v}}}{\Sigma_{\mathrm{g}}} \right) \right]
  - \dot{\Sigma}_{\mathrm{v, wind}},
  \label{eq:vapor_surface_equation}
\end{equation}
where $v_{\mathrm{v}, r}$ is the vapor velocity and is set to be the same as $v_{\mathrm{g}, r}$, $D_{\mathrm{v}} =  \alpha_{\mathrm{turb}} c_{s} h_{\mathrm{g}}$ is the turbulent diffusion coefficient of vapor,
and $\dot{\Sigma}_{\mathrm{v, wind}}$ represents the mass loss of vapor due to the disk wind,
\begin{equation}
  \dot{\Sigma}_{\mathrm{v, wind}} = \frac{3\alpha_{\rm DW}\Sigma_{\rm v} c_{s}^{2}}{4 j \left(\lambda - 1\right)}.
\end{equation}

\subsection{Numerical settings}
\label{sec:method:setup}

We set the computational domain to be from $0.3 \ \mathrm{au}$ to $10^{3} \ \mathrm{au}$ and divide the region into 200 logarithmic divisions.
We solve the differential equations (\ref{eq:gas_surface_evolution}), (\ref{eq:dust_surface_evolution}), (\ref{eq:energy_equation}), and (\ref{eq:vapor_surface_equation}) explicitly. 
To solve the coagulation-fragmentation equation (\ref{eq:coag-frag-equation}), we use the bin method of \citet{2021MNRAS.504.5588K}.
The computational domain of dust radius is in the range $a = 10^{-5}$--$10^{5} \ \mathrm{cm}$, and this range is divided into uniform logarithmic intervals with a resolution of ten bins per decade.
The internal density of dust particles is assumed to be a constant value of $\rho_{\mathrm{di}} = 2.65 \ \mathrm{g \ cm^{-3}}$, although it depends on the composition and structure of the dust particles.
We adopt zero-gradient boundary conditions at the inner boundary and impose the condition that disk material flows out from the boundary only if $v_{\mathrm{g}, r} (v_{\mathrm{d}, r}) < 0$.
We also set the flux at the outer boundary to be zero.
The central density and angular velocity of the molecular cloud core are listed in Table~\ref{table:parameters}.
We start our calculations with a protostellar mass of $M_{\mathrm{star, ini} }= 0.01 \ M_{\sun}$ without a disk.
We set the end of the calculations to $3\times 10^{5}$ years.
We perform calculations with and without the disk wind to investigate the effect of the disk wind on disk evolution.
In the calculations with the disk wind, we set the parameters $\alpha_{\mathrm{\rm DW}}=10^{-2}$ and $\lambda =2$.
The parameters used in this study are summarized in Table~\ref{table:parameters}.

\begin{table*}
  \centering
  \caption{Parameters used in this study.}
  \begin{tabular}{ccc} 
    \hline
    $M_{\mathrm{core}}$      & $1.0 \ M_{\sun}$                       & total mass of cloud core \\ 
    $\Omega_{\mathrm{core}}$ & $2.0\times 10^{-14} \ \mathrm{s^{-1}}$ & angular velocity of cloud core \\ 
    $T_{\mathrm{core}}$      & $10 \ \mathrm{K}$                      & temperature of cloud core \\ 
    $n_{0}$                  & $3\times 10^{5} \ \mathrm{cm^{-3}}$    & central density of cloud core \\
    $f$                      & $1.4$                                  & density enhancement factor of cloud core \\
    $M_{\mathrm{star, ini}}$ & $0.01 \ M_{\sun}$                       & initial stellar mass \\ 

    \hline 
    $T_{\mathrm{MRI}}$         & $1000 \ \mathrm{K}$ & MRI activate temperature \\ 
    $\alpha_{\mathrm{dead}}$   & $10^{-4}$           & alpha parameter for dead zone  \\ 
    $\alpha_{\mathrm{active}}$ & $10^{-2}$           & alpha parameter for MRI active region  \\ 

    \hline
    $\alpha_{\rm DW}$ & $10^{-2}$ & dimensionless parameter for disk wind \\
    $\lambda$     & $2$       & magnetic lever arm parameter \\ 

    \hline
  \end{tabular}
  \label{table:parameters}
\end{table*}

\section{Results}
\label{sec:result}

In this section, we first describe the disk evolution without the disk wind and then present the disk evolution with the disk wind.
Finally, we compare the results between the two cases.

\subsection{Case without disk wind}

\begin{figure}
    \centering
    \includegraphics[width=\columnwidth]{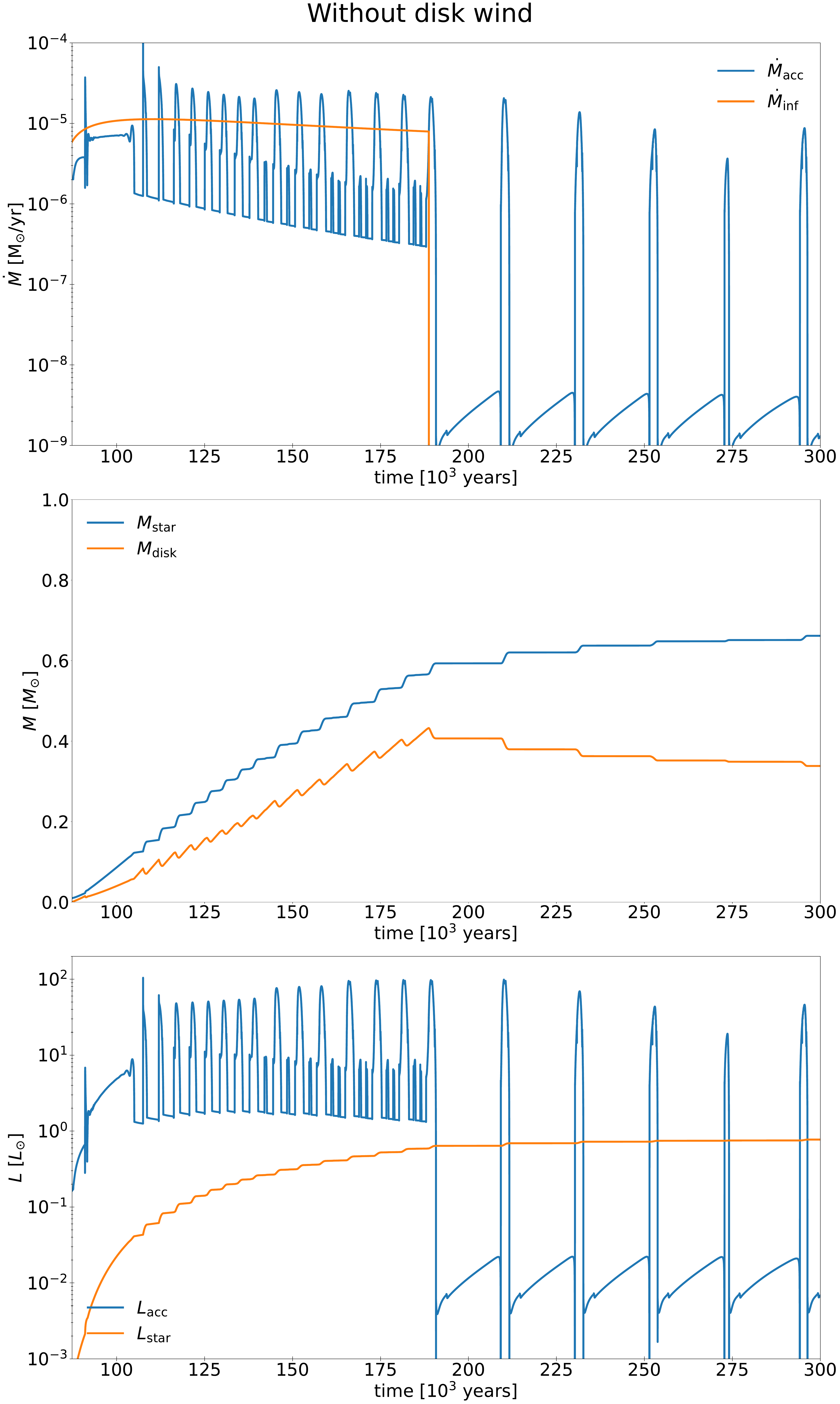}
    \caption{Time evolution of physical quantities in case without disk wind. 
     Panels show the mass accretion rate from the disk to the star $\dot{M}_{\mathrm{acc}}$ and mass infall rate from the infalling envelope to the disk $\dot{M}_{\mathrm{inf}}$ (top);
     Disk $M_{\mathrm{disk}}$ and central stellar $M_{\mathrm{star}}$ masses (middle);
    Stellar luminosity $L_{\star}$ and accretion luminosity $L_{\mathrm{acc}}$ (bottom).}
\label{fig:mdot_nowind}
\end{figure}

This subsection describes the results of gas disk evolution without the disk wind. 
The top panel of Figure~\ref{fig:mdot_nowind} shows the time evolution of the mass infall rate to the entire star--disk system from the core (or infalling envelope) $\dot{M}_{\mathrm{inf}}$ (equation~\ref{eq:mdot_inf}) and the mass accretion rate from the disk to the central star $\dot{M}_{\mathrm{acc}}$, given by
\begin{equation}
  \dot{M}_{\mathrm{acc}} = 
    \left[ -2 \pi \Sigma_{\mathrm{g}}v_{\mathrm{g}, r} 
    - \int 2 \pi \Sigma_{\mathrm{d}}\left( m \right) v_{\mathrm{d}, r} \mathrm{d}m
    -2 \pi \Sigma_{\mathrm{v}} v_{\mathrm{v}, r} 
  \right]_{r = r_{\mathrm{in}}},
\end{equation}
where $r_{\mathrm{in}}$ is the inner disk radius.
We define the period during which mass infall from the cloud core (or the infalling envelope) occurs as the infall phase.
The infall phase lasts until $t \simeq 1.89\times 10^{5} \ \mathrm{year}$ in our setting for the cloud core.
The infall rate in this phase is as high as $\dot{M}_{\mathrm{inf}} \simeq 10^{-5} \ \mathrm{M_{\sun}} \mathrm{yr^{-1}}$.
After the infall phase ($t \gtrsim 1.89\times 10^{5} \ \mathrm{years}$),  the mass infall rate is zero because all the initial core mass has reached the central region.

The mass accretion from the disk to the star $\dot{M}_{\mathrm{acc}}$ exhibits an episodic behavior. 
We call the local peaks of the mass accretion rate outbursts. 
During the infall phase, outbursts have a mass accretion rate of $\dot{M}_{\mathrm{acc}} \simeq 2\times 10^{-5} \ \mathrm{M_{\sun}} \ \mathrm{yr^{-1}}$. 
This accretion rate is larger than the mass infall rate $\dot{M}_{\mathrm{inf}}$.
We confirm that the accretion bursts are thermally driven as proposed by \citet{2013ApJ...764..141B,2014ApJ...795...61B}.

The middle panel of Figure~\ref{fig:mdot_nowind} plots the time evolution of the masses of the disk and the central star.
The masses at the end of the infall phase and the end of the calculation are described in Table~\ref{table:mass_at_end_infall_phase}.
During the infall phase, the disk and stellar masses increase due to the high mass infall and accretion rates.
At the end of the infall phase, the masses of the disk and the central star are $M_{\rm disk} \simeq 0.43 \ \mathrm{M_{\sun}}$ and $M_{\star} \simeq 0.57 \ \mathrm{M_{\sun}}$, respectively.
After the infall phase, the disk mass slowly decreases because no mass reservoir exists outside the disk.
Even after the infall phase, the stellar mass slowly increases due to mass accretion from the disk. 
At the end of the calculation time ($3.0\times 10^{5}$ years), the disk and stellar masses are $M_{\mathrm{disk}} \simeq 0.34 \ \mathrm{M_{\sun}}$ and $M_{\mathrm{star}} \simeq 0.66 \ \mathrm{M_{\sun}}$, respectively.

The bottom panel of Figure~\ref{fig:mdot_nowind} shows the time evolution of the stellar luminosity $L_{\star}$ and the accretion luminosity $L_{\mathrm{acc}}$.
The accretion luminosity shows time variations caused by episodic mass accretions.
During the infall phase, the accretion luminosity is always higher than the stellar luminosity.
After the infall phase, the accretion luminosity decreases as the mass accretion rate to the central star decreases.
However, the accretion luminosity at the peak (outburst epoch) is still higher than the stellar internal luminosity after the infall phase.

\begin{table*}
    \caption{Masses of protostar, disk, and disk wind at end of infall phase and at end of calculation.}
    \begin{tabular}{ccccccc}
      \hline
      \multicolumn{1}{c}{} & \multicolumn{3}{c}{End of infall phase} & \multicolumn{3}{c}{End of calculation} \\
      & $M_{\mathrm{star}} [M_{\sun}]$ & $M_{\mathrm{disk}} [M_{\sun}]$ & $M_{\mathrm{wind}} [M_{\sun}]$ 
           & $M_{\mathrm{star}} [M_{\sun}]$ & $M_{\mathrm{disk}} [M_{\sun}]$ & $M_{\mathrm{wind}} [M_{\sun}]$ \\
      \hline
      Without disk wind  & 0.57 & 0.43 & $-$  & 0.66 & 0.34 & $-$  \\
      With disk wind     & 0.31 & 0.31 & 0.39 & 0.34 & 0.17 & 0.49 \\
      \hline
    \end{tabular}
    \label{table:mass_at_end_infall_phase}
\end{table*}

\subsubsection{Gas evolution without disk wind}

\begin{figure*}
    \centering
    \includegraphics[width=160mm]{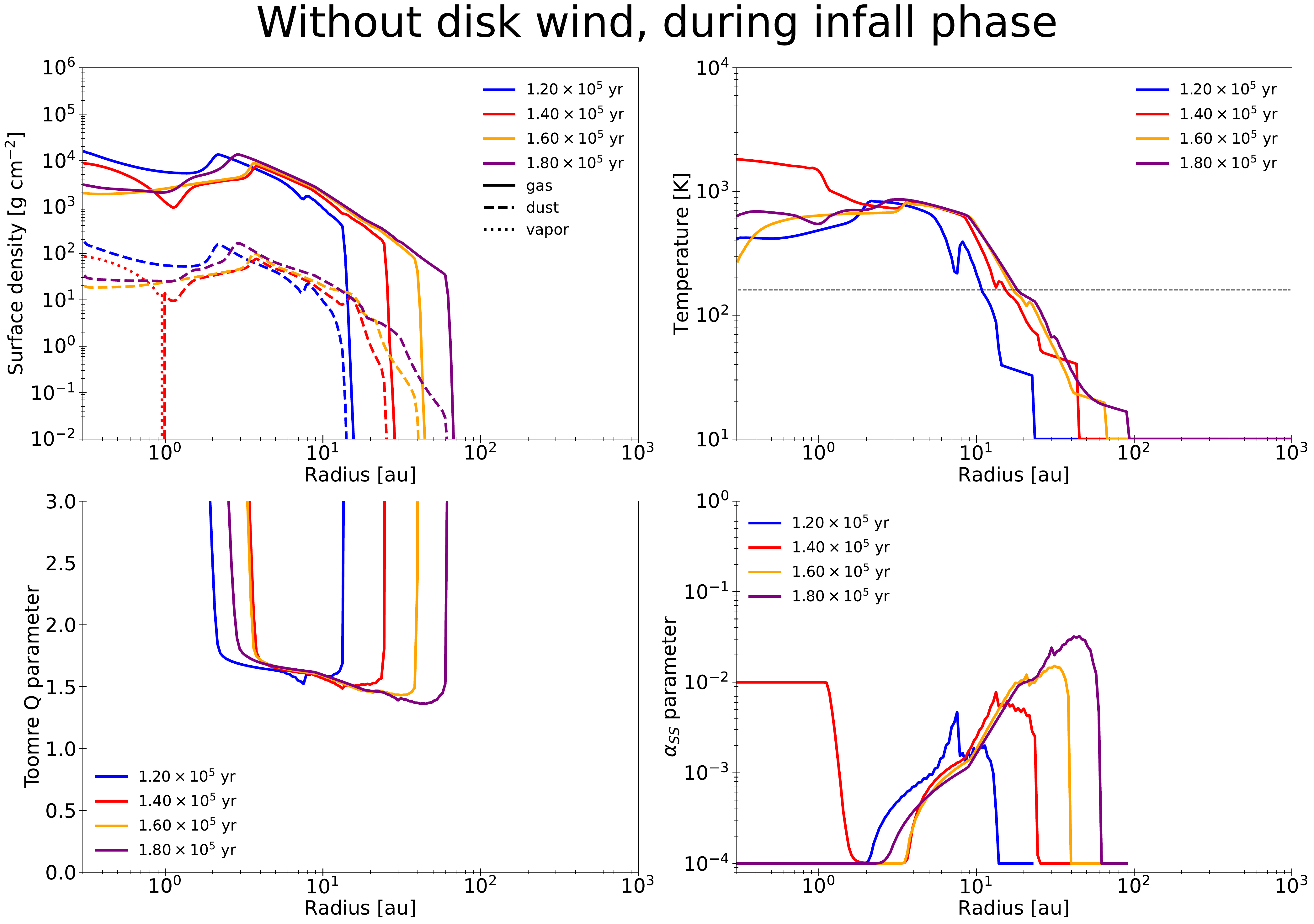}
    \caption{
    Time evolution of radial profiles during the infall phase for the case without disk wind. 
    Panels show the surface density of gas (solid), dust (dashed), and vapor (dotted) (upper left), temperature (upper right), 
    Toomre Q parameter (lower left), and $\alpha$ parameter (lower right).
    Lines in different colors represent different epochs in each panel.
    The horizontal dotted line in the upper right panel represents T=160\,K above which the H$_2$O ice evaporates.
    }
\label{fig:disk_evol_nowind1}
\end{figure*}

Figure~\ref{fig:disk_evol_nowind1} shows the radial profile of the surface density, temperature, Toomre Q parameter, and viscous alpha parameter $\alpha_{\rm SS}$ at four different epochs during the infall phase in the case without the disk wind.
In the upper left panel of Figure~\ref{fig:disk_evol_nowind1}, the gas surface density is shown as solid lines, dust as dashed lines, and vapor as dotted lines.
In the upper right panel, we plot $T = 160 \ \mathrm{K}$ as a horizontal dashed line.
As described in Section~\ref{sec:method:dust_growth}, dust particles are regarded as silicate particles in the disk regions where the temperature is higher than $160 \ \mathrm{T}$,
while in the disk region where $T > 160 \ \mathrm{K}$, dust particles exist as icy particles.
Thus, the location of $T = 160 \ \mathrm{K}$ represents the $\mathrm{H_{2}O}$ snowline.
This difference in the composition of the dust particles affects the dust growth. 
The results of the detailed dust evolution are presented in Section~\ref{sec:results:nowind:dust_disk_evolution}.

The upper left of Figure~\ref{fig:disk_evol_nowind1} shows that the gas surface density in the range $\gtrsim 3 \ \mathrm{au}$ increases with time during the infall phase.
The gas surface density becomes very high due to the mass supply from the infalling envelope.
The gas disk is then unstable against gravitational instability.
The lower left panel of Figure~\ref{fig:disk_evol_nowind1} indicates that the Toomre Q parameter becomes $Q \lesssim 1.5$ for the outer disk region $(\gtrsim 3 \ \mathrm{au})$.
When gravitational instability develops in the disk, angular momentum is transported by the gravitational torque.
The lower right panel of Figure~\ref{fig:disk_evol_nowind1} shows that the viscous alpha parameter $\alpha_{\rm SS}$ becomes high for the outer disk region where a gravitationally unstable state $Q \lesssim 1.5$ is realized. 
Thus, the gas disk expands due to gravitational torque.

The upper right panel of Figure~\ref{fig:disk_evol_nowind1} shows that the disk midplane temperature in the range $\gtrsim 3 \ \mathrm{au}$ increases with time during the infall phase.
When the gas surface density is high, viscous heating becomes the dominant heating source of the disk,
increasing the disk temperature as the gas surface density increases.

The upper panels of Figure~\ref{fig:disk_evol_nowind1} show that the surface density and temperature in the inner disk region $(r \lesssim 3 \ \mathrm{au})$ vary with time.
The fluctuation in surface density and temperature is attributed to outbursts \citep{2013ApJ...764..141B,2014ApJ...795...61B}.

\begin{figure*}
    \centering
    \includegraphics[width=160mm]{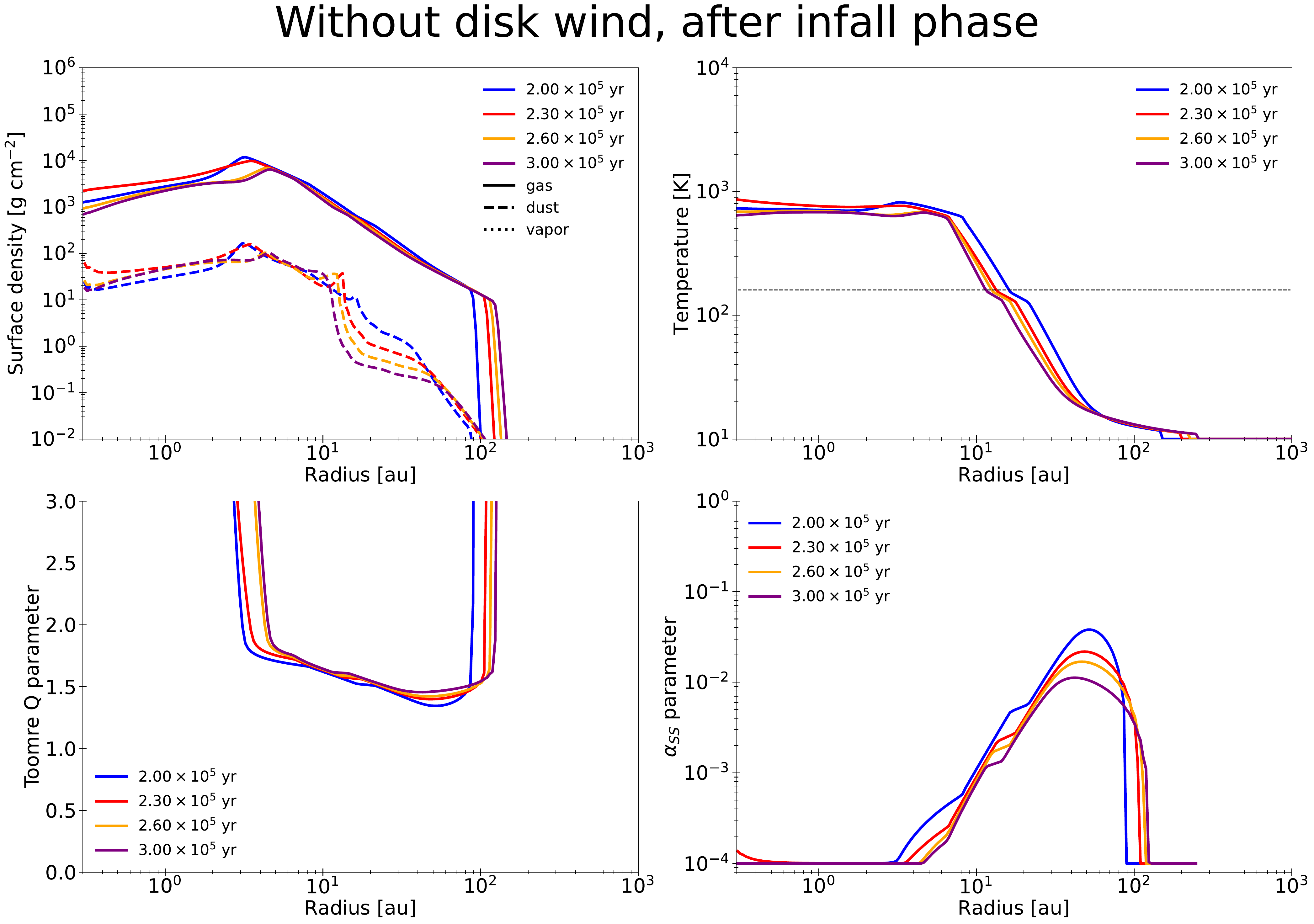}
    \caption{
    Same as Figure~\ref{fig:disk_evol_nowind1}, but for after the infall phase.
    }
\label{fig:disk_evol_nowind2}
\end{figure*}

Figure~\ref{fig:disk_evol_nowind2} shows the radial profile of the surface density, temperature, Toomre Q parameter, and viscous alpha parameter $\alpha_{\rm SS}$ at four different epochs after the infall phase.
The upper left panel of Figure~\ref{fig:disk_evol_nowind2} shows that the gas surface density slightly decreases since there is no mass supply from the infalling envelope.
As shown in the lower panels of Figure~\ref{fig:disk_evol_nowind2}, the outer disk region is still gravitationally unstable, $Q\sim 1.5$--$2$, and the gravitational torque primarily transports the angular momentum. 
As a result, the disk continues to expand after the infall phase.
This implies that the disk mass remains relatively large for $\sim 10^{5}$ years after the infall phase in the case without the disk wind.

The upper right panel of Figure~\ref{fig:disk_evol_nowind2} shows the temperature at different times after the infall phase.
The viscous heating decreases as the gas surface density decreases. 
Thus, the temperature decreases and the snowline moves inward with time.

\subsubsection{Dust evolution without disk wind}
\label{sec:results:nowind:dust_disk_evolution}
As shown in the upper left panel of Figure~\ref{fig:disk_evol_nowind1}, during the infall phase, the dust surface density has a similar profile to the gas surface density, except for the outer disk region. 

When the temperature exceeds $T > 1500 \ \mathrm{K}$, the dust particles evaporate and the vapor increases.
When the temperature is below $1500 \ \mathrm{K}$, the vapor condenses onto the dust particles.
This evaporation and condensation occurs repeatedly due to outbursts in the inner disk region. 
In the upper left panel of Figure~\ref{fig:disk_evol_nowind1}, the vapor is generated within $\lesssim 1 \ \mathrm{au}$ at $t = 1.4\times 10^{5}$ year.

The upper left panel of Figure~\ref{fig:disk_evol_nowind1} shows the dust surface density, exhibiting steeper decreasing trends in the range $\gtrsim 20 \ \mathrm{au}$ compared to the gas surface density.
In the outer disk regions, the dust particles are (marginally) decoupled from the gas particles because of dust particle growth by collisional coagulation and the relatively low gas density.
We confirm that the Stokes number of the grown dust particles with $a \sim 1$--$10$ cm is $\mathrm{St} \sim 0.01$.
Thus, the grown dust particles drift inward and the dust surface density decreases.

\begin{figure*}
  \centering
  \includegraphics[width=160mm]{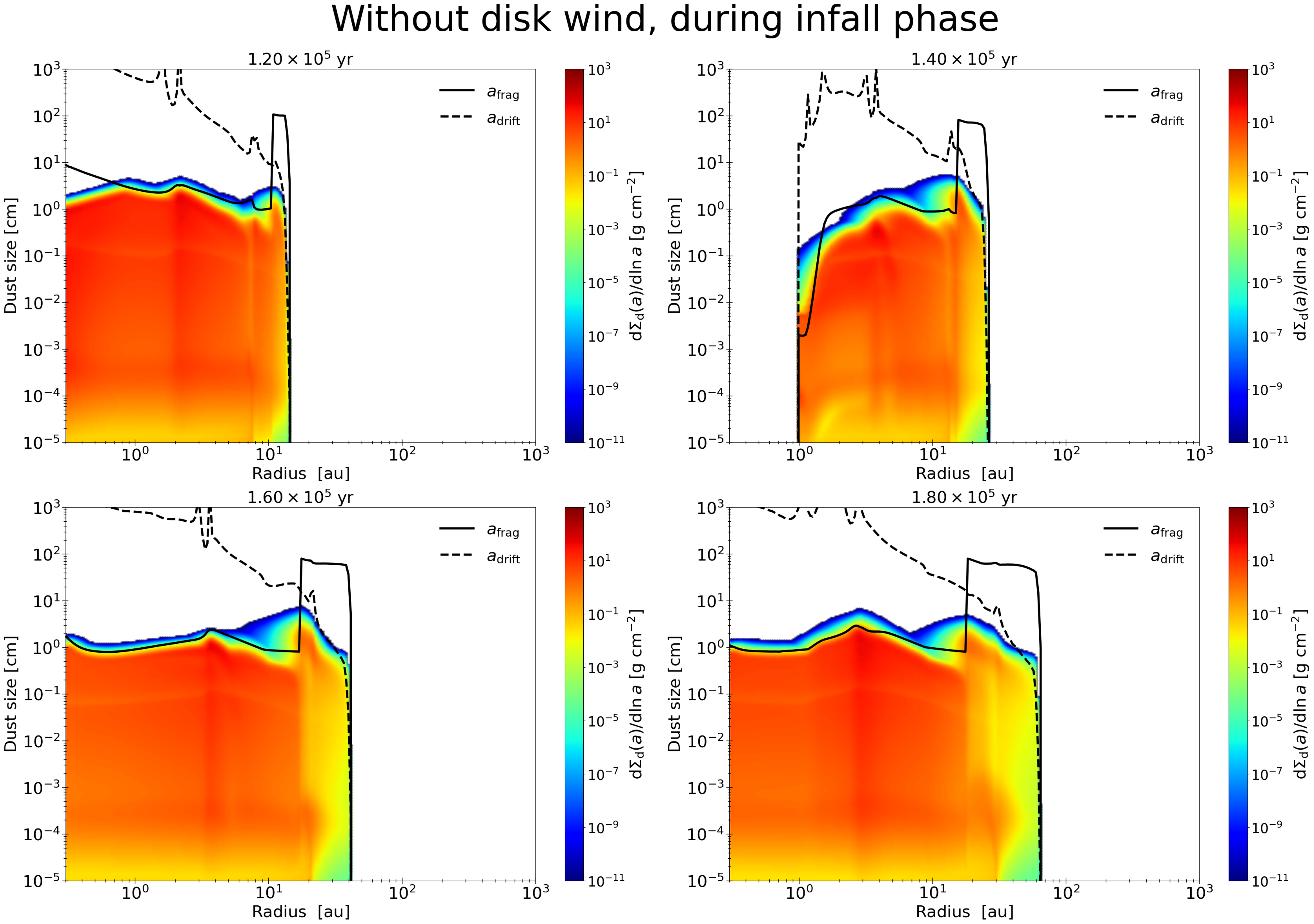}
  \caption{
  Time evolution of dust particle size distribution during infall phase for the case without disk wind. 
  In each panel, dust size distribution is plotted against the radius. 
  The time at each snapshot is indicated at the top part of each panel.  
  The black solid ($a_{\rm frag}$) and dashed ($a_{\rm drift}$) lines represent the dust particle size determined by the fragmentation and radial drift, respectively.
  }
\label{fig:dust_size_distribution_nowind1}
\end{figure*}

Figure~\ref{fig:dust_size_distribution_nowind1} shows the dust particle size distribution at four different epochs during the infall phase in the case without the disk wind.
In the figure, the black solid line shows the dust particle size limited by the collisional fragmentation (the fragmentation barrier) \citep{2012A&A...539A.148B},
\begin{equation}
  a_{\mathrm{frag}} = f_{f} \frac{2}{3\pi} \frac{\Sigma_{\mathrm{g}}}{\rho_{\mathrm{di}} \alpha_{\mathrm{turb}}} \frac{v_{\mathrm{col, crit}}^{2}}{c_{s}^{2}},
\end{equation}
where $f_{f} = 0.77$ is the factor that adjusts the representative size for the largest dust particles in a fragmentation-dominated size distribution \citep{2012A&A...539A.148B}, and the black dashed line shows the dust particle size limited by the dust radial drift (the dust drift barrier) \citep{2012A&A...539A.148B},
\begin{equation}
  a_{\mathrm{drift}} = f_{d} \frac{2\Sigma_{\mathrm{d}}}{\pi \rho_{\mathrm{di}}} \frac{r^{2}\Omega^{2}}{c_{s}^{2}} \left| \frac{\mathrm{d} \ln P}{\mathrm{d} \ln r} \right|^{-1},
\end{equation}
where $f_{d} = 0.55$ is the numerical factor that adjusts the representative size for the largest dust particles limited by the dust drift. 
The region where $a_{\mathrm{frag}}$ increases in a step-function corresponds to the snowline.
Outside the snowline,  $\mathrm{H_{2}O}$ ice particles can be present.
The fragmentation velocity for $\mathrm{H_{2}O}$ ice dust particles is larger than that for the silicate dust particles.
Thus, $a_{\mathrm{frag}}$ outside the snowline is larger than that inside the snowline.
The dust particle can grow up to $a_{\mathrm{frag}}$ as long as they do not suffer radial drift.

Figure~\ref{fig:dust_size_distribution_nowind1} shows that the maximum size of dust particles is determined by collisional fragmentation inside the snowline and is about $a\sim 1$--$10 \ \mathrm{cm}$.
Inside the snowline, the dust particles cannot grow large enough because the fragmentation velocity for silicate dust particles is low.
Outside the snowline, the maximum size of the dust particles is limited by the radial drift.
The line of $a_{\mathrm{frag}}$ in Figure~\ref{fig:dust_size_distribution_nowind1} indicates that the dust particles can grow to a maximum size of $\sim 10^{2} \ \mathrm{cm}$ outside the snowline.
However, the dust particles move inward due to radial drift before they grow to $a_{\mathrm{drift}}$.
Thus, the dust density outside the snowline decreases with time due to the radial drift of the dust particles undergoing growth (see also Figure~\ref{fig:disk_evol_nowind1}).

\begin{figure}
  \includegraphics[width=\columnwidth]{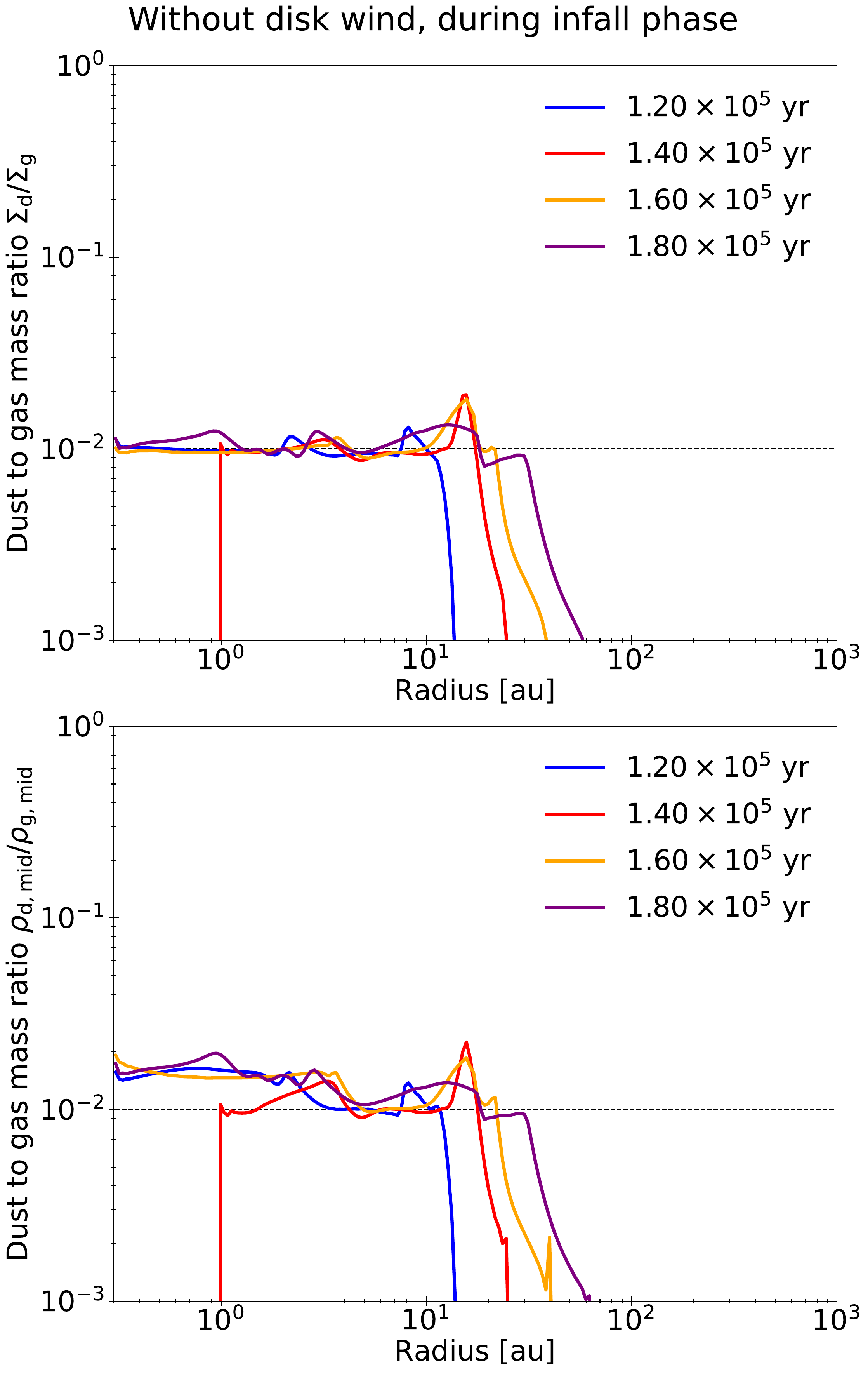}
  \caption{Time evolution of dust-to-gas mass ratio in case without disk wind during infall phase. In the upper panel, the dust-to-gas mass ratios are calculated using the surface densities of gas and dust.
  In the lower panel, the dust-to-gas mass ratio is calculated using the mass density at the disk midplane.
  The dotted line plotted in top and bottom panel corresponds to $\Sigma_{\rm d}/\Sigma_{\rm g}=0.01$ and $\rho_{\rm d,mid}/\rho_{\rm g.mid}=0.01$, respectively.
  }
\label{fig:fdg_nowind1}
\end{figure}

Figure~\ref{fig:fdg_nowind1} shows the dust-to-gas mass ratios at different times during the infall phase.
The upper panel of Figure~\ref{fig:fdg_nowind1} shows the dust-to-gas mass ratios calculated using the surface densities of gas and dust.
In this study, the dust-to-gas mass ratio for the cloud core is set to $0.01$.
During the infall phase, the dust-to-gas surface mass density ratio does not change significantly in the range $\lesssim 20 \ \mathrm{au}$ and maintains a value of $\sim 0.01$.
On the other hand, in the outer disk region $\gtrsim 20 \ \mathrm{au}$, the mass ratios decrease compared to the initial mass ratio of $0.01$ due to the radial drift of the dust particles, as described above.
In the vicinity of the snowline, the mass ratios fluctuate slightly due to the accumulation of dust particles across the snowline.

In the lower panel of Figure~\ref{fig:fdg_nowind1}, the dust-to-gas mass ratios are calculated using the mass densities at the disk midplane of the gas and dust.
Compared to the surface density mass ratio, the mass ratio at the disk midplane increases from the initial value of $0.01$ for the initial cloud core, especially inside the snowline.
This is due to the settling of the growth dust particles onto the disk midplane.

The upper left panel of Figure~\ref{fig:disk_evol_nowind2} shows the time evolution of the dust density after the infall phase (dashed lines) and the vapor surface density (dotted lines).
Note that at the time plotted in this figure, the vapor is not present because the temperature is below $1500 \ \mathrm{K}$ (right panel of Figure~\ref{fig:disk_evol_nowind2}).
In the disk region inside a radius of 10 au, the radial profile of the dust surface density does not change significantly for $\sim 10^{5}$ years after the infall phase.
On the other hand, in the disk region outside 10 au, the dust surface density decreases with time due to radial drift of the grown dust particles.

\begin{figure*}
  \centering
  \includegraphics[width=160mm]{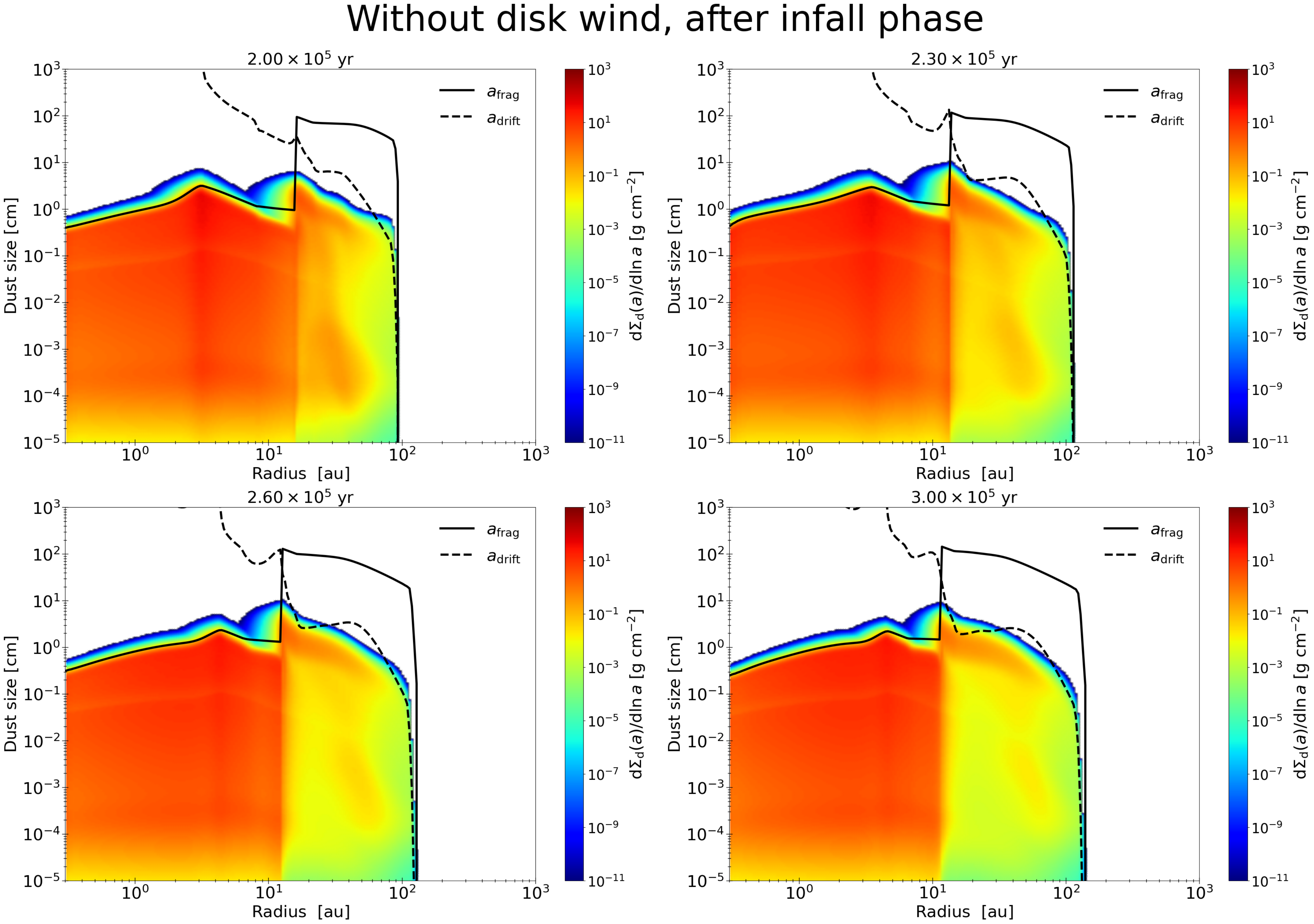}
  \caption{
Same as Figure~\ref{fig:dust_size_distribution_nowind1}, but for after the infall phase.}
\label{fig:dust_size_distribution_nowind2}
\end{figure*}

Figure~\ref{fig:dust_size_distribution_nowind2} shows the time evolution of the dust particle size distribution after the infall phase in the case without the disk wind.
As in the infall phase, the maximum size of dust particles inside the snowline is limited by fragmentation, while outside the snowline, the maximum size of the dust particles is determined by the radial drift.
After the infall phase, since the supply of small dust particles from the infalling envelope halts, small dust particles outside the snowline are depleted over time due to coagulation.
Dust particles that grow outside the snowline drift inward and cross the snowline.
Inside the snowline, the maximum dust particle size is limited by fragmentation and dust radial drift is not effective.

\begin{figure}
  \includegraphics[width=\columnwidth]{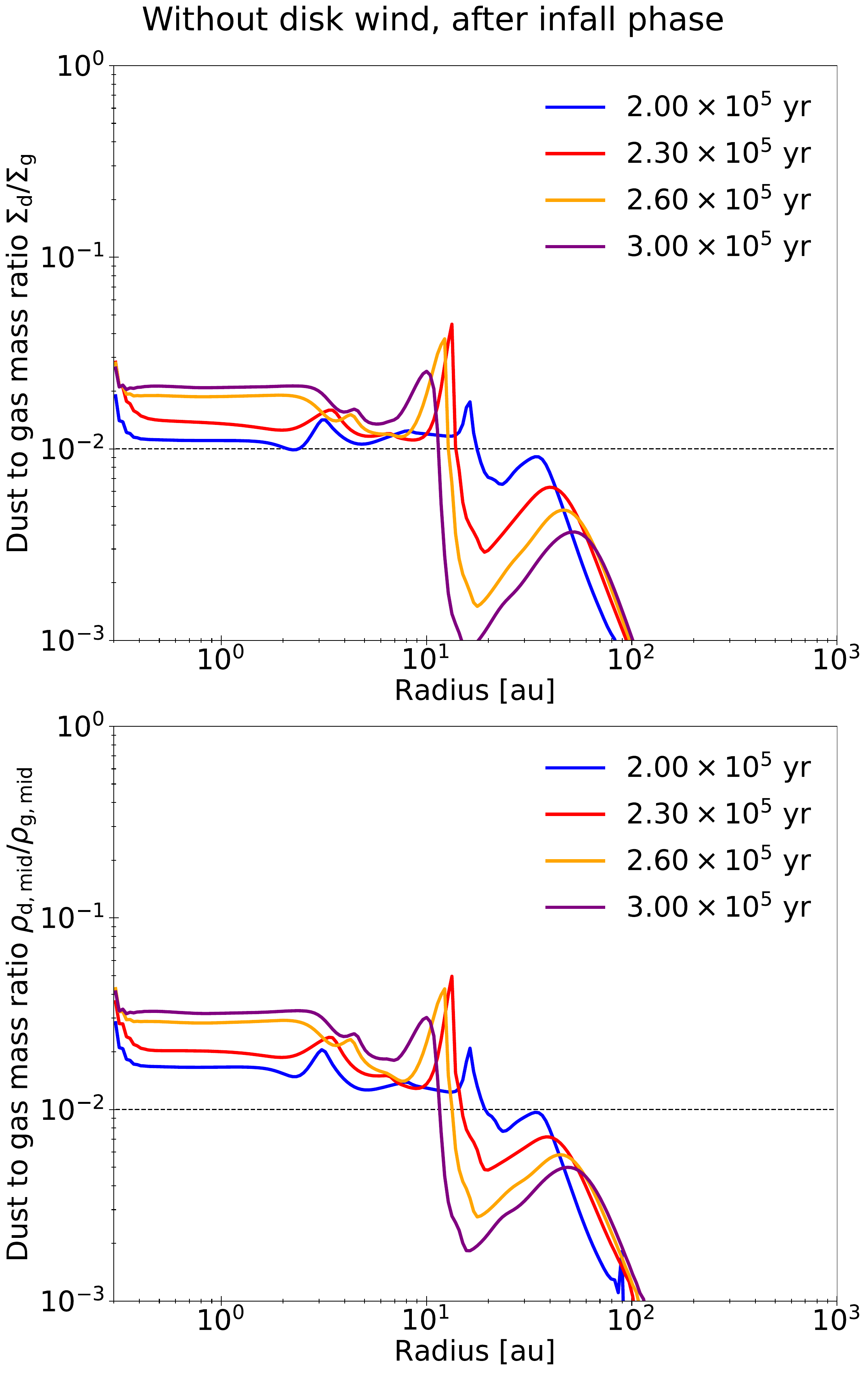}
  \caption{
  Same as Figure~\ref{fig:fdg_nowind1}, but for after the infall phase. 
  }
\label{fig:fdg_nowind2}
\end{figure}

Figure~\ref{fig:fdg_nowind2} shows the dust-to-gas mass ratio calculated using the surface density (upper panel) and the mass density at the disk midplane (lower panel) after the infall phase.
After the infall phase, the dust-to-gas mass ratio outside the snowline decreases with time due to the strong radial drift of the dust particles as described above.
In the vicinity of the snowline, the mass ratio exhibits a peak as a result of the accumulation of dust particles that radially drift from outside the snowline.
The mass ratio inside the snowline increases with time. 
At the end of the calculation, in the disk region within a radius of 3 au, the dust-to-gas mass ratio is doubled from the initial value (0.01) when calculated using the surface density, and increases by a factor of three when calculated using the midplane mass density.

\subsection{Case with disk wind}
\label{sec:results:wind}

\begin{figure}
  \centering
  \includegraphics[width=\columnwidth]{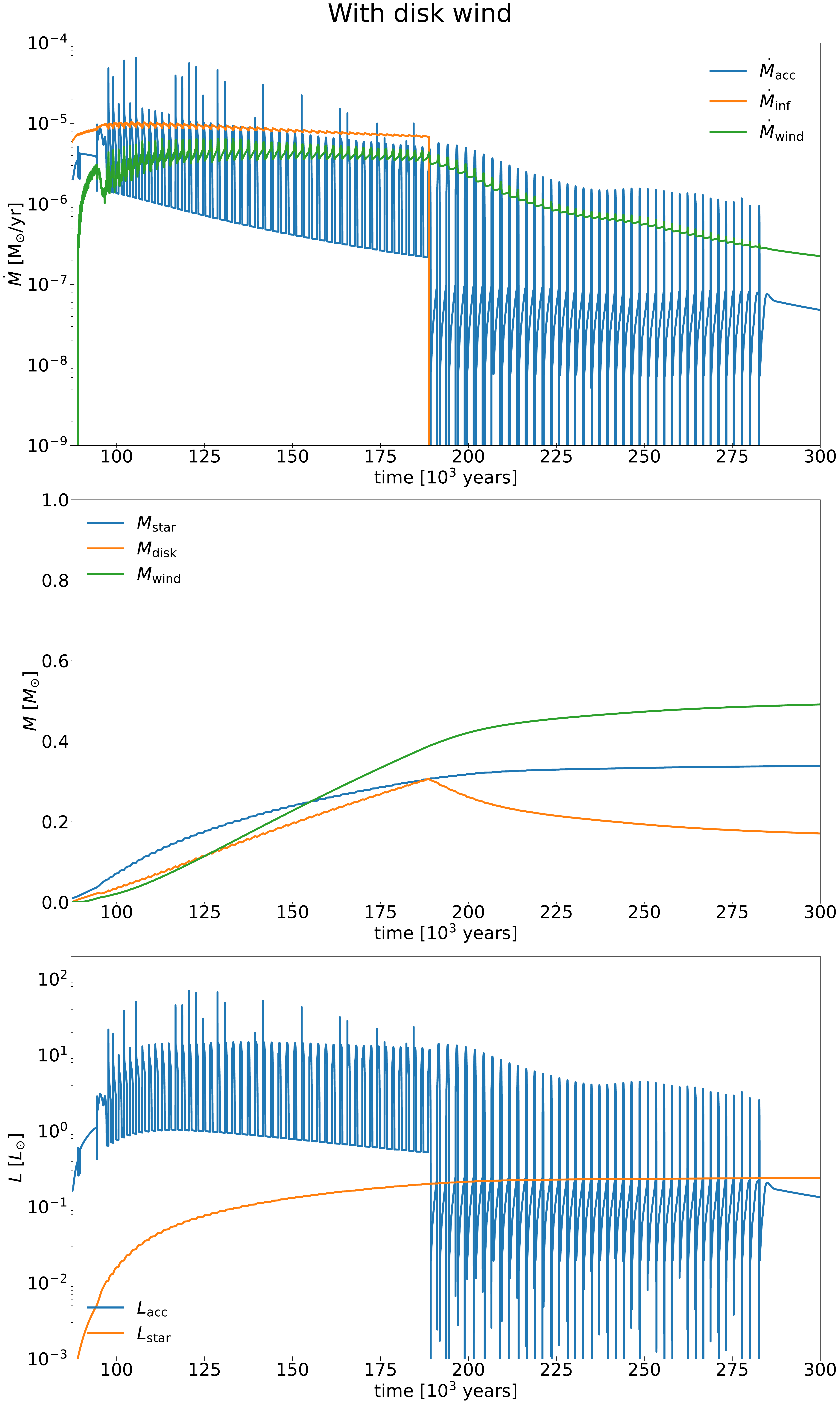}
  \caption{
  Time evolution of physical quantities in case with disk wind. 
  (Top) Mass accretion rate from the disk to the star $\dot{M}_{\mathrm{acc}}$, mass infall rate from the infalling envelope to the disk $\dot{M}_{\mathrm{inf}}$, and mass loss rate from the disk by the disk wind $\dot{M}_{\mathrm{wind}}$. 
  (Middle) Masses of the disk $M_{\mathrm{disk}}$, central star  $M_{\mathrm{star}}$,  and disk wind $M_{\mathrm{wind}}$. 
  (Bottom) Stellar internal luminosity $L_{\star}$ and accretion luminosity $L_{\mathrm{acc}}$.}
\label{fig:mdot_wind}
\end{figure}

This subsection describes the results for disk evolution with the disk wind.
The top panel of Figure~\ref{fig:mdot_wind} shows the time evolution of the mass infall rate from the infalling envelope to the disk  $\dot{M}_{\mathrm{inf}}$, the mass accretion rate from the disk to the star $\dot{M}_{\mathrm{acc}}$, and the mass loss rate from the disk by the disk wind $\dot{M}_{\mathrm{wind}}$.
In the case with disk wind, as described in \S~\ref{sec:method:infall_and_disk_wind_model}, when the disk wind dominants the infall rate ($\dot{\Sigma}_{\mathrm{wind}} > \dot{\Sigma}_{\mathrm{inf}}$), the infalling material is blown away by the disk wind.
Thus, $\dot{M}_{\mathrm{inf}}$ is calculated by subtracting the portion of infalling material blown away by the disk wind from the total mass of infalling material into the star-disk system (equation~(\ref{eq:mdot_inf})).
The mass loss rate from the disk by the disk wind $\dot{M}_{\mathrm{wind}}$ is given by
\begin{align}
  \dot{M}_{\mathrm{wind}} 
  & = \int_{r_{\mathrm{in}}}^{r_{\mathrm{out}}}  2\pi r \dot{\Sigma}_{\mathrm{g, wind}} \mathrm{d}r \notag \\
  & + \int_{r_{\mathrm{in}}}^{r_{\mathrm{out}}}  2\pi r  \left( \int_{\mathrm{St} < 10^{-4}} \dot{\Sigma}_{\mathrm{d, wind}}(m) \mathrm{d}m  \right) \mathrm{d}r \notag \\
  & + \int_{r_{\mathrm{in}}}^{r_{\mathrm{out}}}  2\pi r \dot{\Sigma}_{\mathrm{v, wind}} \mathrm{d}r, \notag \\
  & \text{(if $\dot{\Sigma}_{\mathrm{wind}} > \dot{\Sigma}_{\mathrm{inf}}$ )}.
\end{align}
The mass accretion rate from the disk to the central star $\dot{M}_{\mathrm{acc}}$ shows episodic behavior, as in the case without the disk wind (Figure~\ref{fig:mdot_nowind}).
The magnitude of the peak of the outbursts is lower with the disk wind than in the case without the disk wind.
The outburst interval with the disk wind is as short as $\sim 700$--$800$ years, which is shorter than in the case without the disk wind.
In contrast to the case without the disk wind, no outbursts occur for $t \gtrsim 2.8\times 10^{5}$ years with the disk wind.

During the infall phase, the wind mass loss rate maintains a value of $\sim 4\times 10 ^{-6} \ \mathrm{M_{\sun} yr^{-1}}$ because the gas surface density is maintained by the mass supply from the infalling envelope.
After the infall phase, since there is no mass infalling from the core and the surface density decreases, the wind mass loss rate decreases with time.

The middle panel of Figure~\ref{fig:mdot_wind} shows the mass evolution of the disk, star, and wind.
During the infall phase, the masses of the disk and the star are lower in the case with the disk wind than without the disk wind.
At the end of the infall phase, the star, disk, and wind masses are $M_{\mathrm{star}} \simeq 0.31 \ \mathrm{M_{\sun}}$, $M_{\mathrm{disk}} \simeq 0.31 \ \mathrm{M_{\sun}}$, and $M_{\mathrm{wind}} \simeq 0.39 \ \mathrm{M_{\sun}}$, respectively, as described in Table~\ref{table:mass_at_end_infall_phase}.
After the infall phase, the rates of increase of the masses of the star and wind are small as the disk mass decreases.
At the end of the calculation, the masses of the star, disk, and wind are $M_{\mathrm{star}} \simeq 0.34 \ \mathrm{M_{\sun}}$, $M_{\mathrm{disk}} \simeq 0.17 \ \mathrm{M_{\sun}}$ and $M_{\mathrm{wind}} \simeq 0.49 \ \mathrm{M_{\sun}}$, respectively.
Thus, about 50\% of the mass of the prestellar core is blown away from the central region by the disk wind, which means that the star formation efficiency is about 50\%.

The bottom panel of Figure~\ref{fig:mdot_wind} shows the time evolution of the stellar luminosity $L_{\star}$ and the accretion luminosity $L_{\mathrm{acc}}$ in the case with the disk wind.
During the infall phase, the accretion luminosity dominates the stellar luminosity. 
The accretion luminosity is lower in the case with the disk wind than without the disk wind.
After the infall phase, the accretion luminosity decreases as the mass accretion rate decreases. 
After $t \sim 2.8\times 10^{5}$ year, the accretion luminosity is lower than the stellar luminosity because outbursts, which enhance the accretion luminosity, do not occur.

\subsubsection{Gas evolution with disk wind}
\label{sec:results:wind:gas_evolution}

\begin{figure*}
  \centering
  \includegraphics[width=160mm]{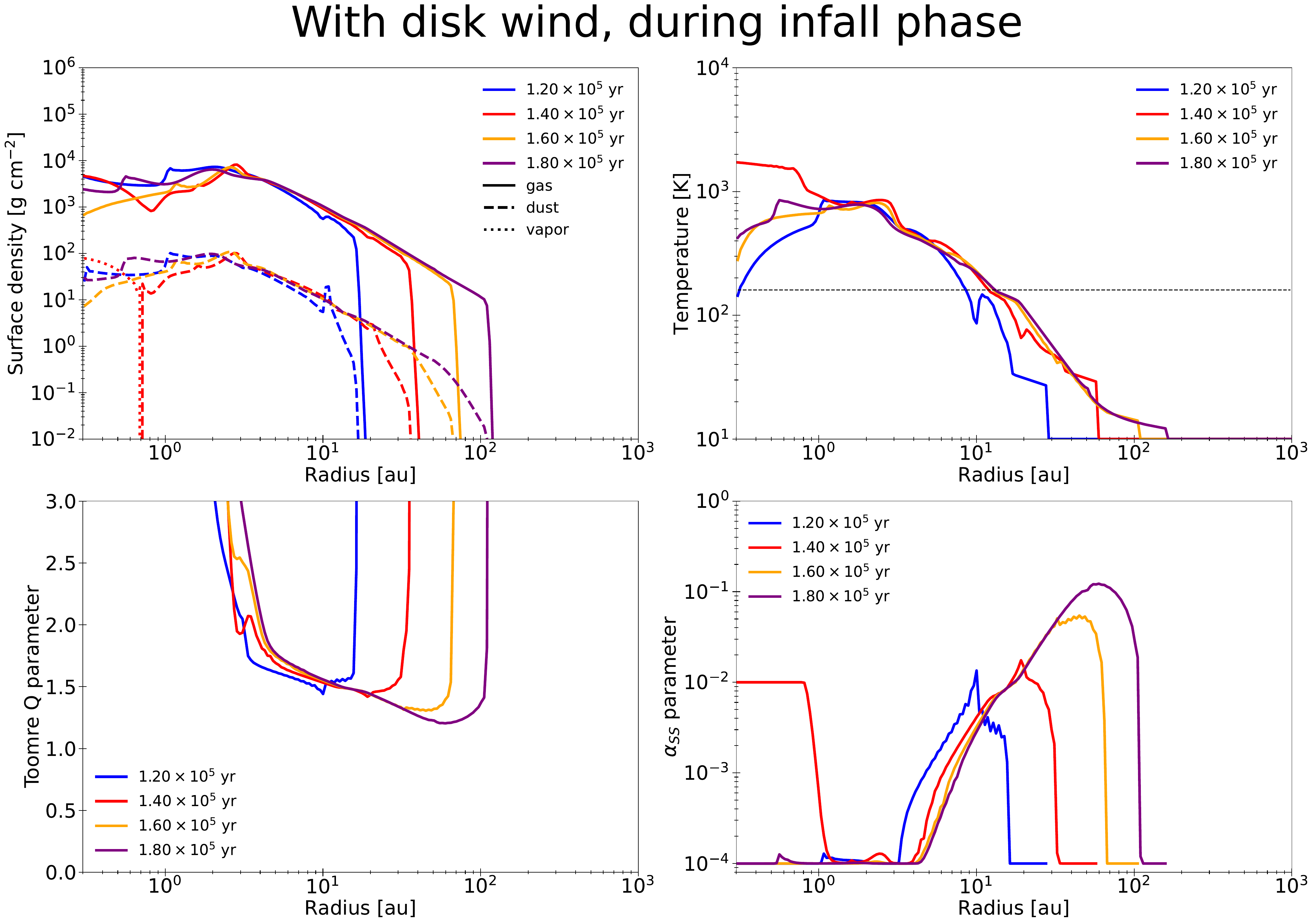}
  \caption{
Same as Figure~\ref{fig:disk_evol_nowind1}, but with disk wind.
}
\label{fig:disk_evol_wind1}
\end{figure*}

\begin{figure}
  \centering
  \includegraphics[width=\columnwidth]{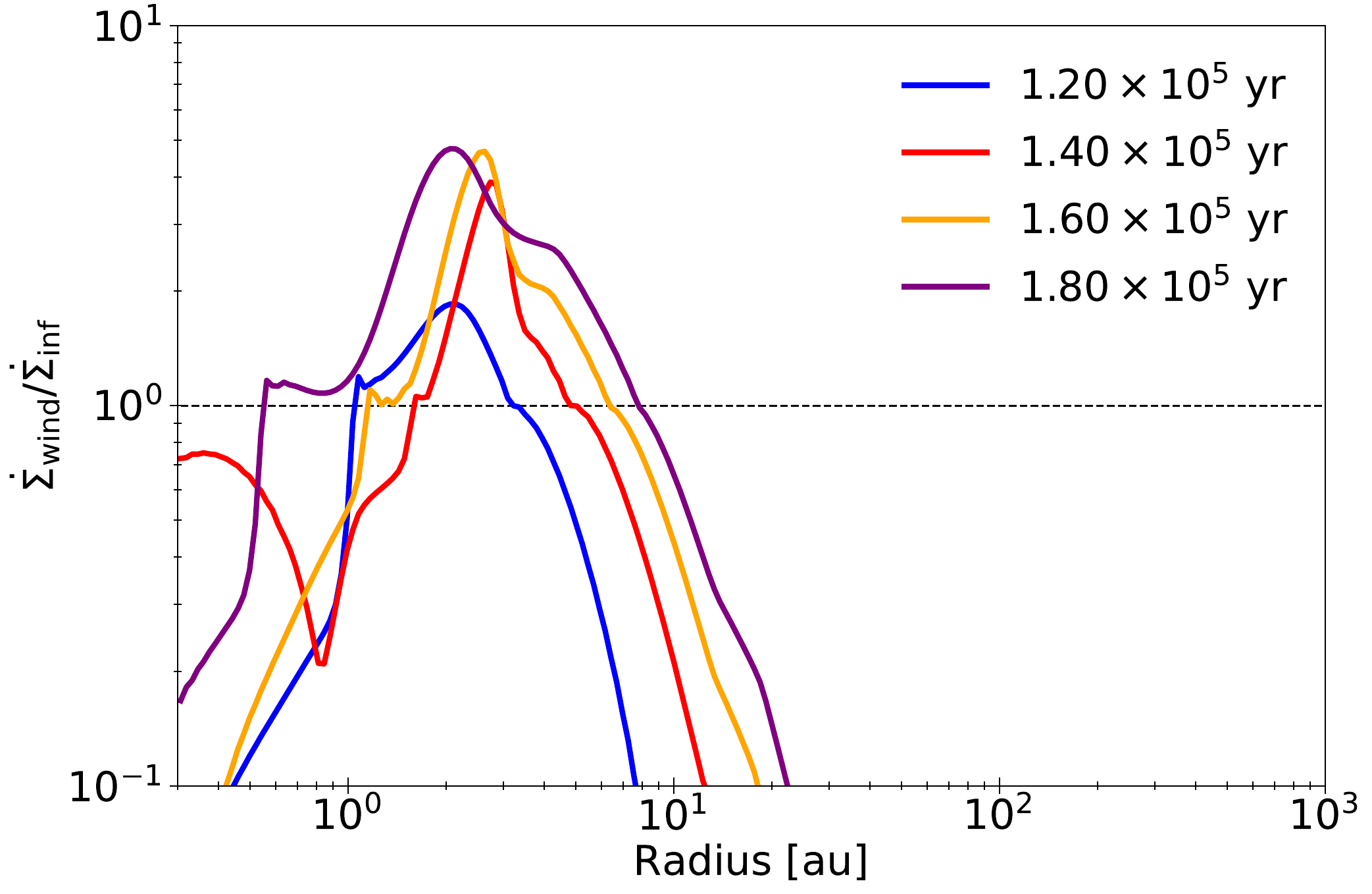}
  \caption{Ratio of wind mass loss rate $\dot{\Sigma}_{\mathrm{wind}}$ to mass infall rate $\dot{\Sigma}_{\mathrm{inf}}$ against radius.
  The ratio $\dot{\Sigma}_{\mathrm{wind}}/\dot{\Sigma}_{\mathrm{inf}}=1$ is plotted as a dotted line.
  }
\label{fig:sigma_wind_sigma_inf}
\end{figure}

Figure~\ref{fig:disk_evol_wind1} shows the radial profile of the surface density, temperature, Toomre Q parameter, and viscous alpha parameter $\alpha_{\rm SS}$ in the case with the disk wind at different epochs during the infall phase.
Figure~\ref{fig:sigma_wind_sigma_inf} shows the ratio of the wind mass loss rate to the mass infall rate $\dot{\Sigma}_{\mathrm{wind}}/ \dot{\Sigma}_{\mathrm{inf}}$ against the disk radius.
As described in \S\ref{sec:method:infall_and_disk_wind_model}, if $\dot{\Sigma}_{\mathrm{wind}} / \dot{\Sigma}_{\mathrm{inf}} > 1.0$ is realized, the disk wind is driven.
The disk wind is present in the range $\sim0.3$--$10$\,au, where  the condition $\dot{\Sigma}_{\mathrm{wind}} >\dot{\Sigma}_{\mathrm{inf}}$ is satisfied, and the wind driving region extends outward with time, as shown in Figure~\ref{fig:sigma_wind_sigma_inf}.
In the region where the disk wind occurs, the disk wind transports both mass and angular momentum. 
Thus, the wind torque also promotes mass accretion in addition to the viscous torque.

The decrease in the gas surface density due to the disk wind affects the disk midplane temperature.
The upper right panel of Figure~\ref{fig:disk_evol_wind1} shows the disk midplane temperature during the infall phase.
In the range of $\sim1$--$10$\,au, the surface density is low due to mass loss by the disk wind and mass accretion by the wind torque (upper left panel of Figure~\ref{fig:disk_evol_wind1}).
The decrease in surface density suppresses viscous heating, which lowers the disk temperature (upper right panel of Figure~\ref{fig:disk_evol_wind1}) and reduces the region where MRI is active (lower right panel of Figure~\ref{fig:disk_evol_wind1}).
Thus, the disk evolution in the region $\sim1$--$10$\,au is controlled by the wind torque.

In the outer disk region ($r \gtrsim 10$\,au), the disk wind is absent (Figure~\ref{fig:sigma_wind_sigma_inf}).
Thus, the lower two panels of Figure~\ref{fig:disk_evol_wind1} indicate that in this region, the angular momentum is redistributed primarily by the torque caused by the gravitational instability.
Therefore, the disk evolves while maintaining a Toomre Q parameter of $Q \sim 1$--$1.5$ in the outer region, as in the case without the disk wind.

\begin{figure*}
  \includegraphics[width=160mm]{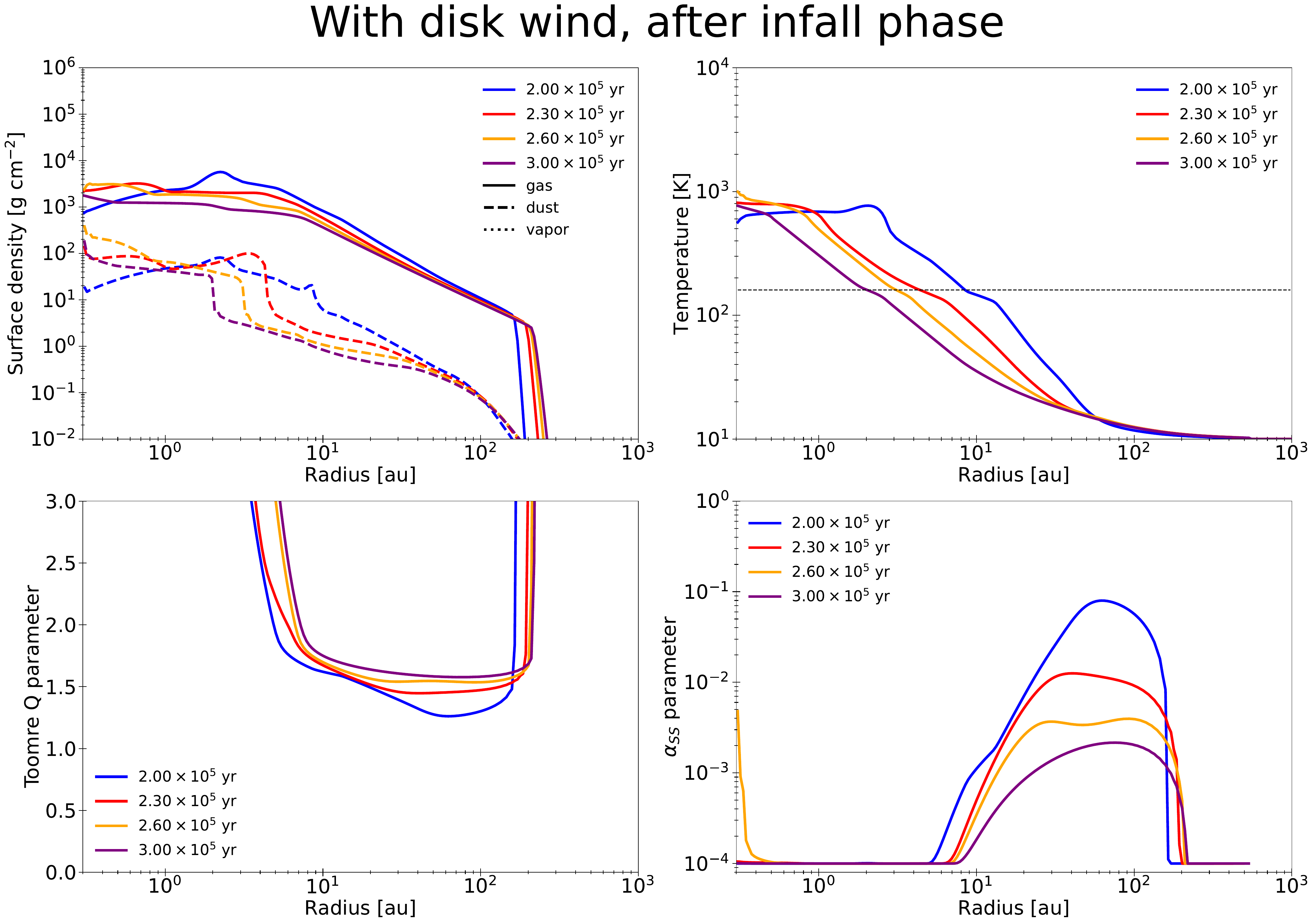}
  \caption{
  Same as Figure~\ref{fig:disk_evol_nowind1}, but with disk wind and for after the infall phase.
}
\label{fig:disk_evol_wind2}
\end{figure*}

Figure~\ref{fig:disk_evol_wind2} shows the radial profiles of the surface density, temperature, Toomre Q parameter, and viscosity parameter $\alpha_{\rm SS}$ at different epochs after the infall phase in the case with the disk wind.
After the infall phase, the disk wind is present throughout the disk because the infalling gas, which suppresses the disk wind, has already dissipated. 
Thus, the upper left panel of Figure~\ref{fig:disk_evol_wind2} shows that the gas surface density decreases with time due to the disk wind.
The effect of the disk wind tends to be more efficient in the inner region of the disk where the gas density and the temperature are higher, as described in equation (\ref{eq:sigma_dot_wind}).
Thus, as shown in the upper left panel of Figure~\ref{fig:disk_evol_wind2}, the gas surface density for the inner disk is greatly reduced by the disk wind compared to the outer disk region.

The upper right panel of Figure~\ref{fig:disk_evol_wind2} shows that after the infall phase, the disk midplane temperature is lower in the case with the disk wind than without the disk wind.
This is because viscous heating is suppressed by the decrease in the gas surface density due to the disk wind.
We confirm that the viscous heating is dominant only in the disk region inside 10 au, while irradiation heating is dominant outside 10 au at the end of the calculation $t = 3\times 10^{5}$ year.
In contrast to the case without the disk wind, the snowline, $T = 160 \ \mathrm{K}$, moves to within 10 au.
This difference in the location of the snowline due to the presence of the disk wind affects the growth of dust particles after the infall phase (see Section~\ref{sec:result:wind:dust_size_evolution}).

The lower left panel of Figure~\ref{fig:disk_evol_wind2} shows that the Toomre Q parameter is $Q \lesssim 1.5$ in the outer region of the disk.
The lower right panel of Figure~\ref{fig:disk_evol_wind2} shows that the viscous alpha parameter $\alpha_{SS}$ in the outer region of the disk is enhanced by the gravitational instability.
Thus, the disk continues to expand after the infall phase even in the case with the disk wind.

\subsubsection{Dust evolution with disk wind}
\label{sec:result:wind:dust_size_evolution}
The upper left panel of Figure~\ref{fig:disk_evol_wind1} shows that the radial profile of the dust surface density is similar to that of the gas surface density.
When outbursts occur and the disk temperature exceeds $1500 \ \mathrm{K}$, the dust particles evaporate and the vapor increases.

\begin{figure*}
  \centering
  \includegraphics[width=160mm]{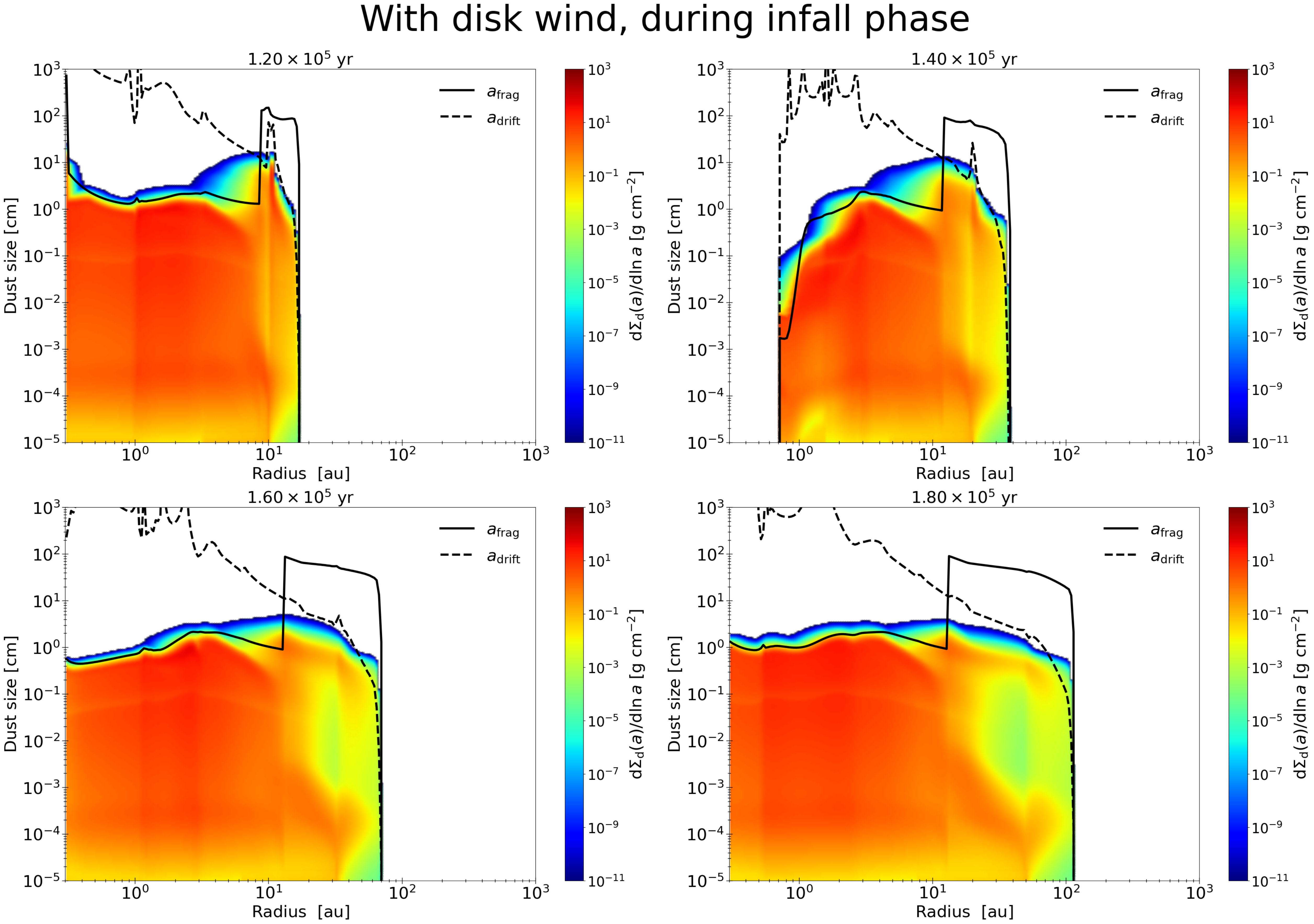}
  \caption{
  Same as Figure~\ref{fig:dust_size_distribution_nowind1}, but with disk wind.
}
\label{fig:dust_size_distribution_wind1}
\end{figure*}

Figure~\ref{fig:dust_size_distribution_wind1} shows the time evolution during the infall phase of the dust particle size distribution in the case with the disk wind.
As in the case without the disk wind, the maximum size of dust particles is limited by fragmentation inside the snowline and by the radial drift outside the snowline.
There is no significant difference in the evolution of the dust particle size distribution in the cases with and without the disk wind during the infall phase.
This is because the gas density is maintained due to the supply of gas to the disk even with the disk wind present.
Additionally, the fact that the snowline is located beyond 10 au during the infall phase is also a factor causing the lack of significant differences in the dust particle size distribution.

\begin{figure}
  \centering
  \includegraphics[width=\columnwidth]{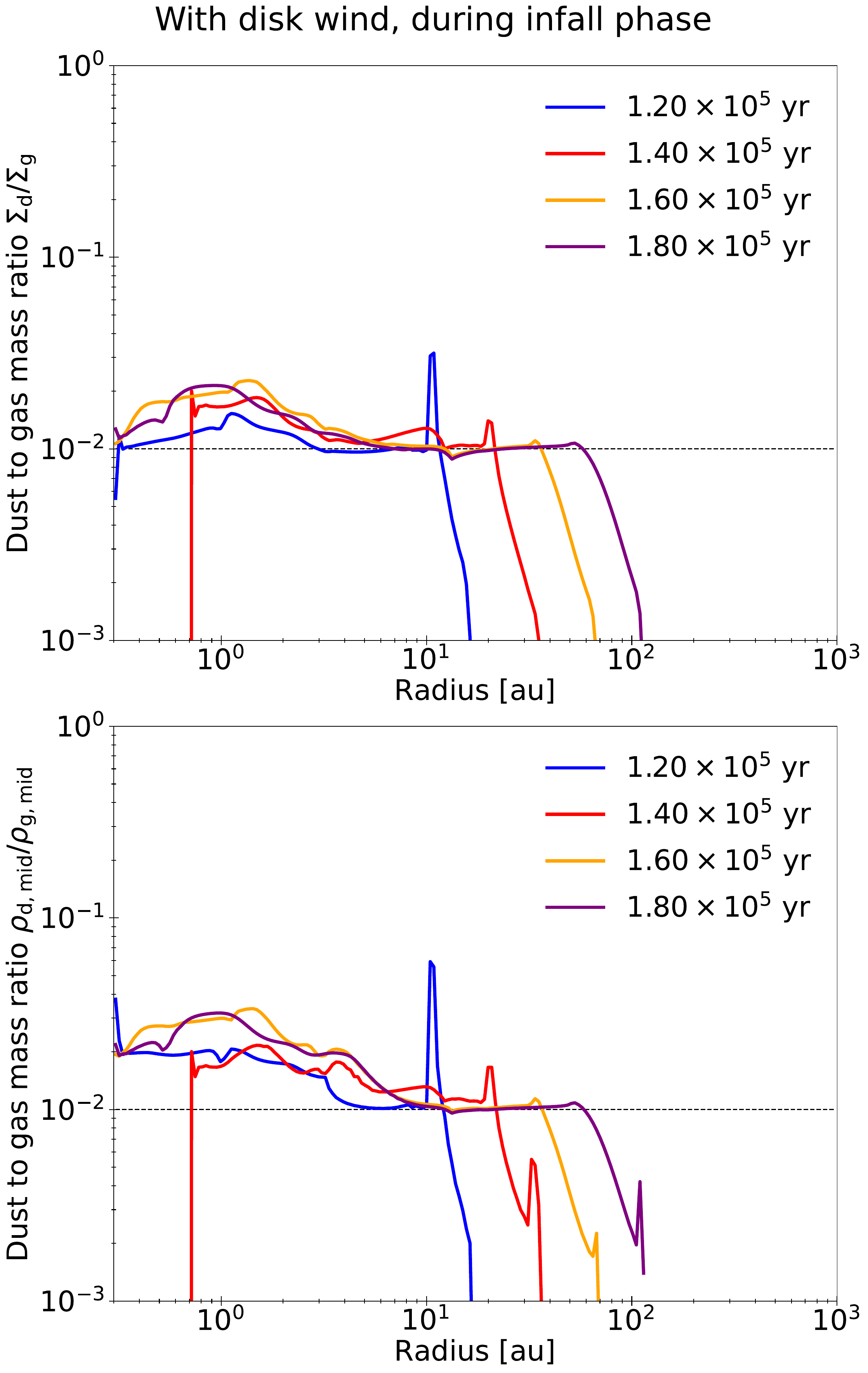}
  \caption{
  Same as Figure~\ref{fig:fdg_nowind1}, but with disk wind.
  }
\label{fig:fdg_wind1}
\end{figure}

Figure~\ref{fig:fdg_wind1} shows the dust-to-gas mass ratio calculated using the surface density (upper panel) and the mass density at the disk midplane (lower panel) in the case with the disk wind.
As described in Section~\ref{sec:results:wind:gas_evolution}, the disk wind is driven in the inner disk region of $\lesssim 10 \ \mathrm{au}$.
In this study, only small dust particles that satisfy $\mathrm{St} < 10^{-4}$ are assumed to be removed with the gas from the disk by the disk wind.
The contribution of these small dust particles to the total dust mass density in the disk is small.
In addition, the mass loss of gas due to the disk wind is larger than that of dust particles.
As a result, in the disk region where the disk wind is driven, the dust-to-mass mass ratio becomes larger in the case with the disk wind than without the disk wind.
At $t = 1.2 \times 10^{5}$ year, a sharp enhancement of the mass ratio is seen at $\sim 10$ au.
This is probably due to the effect of the temperature maximum at $\sim 10 \ \mathrm{au}$ (upper right panel of Figure~\ref{fig:disk_evol_wind1}).
In the subsequent evolution, the enhancement disappears.

The upper left panel of Figure~\ref{fig:disk_evol_wind2} plots the dust surface density at four different epochs after the infall phase.
As described above, the location of the snowline migrates inward with time after the infall phase.
In the case with the disk wind, the snowline exists at a smaller radius compared to the case without the disk wind. 
Outside the snowline, the dust particles can grow to a larger size and radially drift inward.
Consequently, the dust surface density outside the snowline decreases with time.

\begin{figure*}
  \centering
  \includegraphics[width=160mm]{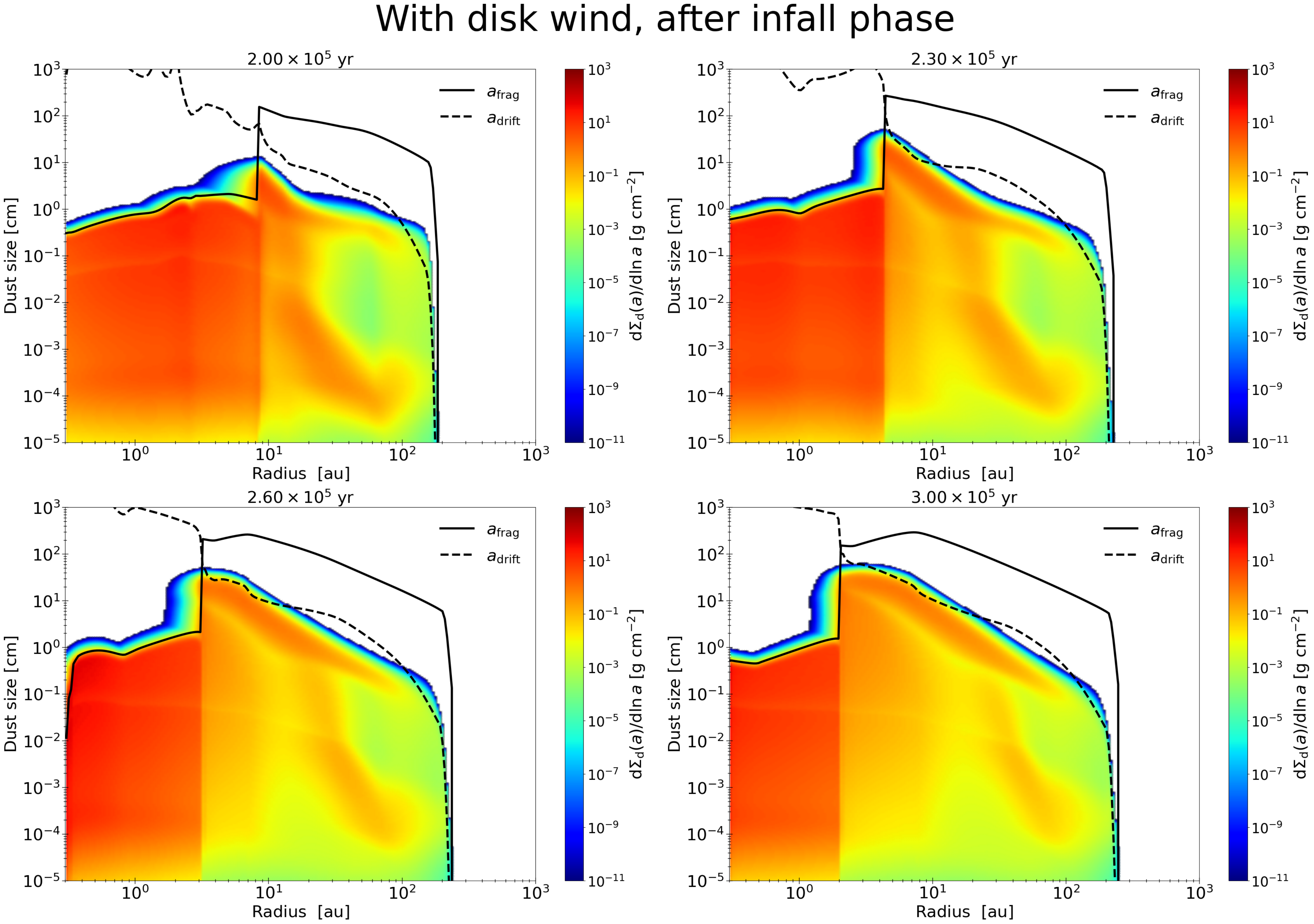}
  \caption{
Same as Figure~\ref{fig:dust_size_distribution_nowind1}, but with disk wind and for after the infall phase.
  }
\label{fig:dust_size_distribution_wind2}

\end{figure*}

Figure~\ref{fig:dust_size_distribution_wind2} shows the time evolution of the dust particle size distribution after the infall phase in the case with the disk wind.
Outside the snowline, dust particles can grow to a larger size.
As described above, the snowline migrates toward the center.
Inside the snow line, the fragmentation velocity of dust is higher, making  coagulation more likely.
Additionally, due to the higher density in the inner regions of the disk, the collision frequency of dust particles increases, which facilitates their growth.
As a result, the maximum size of dust particles becomes larger.
At the end of the calculation $(t = 3\times 10^{5} \ \mathrm{year})$, the maximum size of dust particles is approximately $\sim 50 \ \mathrm{cm}$ near  the snowline ($\sim 2 \ \mathrm{au}$). 
While dust particles can grow significantly outside the snowline, the maximum size of dust particles is still constrained by the radial drift. 
Outside the snowline, small dust particles are depleted due to coagulation.

The maximum size of dust particles inside the snowline is restricted by collisional fragmentation, typically around $\sim 1 \ \mathrm{cm}$. 
Small dust particles are well coupled with the gas, preventing significant inward drift toward the center. 
As a result, the dust surface density inside the snowline is higher than outside the snowline.

\begin{figure}
  \centering
  \includegraphics[width=\columnwidth]{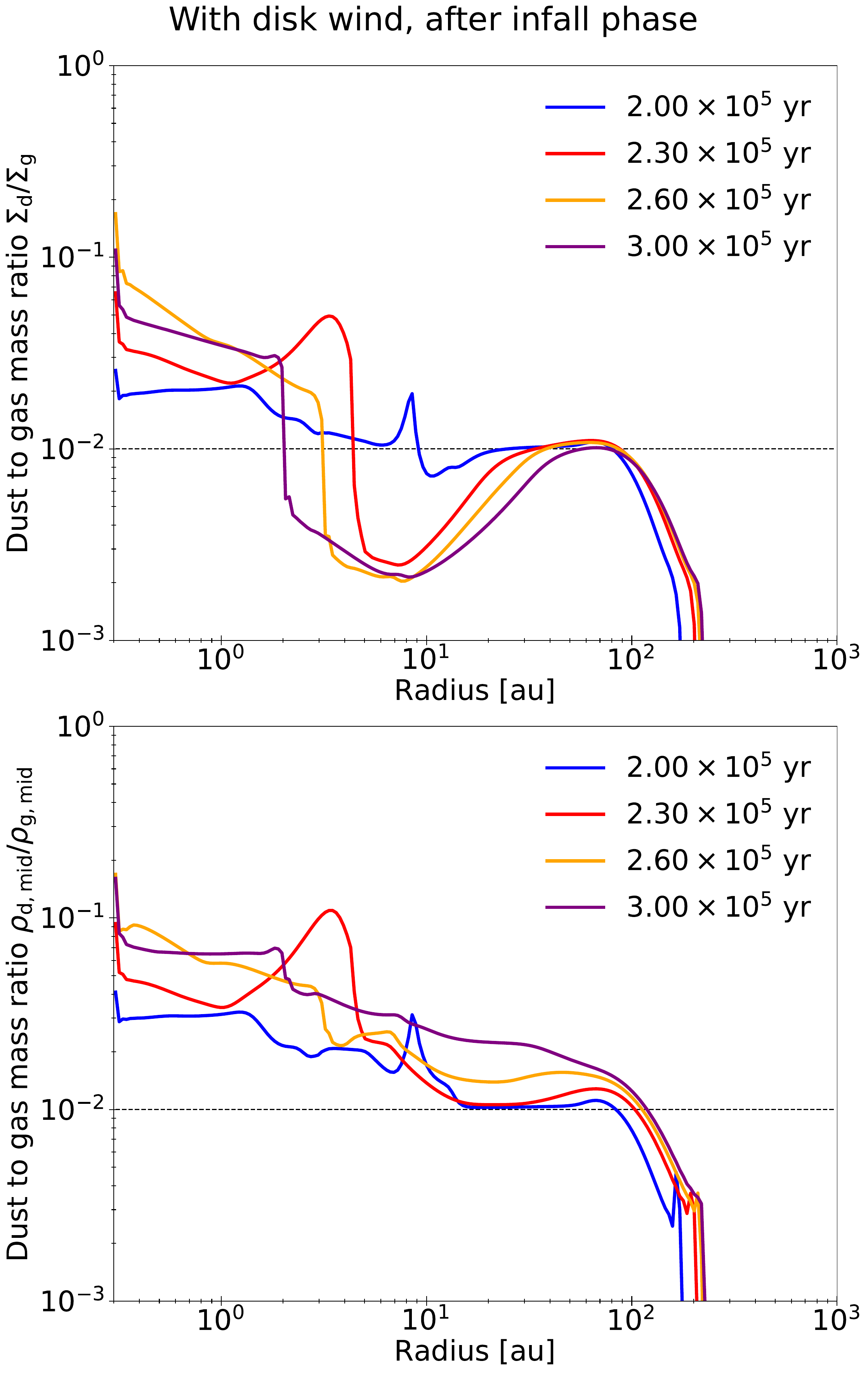}
  \caption{
  Same as Figure~\ref{fig:fdg_nowind1}, but with disk wind and for after the infall phase.
  }
\label{fig:fdg_wind2}
\end{figure}

The upper panel of Figure~\ref{fig:fdg_wind2} shows the time evolution of the dust-to-gas surface density ratio.
The surface density ratio falls below 0.01 outside the snowline. 
Inside the snowline, the mass ratio is significantly higher compared to the case without the disk wind.
In the inner edge region ($\sim 0.3 \ \mathrm{au}$), the mass ratio reaches around $\sim 0.1$.

The lower panel of Figure~\ref{fig:fdg_wind2} shows the time evolution of the dust-to-gas-mass ratio at the disk midplane.
In contrast to the upper panel of Figure~\ref{fig:fdg_wind2}, the mass ratio, excluding the outermost edge of the disk, exceeds 0.01.
The increase in mass ratio is attributed to both the mass loss due to the disk wind and the sedimentation of growth dust particles to the disk midplane.
There is a tendency for the mass ratio to be higher toward the inner regions of the disk, with the inner edge of disk exceeding a mass ratio of $0.1$.
These results suggest that the disk wind is expected to sufficiently enhance the dust-to-gas mass ratio after the infall phase.

\section{Discussion}
\subsection{Possibility of streaming instability}
In this section, we explore the application of the results obtained in this study to the process of planet formation.
Streaming instability is considered as the mechanism for the accumulation of dust in the disk \citep{2005ApJ...620..459Y,2007ApJ...662..627J, 2015A&A...579A..43C,2017A&A...606A..80Y}. 
This instability arises from the aerodynamic interaction between gas and dust and the effects of rotation \citep{2023ASPC..534..465L}. 
The results of multidimensional numerical simulations confirm the local accumulation of dust due to this instability.
 
Since streaming instability plays a crucial role in planetesimal formation, many authors have investigated it from various perspectives.
\citet{2020ApJ...891..132C} developed a linear theory of streaming instability, incorporating external turbulence due to gas viscosity and particle diffusion, and found that streaming instability is sensitive to turbulence, with small particles being stabilized by turbulent effects.
The influence of magnetic fields on streaming instability was explored by \citet{2022ApJ...926...14L}.
In addition, the Hall effect, a non-ideal MHD effect, has been shown to significantly impact streaming instability \citep{2024ApJ...962..173W}.
Although we need to carefully consider the conditions affecting the onset and growth of streaming instability, a comprehensive analysis of its development under various effects and settings is beyond the scope of this study.
While a detailed investigation of streaming instability remains important, we will address this in future studies.
Thus, we provide only a brief discussion of streaming instability below. 

The conditions for streaming instability are evaluated based on the mass ratio between dust and gas surface density,
$Z = \Sigma_{\mathrm{d}} / \Sigma_{\mathrm{g}}$.
\citet{2021ApJ...919..107L} expresses the conditions under which streaming instability occurs based on the results of numerical simulations with the following formula,
\begin{equation}
  \log{\left( \frac{Z_{\mathrm{crit}}}{\Pi} \right)} = 
  \begin{cases}
    0.1 \left( \log{\mathrm{St}} \right)^{2} + 0.32 \log{\mathrm{St}} - 0.24 \\ 
        \ \ \ \ \ \ \ \ \left( \mathrm{St} < 0.015 \right), \\ 
    0.13 \left( \log{\mathrm{St}} \right)^{2} + 0.1 \log{\mathrm{St}} - 1.07 \\ 
        \ \ \ \ \ \ \ \ \left( \mathrm{St} > 0.015 \right),
  \end{cases}
  \label{eq:SI_condition}
\end{equation}
where $Z_{\mathrm{crit}}$ is the critical mass ratio and $\Pi$ represents the effect of the pressure gradient force and is described as $\Pi = \eta r\Omega / c_{s}$.
If $Z > Z_{\mathrm{crit}}$, streaming instability occurs.
Applying these criteria to the results of this study, we evaluate whether streaming instability occurs in the disk.
Using the results for the disk evolution from this study, we can calculate the dust-to-gas mass ratio $(Z)$ and average Stokes number at each time for radii of 5 au, 10 au, and 20 au.
The average Stokes number is determined based on the average dust particle mass.
Then, we plot the evolution curves for each radius (5, 10, 20 au) on a plane with mass ratio and Stokes number as axes, evaluating the possibility of streaming instability.

\begin{figure*}
  \centering
  \includegraphics[width=160mm]{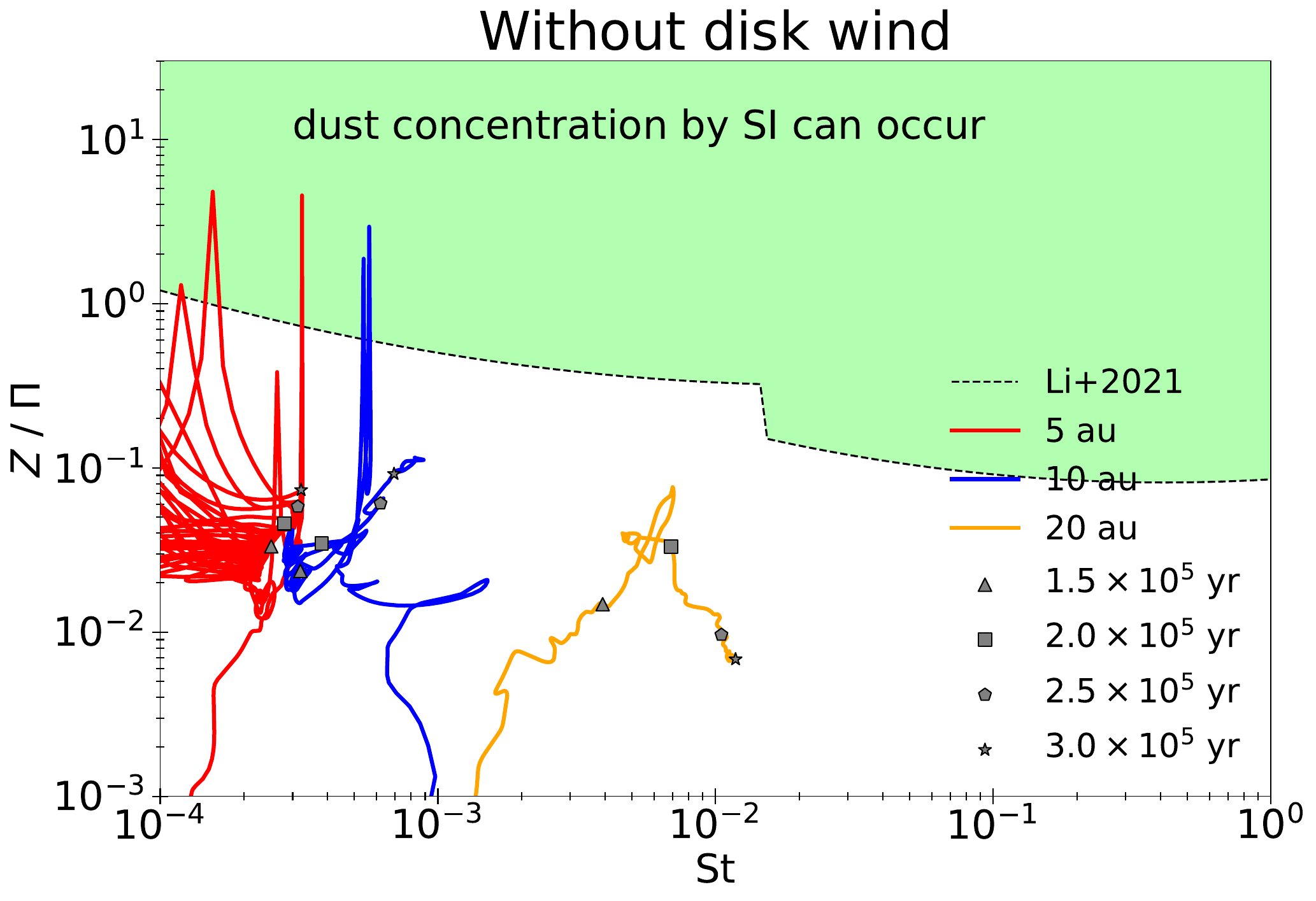}
  \caption{Evolution curves at each time for radii of 5 au, 10 au, and 20 au in calculation without disk wind.
  The horizontal and vertical axes represent the Stokes number and gas-to-dust mass ratio divided by $\Pi$, respectively.
  The dashed lines represent the critical condition for streaming instability (equation~\ref{eq:SI_condition}).}
\label{fig:SI_condition_nowind}
\end{figure*}

Figure~\ref{fig:SI_condition_nowind} plots the evolution curves at each time for radii of 5 au, 10 au, and 20 au in the case without the disk wind.
The horizontal axis of the figure represents the Stokes number, and the vertical axis represents the dust-to-gas mass ratio divided by $\Pi$.
The dashed lines represent the critical condition for streaming instability (equation~(\ref{eq:SI_condition})).
As time references, markers are plotted on the evolution curves at $1.5\times 10^{5}$ years, $2.0\times 10^{5}$ years, $2.5\times 10^{5}$ years, and $3.0\times 10^{5}$ years, respectively.
If each evolution curve remains in the region above the dashed lines for a sufficient duration, it is considered that there is a possibility of streaming instability occurring.

In the case without the disk wind, the evolution curves for 5 au and 10 au exist in regions where streaming instability could occur.
We confirm that the curves spend about 50 years in the region where streaming instability is likely to occur.
On the other hand, the time for dust accumulation due to streaming instability is estimated to be $10^{2}$--$10^{3} \ \Omega^{-1} \sim 10^{2}$--$10^{3}$ years \citep{2021ApJ...919..107L,2023ApJ...958..168T}.
In other words, the time during which the evolution curves exist in the region where streaming instability could occur is shorter than the time it takes for streaming instability to grow and dust accumulation to occur.
Therefore, in the case without the disk wind, it is considered that dust accumulation due to streaming instability does not occur within the calculation time of this study, and planet formation is not promoted.

\begin{figure*}
  \centering
  \includegraphics[width=160mm]{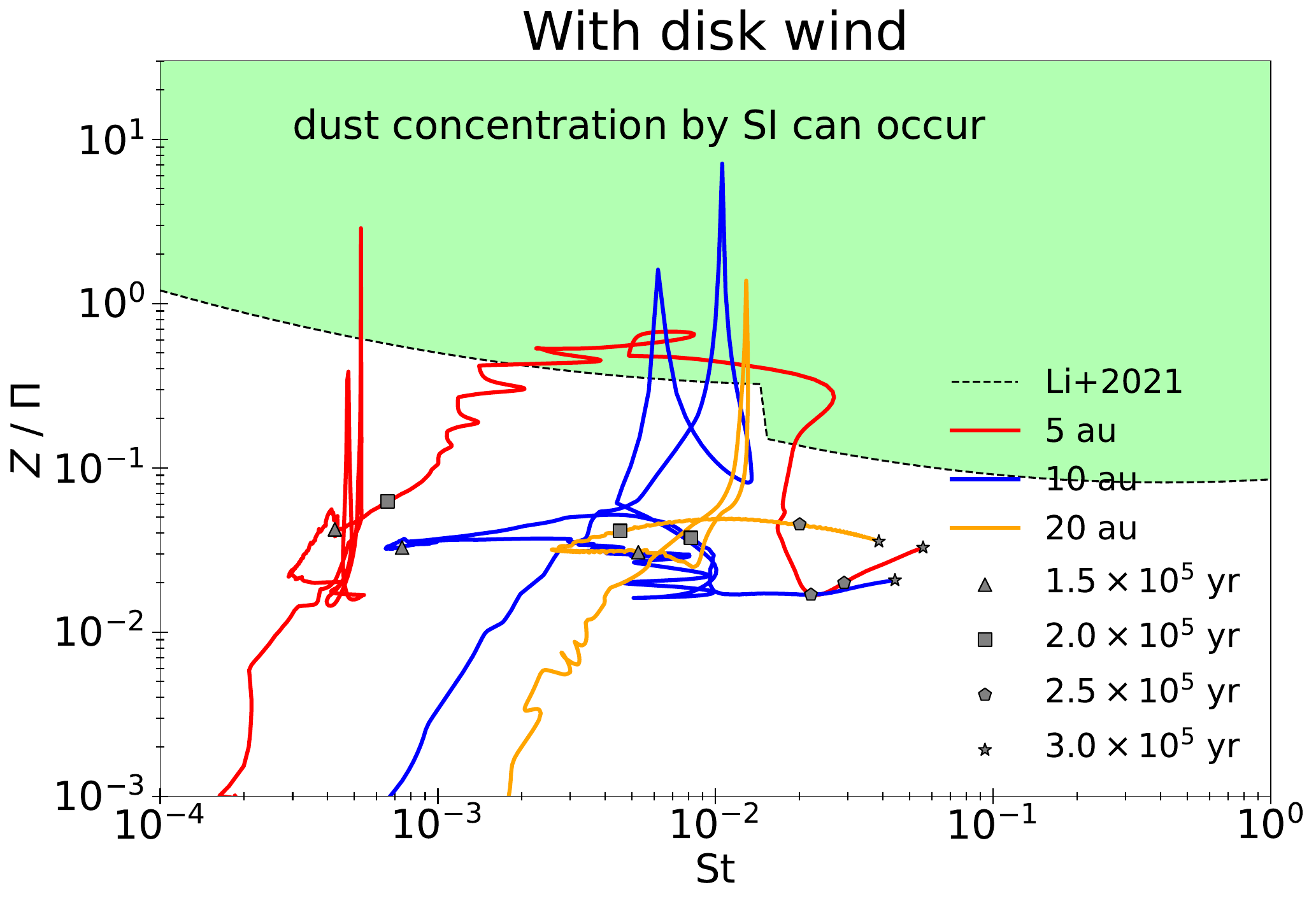}
  \caption{Same as Figure~\ref{fig:SI_condition_nowind} but for case with disk wind.}
\label{fig:SI_condition_wind}
\end{figure*}

Figure~\ref{fig:SI_condition_wind} is the same as Figure~\ref{fig:SI_condition_nowind} but for the case with the disk wind.
Each evolution curve at different radii exists within the region where streaming instability can occur. 
For the evolution curves of 10 au and 20 au, it is difficult for dust accumulation to occur due to streaming instability for the same reasons as in the case without the disk wind.
On the other hand, the evolution curve for 5 au spends $\sim 5\times 10^{3}$ years in the region where streaming instability occurs.  
This duration is longer than that required for dust accumulation due to streaming instability.
In other words, at 5 au, there is a high possibility of dust accumulation due to streaming instability.
Therefore, it is considered that the calculation with the disk wind describes circumstances more likely to lead to planet formation.

\subsection{Comparison with other studies}

This section compares the results of this study with previous studies on the evolution of disk and dust growth.

\citet{2018ApJ...865..102T} calculated the evolution of a disk composed of gas and dust starting from the disk formation phase, including the MHD disk wind.
They showed that the MHD disk wind removes gas from the inner disk region and eventually forms a hole structure in a protoplanetary disk.
The dust particles then accumulate around the pressure maxima, leading to the formation of dust rings.
However, while the wind model adopted in \citet{2018ApJ...865..102T} considered mass loss from the disk, removal of angular momentum from the disk by the wind torque was ignored \citep[see also][]{2010ApJ...718.1289S, 2016A&A...596A..81P}.

In this study, we examined the disk evolution considering both mass loss due to the MHD disk wind and angular momentum transport due to the wind torque.
The surface density decreases due to the MHD disk wind, while the mass accretion driven by the wind torque replenishes the surface density.
Therefore, the formation of pressure maxima is less likely when angular momentum transport by the disk wind is considered. 
However, in this study, the strength of the MHD disk wind was assumed to be constant, 
whereas in reality, the wind strength is determined by the strength and configuration of the magnetic field and the degree of coupling between the gas and the magnetic field at each radius.
If the wind strength varies across the disk, it should be possible for local pressure maxima to form.

\citet{2017ApJ...838..151T} calculated the dust growth in a steady gas disk using the single-size approximation \citep{2016A&A...589A..15S,2016ApJ...821...82O}.
Their gas disk model is assumed to be gravitationally unstable.
They showed that icy dust particles can significantly grow in the inner disk regions and the surface density of dust decreases due to radial drift.
While our results differ, it is important to note that our calculations consider the gas evolution including the mass supply onto the disk from the infalling envelope.
In our calculations, during the infall phase, icy dust particles outside the snowline have their size limited by radial drift before collisional growth.
Inside the snowline, the growth of dust particles is prevented by collisional fragmentation.
Furthermore, \citet{2018ApJ...868..118H} notes that the treatment of the frictional laws for dust particles can alter the time-scale of radial drift, thereby affecting dust growth. 
It is therefore essential to pay attention to the models of disk and dust motion in the evolution of the disk.

\citet{2023ApJ...946...94X} calculated dust particle growth in a steady disk assumed to be gravitationally unstable.
Unlike \citet{2017ApJ...838..151T}, they solved for the evolution of the dust particle size distribution, considering fragmentation. 
Their fragmentation model is constructed based on the model proposed by \citet{2012A&A...540A..73W}, which is derived from laboratory experiments.
They demonstrated that the mass increase by collisions between larger and smaller dust particles at high velocities allows dust particles to overcome the fragmentation barrier and grow to centimeter sizes.
In contrast, in this study, we construct the fragmentation model based on collisional numerical calculations \citep{2013A&A...559A..62W}. 
Despite this difference in the fragmentation model, both studies agree that dust particles grow within the disk during the infall phase (Class 0/I phase).

\citet{2018ApJ...868..118H} calculated the gas and dust evolution from the disk formation stage.
They considered dust growth including porosity evolution of dust particles based on \citet{2012ApJ...752..106O}.
They argued that in the building phase of the disk, since the growth of icy dust particles is prevented by radial drift, planetesimal formation via direct collisional growth of dust particles is challenging.
While the porosity evolution of dust particles is not considered in this study, in the region outside the snowline, the growth of icy dust particles is inhibited by radial drift. 
This result is consistent with \citet{2018ApJ...868..118H}. 
In our calculations, during the infall phase, there is little impact of the MHD disk wind on dust particle growth.
On the other hand, differences in dust particle growth between the cases without and with the MHD disk wind are seen after the infall phase. 
In the case with the MHD disk wind, the region where icy dust particles can exist expands with decreasing disk temperature. 
Therefore, the consideration of porosity evolution of dust particles is expected to become more crucial for collisional coagulation after the infall phase, especially in the case with the MHD disk wind.

\section{Conclusions}

In this study, we investigated the early evolutionary stage of protoplanetary disk and dust growth, considering the MHD disk wind.
We constructed a one-dimensional disk evolution model of gas and dust beginning from the disk formation stage, mainly based on \citet{2013ApJ...770...71T}.
In addition to infall onto the disk from the core or the infalling envelope, we incorporated the magnetically driven disk wind model proposed by \citet{2022MNRAS.512.2290T} into the disk evolution model.
We calculated both the spatial evolution of dust particles and the evolution of the dust particle size distribution.
We calculated the cases with and without the MHD disk wind and compared the disk evolution and dust growth between the two cases.

When the disk wind is not considered, episodic accretions (or outbursts) occur and continue until the end of the calculation ($3.0\times 10^{5}$ years).
The disk surface density and temperature profile vary significantly in the inner disk region due to the recurrent outbursts.
For the outer disk region, the angular momentum is transported by gravitational torque, and the disk extends outward with time.
The disk mass is relatively massive even at the end of the calculation in the case without the disk wind.

In the calculation with the MHD disk wind, episodic accretions (or outbursts) also occur recurrently as in the case without the disk wind.
The interval between outbursts is shorter than in the case without the disk wind.
After the infall phase, the disk mass gradually decreases and the outbursts are quenched.
The MHD disk wind extracts mass from the disk and thus has a significant impact on the evolution of stellar mass.
The star and disk mass are smaller than in the case without the disk wind.

During the infall phase, the presence of the MHD disk wind did not appear to create a significant difference in dust particle size evolution compared to the case without the disk wind.
Dust particles grow to approximately 1--10 cm in the entire disk during the infall phase. 
The maximum size of dust particles was determined by different constraints inside and outside the $\mathrm{H_{2}O}$ snowline.
Inside the $\mathrm{H_{2}O}$ snowline, the maximum size of dust particles is limited by collisional fragmentation.
On the other hand, outside the snowline, the maximum dust particle size is constrained by the radial drift.

After the infall phase, the presence or absence of the MHD disk wind resulted in differences in dust particle size evolution.
In the case without the disk wind, the snowline remained outside 10 au, and there was little change in the dust particle size during the infall phase.
In the case with the MHD disk wind, the wind decreases the disk temperature and the snowline migrates inward. 
As a result, the dust particles can grow larger than 10 cm.
Therefore, it is found that the MHD disk wind is important for dust growth or planet formation after the infall phase.

\vskip\baselineskip

We thank Shahram Abbassi for very useful comments on this paper.
This study was supported by JSPS KAKENHI Grant (JP21H00046, JP21K03617: MNM).
This work was also supported by a NAOJ ALMA Scientific Research grant (No. 2022-22B). 

\bibliography{main}{}

\begin{thebibliography}{}
\expandafter\ifx\csname natexlab\endcsname\relax\def\natexlab#1{#1}\fi
\providecommand{\url}[1]{\href{#1}{#1}}
\providecommand{\dodoi}[1]{doi:~\href{http://doi.org/#1}{\nolinkurl{#1}}}
\providecommand{\doeprint}[1]{\href{http://ascl.net/#1}{\nolinkurl{http://ascl.net/#1}}}
\providecommand{\doarXiv}[1]{\href{https://arxiv.org/abs/#1}{\nolinkurl{https://arxiv.org/abs/#1}}}

\bibitem[{{Adachi} {et~al.}(1976){Adachi}, {Hayashi}, \&
  {Nakazawa}}]{1976PThPh..56.1756A}
{Adachi}, I., {Hayashi}, C., \& {Nakazawa}, K. 1976, Progress of Theoretical
  Physics, 56, 1756, \dodoi{10.1143/PTP.56.1756}

\bibitem[{{Alexander} {et~al.}(2006){Alexander}, {Clarke}, \&
  {Pringle}}]{2006MNRAS.369..229A}
{Alexander}, R.~D., {Clarke}, C.~J., \& {Pringle}, J.~E. 2006, \mnras, 369,
  229, \dodoi{10.1111/j.1365-2966.2006.10294.x}

\bibitem[{{Anderson} {et~al.}(2003){Anderson}, {Li}, {Krasnopolsky}, \&
  {Blandford}}]{2003ApJ...590L.107A}
{Anderson}, J.~M., {Li}, Z.-Y., {Krasnopolsky}, R., \& {Blandford}, R.~D. 2003,
  \apjl, 590, L107, \dodoi{10.1086/376824}

\bibitem[{{Andr{\'e}} {et~al.}(2010){Andr{\'e}}, {Men'shchikov}, {Bontemps},
  {K{\"o}nyves}, {Motte}, {Schneider}, {Didelon}, {Minier}, {Saraceno},
  {Ward-Thompson}, {di Francesco}, {White}, {Molinari}, {Testi}, {Abergel},
  {Griffin}, {Henning}, {Royer}, {Mer{\'\i}n}, {Vavrek}, {Attard},
  {Arzoumanian}, {Wilson}, {Ade}, {Aussel}, {Baluteau}, {Benedettini},
  {Bernard}, {Blommaert}, {Cambr{\'e}sy}, {Cox}, {di Giorgio}, {Hargrave},
  {Hennemann}, {Huang}, {Kirk}, {Krause}, {Launhardt}, {Leeks}, {Le Pennec},
  {Li}, {Martin}, {Maury}, {Olofsson}, {Omont}, {Peretto}, {Pezzuto}, {Prusti},
  {Roussel}, {Russeil}, {Sauvage}, {Sibthorpe}, {Sicilia-Aguilar}, {Spinoglio},
  {Waelkens}, {Woodcraft}, \& {Zavagno}}]{2010A&A...518L.102A}
{Andr{\'e}}, P., {Men'shchikov}, A., {Bontemps}, S., {et~al.} 2010, \aap, 518,
  L102, \dodoi{10.1051/0004-6361/201014666}

\bibitem[{{Andrews} {et~al.}(2018){Andrews}, {Huang}, {P{\'e}rez}, {Isella},
  {Dullemond}, {Kurtovic}, {Guzm{\'a}n}, {Carpenter}, {Wilner}, {Zhang}, {Zhu},
  {Birnstiel}, {Bai}, {Benisty}, {Hughes}, {{\"O}berg}, \&
  {Ricci}}]{2018ApJ...869L..41A}
{Andrews}, S.~M., {Huang}, J., {P{\'e}rez}, L.~M., {et~al.} 2018, \apjl, 869,
  L41, \dodoi{10.3847/2041-8213/aaf741}

\bibitem[{{Arakawa} {et~al.}(2023{\natexlab{a}}){Arakawa}, {Okuzumi},
  {Tatsuuma}, {Tanaka}, {Kokubo}, {Nishiura}, {Furuichi}, \&
  {Nakamoto}}]{2023ApJ...951L..16A}
{Arakawa}, S., {Okuzumi}, S., {Tatsuuma}, M., {et~al.} 2023{\natexlab{a}},
  \apjl, 951, L16, \dodoi{10.3847/2041-8213/acdb5f}

\bibitem[{{Arakawa} {et~al.}(2023{\natexlab{b}}){Arakawa}, {Tanaka}, {Kokubo},
  {Nishiura}, \& {Furuichi}}]{2023A&A...670L..21A}
{Arakawa}, S., {Tanaka}, H., {Kokubo}, E., {Nishiura}, D., \& {Furuichi}, M.
  2023{\natexlab{b}}, \aap, 670, L21, \dodoi{10.1051/0004-6361/202345887}

\bibitem[{{Bae} {et~al.}(2013){Bae}, {Hartmann}, {Zhu}, \&
  {Gammie}}]{2013ApJ...764..141B}
{Bae}, J., {Hartmann}, L., {Zhu}, Z., \& {Gammie}, C. 2013, \apj, 764, 141,
  \dodoi{10.1088/0004-637X/764/2/141}

\bibitem[{{Bae} {et~al.}(2014){Bae}, {Hartmann}, {Zhu}, \&
  {Nelson}}]{2014ApJ...795...61B}
{Bae}, J., {Hartmann}, L., {Zhu}, Z., \& {Nelson}, R.~P. 2014, \apj, 795, 61,
  \dodoi{10.1088/0004-637X/795/1/61}

\bibitem[{{Bai}(2011)}]{2011ApJ...739...50B}
{Bai}, X.-N. 2011, \apj, 739, 50, \dodoi{10.1088/0004-637X/739/1/50}

\bibitem[{{Bai}(2016)}]{2016ApJ...821...80B}
---. 2016, \apj, 821, 80, \dodoi{10.3847/0004-637X/821/2/80}

\bibitem[{{Balbus} \& {Hawley}(1992)}]{1992ApJ...400..610B}
{Balbus}, S.~A., \& {Hawley}, J.~F. 1992, \apj, 400, 610,
  \dodoi{10.1086/172022}

\bibitem[{{Birnstiel} {et~al.}(2010){Birnstiel}, {Dullemond}, \&
  {Brauer}}]{2010A&A...513A..79B}
{Birnstiel}, T., {Dullemond}, C.~P., \& {Brauer}, F. 2010, \aap, 513, A79,
  \dodoi{10.1051/0004-6361/200913731}

\bibitem[{{Birnstiel} {et~al.}(2012){Birnstiel}, {Klahr}, \&
  {Ercolano}}]{2012A&A...539A.148B}
{Birnstiel}, T., {Klahr}, H., \& {Ercolano}, B. 2012, \aap, 539, A148,
  \dodoi{10.1051/0004-6361/201118136}

\bibitem[{{Blandford} \& {Payne}(1982)}]{1982MNRAS.199..883B}
{Blandford}, R.~D., \& {Payne}, D.~G. 1982, \mnras, 199, 883,
  \dodoi{10.1093/mnras/199.4.883}

\bibitem[{{Blum} \& {M{\"u}nch}(1993)}]{1993Icar..106..151B}
{Blum}, J., \& {M{\"u}nch}, M. 1993, \icarus, 106, 151,
  \dodoi{10.1006/icar.1993.1163}

\bibitem[{{Bonnor}(1956)}]{1956MNRAS.116..351B}
{Bonnor}, W.~B. 1956, \mnras, 116, 351, \dodoi{10.1093/mnras/116.3.351}

\bibitem[{{Booth} {et~al.}(2018){Booth}, {Meru}, {Lee}, \&
  {Clarke}}]{2018MNRAS.475..167B}
{Booth}, R.~A., {Meru}, F., {Lee}, M.~H., \& {Clarke}, C.~J. 2018, \mnras, 475,
  167, \dodoi{10.1093/mnras/stx3084}

\bibitem[{{Brauer} {et~al.}(2008){Brauer}, {Dullemond}, \&
  {Henning}}]{2008A&A...480..859B}
{Brauer}, F., {Dullemond}, C.~P., \& {Henning}, T. 2008, \aap, 480, 859,
  \dodoi{10.1051/0004-6361:20077759}

\bibitem[{{Carrera} {et~al.}(2015){Carrera}, {Johansen}, \&
  {Davies}}]{2015A&A...579A..43C}
{Carrera}, D., {Johansen}, A., \& {Davies}, M.~B. 2015, \aap, 579, A43,
  \dodoi{10.1051/0004-6361/201425120}

\bibitem[{{Cassen} \& {Moosman}(1981)}]{1981Icar...48..353C}
{Cassen}, P., \& {Moosman}, A. 1981, \icarus, 48, 353,
  \dodoi{10.1016/0019-1035(81)90051-8}

\bibitem[{{Coradini} {et~al.}(1981){Coradini}, {Magni}, \&
  {Federico}}]{1981A&A....98..173C}
{Coradini}, A., {Magni}, G., \& {Federico}, C. 1981, \aap, 98, 173

\bibitem[{{Delage} {et~al.}(2023){Delage}, {G{\'a}rate}, {Okuzumi}, {Yang},
  {Pinilla}, {Flock}, {Stammler}, \& {Birnstiel}}]{2023A&A...674A.190D}
{Delage}, T.~N., {G{\'a}rate}, M., {Okuzumi}, S., {et~al.} 2023, \aap, 674,
  A190, \dodoi{10.1051/0004-6361/202244731}

\bibitem[{{Desch} \& {Turner}(2015)}]{2015ApJ...811..156D}
{Desch}, S.~J., \& {Turner}, N.~J. 2015, \apj, 811, 156,
  \dodoi{10.1088/0004-637X/811/2/156}

\bibitem[{{Dominik} \& {Tielens}(1997)}]{1997ApJ...480..647D}
{Dominik}, C., \& {Tielens}, A.~G.~G.~M. 1997, \apj, 480, 647,
  \dodoi{10.1086/303996}

\bibitem[{{Ebert}(1955)}]{1955ZA.....37..217E}
{Ebert}, R. 1955, \zap, 37, 217

\bibitem[{{Flock} {et~al.}(2016){Flock}, {Fromang}, {Turner}, \&
  {Benisty}}]{2016ApJ...827..144F}
{Flock}, M., {Fromang}, S., {Turner}, N.~J., \& {Benisty}, M. 2016, \apj, 827,
  144, \dodoi{10.3847/0004-637X/827/2/144}

\bibitem[{{Gammie}(1996)}]{1996ApJ...457..355G}
{Gammie}, C.~F. 1996, \apj, 457, 355, \dodoi{10.1086/176735}

\bibitem[{{Garaud} {et~al.}(2013){Garaud}, {Meru}, {Galvagni}, \&
  {Olczak}}]{2013ApJ...764..146G}
{Garaud}, P., {Meru}, F., {Galvagni}, M., \& {Olczak}, C. 2013, \apj, 764, 146,
  \dodoi{10.1088/0004-637X/764/2/146}

\bibitem[{{Glassgold} {et~al.}(1997){Glassgold}, {Najita}, \&
  {Igea}}]{1997ApJ...480..344G}
{Glassgold}, A.~E., {Najita}, J., \& {Igea}, J. 1997, \apj, 480, 344,
  \dodoi{10.1086/303952}

\bibitem[{{Harada} {et~al.}(2023){Harada}, {Tokuda}, {Yamasaki}, {Sato},
  {Omura}, {Hirano}, {Onishi}, {Tachihara}, \& {Machida}}]{2023ApJ...945...63H}
{Harada}, N., {Tokuda}, K., {Yamasaki}, H., {et~al.} 2023, \apj, 945, 63,
  \dodoi{10.3847/1538-4357/acb930}

\bibitem[{{Hasegawa} {et~al.}(2021){Hasegawa}, {Suzuki}, {Tanaka}, {Kobayashi},
  \& {Wada}}]{2021ApJ...915...22H}
{Hasegawa}, Y., {Suzuki}, T.~K., {Tanaka}, H., {Kobayashi}, H., \& {Wada}, K.
  2021, \apj, 915, 22, \dodoi{10.3847/1538-4357/abf6cf}

\bibitem[{{Hasegawa} {et~al.}(2023){Hasegawa}, {Suzuki}, {Tanaka}, {Kobayashi},
  \& {Wada}}]{2023ApJ...944...38H}
---. 2023, \apj, 944, 38, \dodoi{10.3847/1538-4357/acadda}

\bibitem[{{Homma} \& {Nakamoto}(2018)}]{2018ApJ...868..118H}
{Homma}, K., \& {Nakamoto}, T. 2018, \apj, 868, 118,
  \dodoi{10.3847/1538-4357/aae0fb}

\bibitem[{{Hueso} \& {Guillot}(2005)}]{2005A&A...442..703H}
{Hueso}, R., \& {Guillot}, T. 2005, \aap, 442, 703,
  \dodoi{10.1051/0004-6361:20041905}

\bibitem[{{Igea} \& {Glassgold}(1999)}]{1999ApJ...518..848I}
{Igea}, J., \& {Glassgold}, A.~E. 1999, \apj, 518, 848, \dodoi{10.1086/307302}

\bibitem[{{Inutsuka}(2012)}]{2012PTEP.2012aA307I}
{Inutsuka}, S.-i. 2012, Progress of Theoretical and Experimental Physics, 2012,
  01A307, \dodoi{10.1093/ptep/pts024}

\bibitem[{{Johansen} {et~al.}(2007){Johansen}, {Oishi}, {Mac Low}, {Klahr},
  {Henning}, \& {Youdin}}]{2007Natur.448.1022J}
{Johansen}, A., {Oishi}, J.~S., {Mac Low}, M.-M., {et~al.} 2007, \nat, 448,
  1022, \dodoi{10.1038/nature06086}

\bibitem[{{Johansen} \& {Youdin}(2007)}]{2007ApJ...662..627J}
{Johansen}, A., \& {Youdin}, A. 2007, \apj, 662, 627, \dodoi{10.1086/516730}

\bibitem[{{Kataoka} {et~al.}(2013){Kataoka}, {Tanaka}, {Okuzumi}, \&
  {Wada}}]{2013A&A...557L...4K}
{Kataoka}, A., {Tanaka}, H., {Okuzumi}, S., \& {Wada}, K. 2013, \aap, 557, L4,
  \dodoi{10.1051/0004-6361/201322151}

\bibitem[{{Kawasaki} {et~al.}(2021){Kawasaki}, {Koga}, \&
  {Machida}}]{2021MNRAS.504.5588K}
{Kawasaki}, Y., {Koga}, S., \& {Machida}, M.~N. 2021, \mnras, 504, 5588,
  \dodoi{10.1093/mnras/stab1224}

\bibitem[{{Kawasaki} {et~al.}(2022){Kawasaki}, {Koga}, \&
  {Machida}}]{2022MNRAS.515.2072K}
---. 2022, \mnras, 515, 2072, \dodoi{10.1093/mnras/stac1919}

\bibitem[{{Kawasaki} \& {Machida}(2023)}]{2023MNRAS.522.3679K}
{Kawasaki}, Y., \& {Machida}, M.~N. 2023, \mnras, 522, 3679,
  \dodoi{10.1093/mnras/stad1241}

\bibitem[{{Kobayashi} \& {Tanaka}(2010)}]{2010Icar..206..735K}
{Kobayashi}, H., \& {Tanaka}, H. 2010, \icarus, 206, 735,
  \dodoi{10.1016/j.icarus.2009.10.004}

\bibitem[{{Kobayashi} \& {Tanaka}(2021)}]{2021ApJ...922...16K}
---. 2021, \apj, 922, 16, \dodoi{10.3847/1538-4357/ac289c}

\bibitem[{{Koga} {et~al.}(2022){Koga}, {Kawasaki}, \&
  {Machida}}]{2022MNRAS.515.6073K}
{Koga}, S., {Kawasaki}, Y., \& {Machida}, M.~N. 2022, \mnras, 515, 6073,
  \dodoi{10.1093/mnras/stac2115}

\bibitem[{{Koga} \& {Machida}(2023)}]{2023MNRAS.519.3595K}
{Koga}, S., \& {Machida}, M.~N. 2023, \mnras, 519, 3595,
  \dodoi{10.1093/mnras/stac3503}

\bibitem[{{Koga} {et~al.}(2019){Koga}, {Tsukamoto}, {Okuzumi}, \&
  {Machida}}]{2019MNRAS.484.2119K}
{Koga}, S., {Tsukamoto}, Y., {Okuzumi}, S., \& {Machida}, M.~N. 2019, \mnras,
  484, 2119, \dodoi{10.1093/mnras/sty3524}

\bibitem[{{K{\"o}nyves} {et~al.}(2015){K{\"o}nyves}, {Andr{\'e}},
  {Men'shchikov}, {Palmeirim}, {Arzoumanian}, {Schneider}, {Roy}, {Didelon},
  {Maury}, {Shimajiri}, {Di Francesco}, {Bontemps}, {Peretto}, {Benedettini},
  {Bernard}, {Elia}, {Griffin}, {Hill}, {Kirk}, {Ladjelate}, {Marsh}, {Martin},
  {Motte}, {Nguy{\^e}n Luong}, {Pezzuto}, {Roussel}, {Rygl}, {Sadavoy},
  {Schisano}, {Spinoglio}, {Ward-Thompson}, \& {White}}]{2015A&A...584A..91K}
{K{\"o}nyves}, V., {Andr{\'e}}, P., {Men'shchikov}, A., {et~al.} 2015, \aap,
  584, A91, \dodoi{10.1051/0004-6361/201525861}

\bibitem[{{Kudoh} \& {Shibata}(1997)}]{1997ApJ...474..362K}
{Kudoh}, T., \& {Shibata}, K. 1997, \apj, 474, 362, \dodoi{10.1086/303437}

\bibitem[{{Langkowski} {et~al.}(2008){Langkowski}, {Teiser}, \&
  {Blum}}]{2008ApJ...675..764L}
{Langkowski}, D., {Teiser}, J., \& {Blum}, J. 2008, \apj, 675, 764,
  \dodoi{10.1086/525841}

\bibitem[{{Lebreuilly} {et~al.}(2019){Lebreuilly}, {Commer{\c{c}}on}, \&
  {Laibe}}]{2019A&A...626A..96L}
{Lebreuilly}, U., {Commer{\c{c}}on}, B., \& {Laibe}, G. 2019, \aap, 626, A96,
  \dodoi{10.1051/0004-6361/201834147}

\bibitem[{{Lebreuilly} {et~al.}(2020){Lebreuilly}, {Commer{\c{c}}on}, \&
  {Laibe}}]{2020A&A...641A.112L}
---. 2020, \aap, 641, A112, \dodoi{10.1051/0004-6361/202038174}

\bibitem[{{Lesur}(2021)}]{2021JPlPh..87a2001P}
{Lesur}, G. 2021, Journal of Plasma Physics, 87, 205870101,
  \dodoi{10.1017/S0022377820001002}

\bibitem[{{Lesur} {et~al.}(2023){Lesur}, {Flock}, {Ercolano}, {Lin}, {Yang},
  {Barranco}, {Benitez-Llambay}, {Goodman}, {Johansen}, {Klahr}, {Laibe},
  {Lyra}, {Marcus}, {Nelson}, {Squire}, {Simon}, {Turner}, {Umurhan}, \&
  {Youdin}}]{2023ASPC..534..465L}
{Lesur}, G., {Flock}, M., {Ercolano}, B., {et~al.} 2023, in Astronomical
  Society of the Pacific Conference Series, Vol. 534, Astronomical Society of
  the Pacific Conference Series, ed. S.~{Inutsuka}, Y.~{Aikawa}, T.~{Muto},
  K.~{Tomida}, \& M.~{Tamura}, 465

\bibitem[{{Li} \& {Youdin}(2021)}]{2021ApJ...919..107L}
{Li}, R., \& {Youdin}, A.~N. 2021, \apj, 919, 107,
  \dodoi{10.3847/1538-4357/ac0e9f}

\bibitem[{{Liu} {et~al.}(2023){Liu}, {Takahashi}, {Machida}, {Tomisaka},
  {Girart}, {Ho}, {Nakanishi}, \& {Sato}}]{2023arXiv231213573L}
{Liu}, Y., {Takahashi}, S., {Machida}, M., {et~al.} 2023, arXiv e-prints,
  arXiv:2312.13573, \dodoi{10.48550/arXiv.2312.13573}

\bibitem[{{Machida} \& {Basu}(2019)}]{2019ApJ...876..149M}
{Machida}, M.~N., \& {Basu}, S. 2019, \apj, 876, 149,
  \dodoi{10.3847/1538-4357/ab18a7}

\bibitem[{{Machida} \& {Hosokawa}(2020)}]{2020MNRAS.499.4490M}
{Machida}, M.~N., \& {Hosokawa}, T. 2020, \mnras, 499, 4490,
  \dodoi{10.1093/mnras/staa3139}

\bibitem[{{Machida} {et~al.}(2010){Machida}, {Inutsuka}, \&
  {Matsumoto}}]{2010ApJ...724.1006M}
{Machida}, M.~N., {Inutsuka}, S.-i., \& {Matsumoto}, T. 2010, \apj, 724, 1006,
  \dodoi{10.1088/0004-637X/724/2/1006}

\bibitem[{{Machida} \& {Matsumoto}(2011)}]{2011MNRAS.413.2767M}
{Machida}, M.~N., \& {Matsumoto}, T. 2011, \mnras, 413, 2767,
  \dodoi{10.1111/j.1365-2966.2011.18349.x}

\bibitem[{{Machida} \& {Matsumoto}(2012)}]{2012MNRAS.421..588M}
---. 2012, \mnras, 421, 588, \dodoi{10.1111/j.1365-2966.2011.20336.x}

\bibitem[{{Machida} {et~al.}(2006){Machida}, {Matsumoto}, {Hanawa}, \&
  {Tomisaka}}]{2006ApJ...645.1227M}
{Machida}, M.~N., {Matsumoto}, T., {Hanawa}, T., \& {Tomisaka}, K. 2006, \apj,
  645, 1227, \dodoi{10.1086/504423}

\bibitem[{{Marchand} {et~al.}(2023){Marchand}, {Lebreuilly}, {Mac Low}, \&
  {Guillet}}]{2023A&A...670A..61M}
{Marchand}, P., {Lebreuilly}, U., {Mac Low}, M.~M., \& {Guillet}, V. 2023,
  \aap, 670, A61, \dodoi{10.1051/0004-6361/202244291}

\bibitem[{{Myers} {et~al.}(2023){Myers}, {Dunham}, \&
  {Stephens}}]{2023ApJ...949...19M}
{Myers}, P.~C., {Dunham}, M.~M., \& {Stephens}, I.~W. 2023, \apj, 949, 19,
  \dodoi{10.3847/1538-4357/acca74}

\bibitem[{{Nakamoto} \& {Nakagawa}(1994)}]{1994ApJ...421..640N}
{Nakamoto}, T., \& {Nakagawa}, Y. 1994, \apj, 421, 640, \dodoi{10.1086/173678}

\bibitem[{{Ohashi} {et~al.}(2023{\natexlab{a}}){Ohashi}, {Tobin},
  {J{\o}rgensen}, {Takakuwa}, {Sheehan}, {Aikawa}, {Li}, {Looney}, {Williams},
  {Aso}, {Sharma}, {Sai}, {Yamato}, {Lee}, {Tomida}, {Yen}, {Encalada},
  {Flores}, {Gavino}, {Kido}, {Han}, {Lin}, {Narayanan}, {Phuong},
  {Santamar{\'\i}a-Miranda}, {Thieme}, {van't Hoff}, {de Gregorio-Monsalvo},
  {Koch}, {Kwon}, {Lai}, {Lee}, {Plunkett}, {Saigo}, {Hirano}, {Lam}, \&
  {Mori}}]{2023ApJ...951....8O}
{Ohashi}, N., {Tobin}, J.~J., {J{\o}rgensen}, J.~K., {et~al.}
  2023{\natexlab{a}}, \apj, 951, 8, \dodoi{10.3847/1538-4357/acd384}

\bibitem[{{Ohashi} {et~al.}(2021){Ohashi}, {Kobayashi}, {Nakatani}, {Okuzumi},
  {Tanaka}, {Murakawa}, {Zhang}, {Liu}, \& {Sakai}}]{2021ApJ...907...80O}
{Ohashi}, S., {Kobayashi}, H., {Nakatani}, R., {et~al.} 2021, \apj, 907, 80,
  \dodoi{10.3847/1538-4357/abd0fa}

\bibitem[{{Ohashi} {et~al.}(2023{\natexlab{b}}){Ohashi}, {Momose}, {Kataoka},
  {Higuchi}, {Tsukagoshi}, {Ueda}, {Codella}, {Podio}, {Hanawa}, {Sakai},
  {Kobayashi}, {Okuzumi}, \& {Tanaka}}]{2023ApJ...954..110O}
{Ohashi}, S., {Momose}, M., {Kataoka}, A., {et~al.} 2023{\natexlab{b}}, \apj,
  954, 110, \dodoi{10.3847/1538-4357/ace9b9}

\bibitem[{{Okuzumi} {et~al.}(2016){Okuzumi}, {Momose}, {Sirono}, {Kobayashi},
  \& {Tanaka}}]{2016ApJ...821...82O}
{Okuzumi}, S., {Momose}, M., {Sirono}, S.-i., {Kobayashi}, H., \& {Tanaka}, H.
  2016, \apj, 821, 82, \dodoi{10.3847/0004-637X/821/2/82}

\bibitem[{{Okuzumi} {et~al.}(2012){Okuzumi}, {Tanaka}, {Kobayashi}, \&
  {Wada}}]{2012ApJ...752..106O}
{Okuzumi}, S., {Tanaka}, H., {Kobayashi}, H., \& {Wada}, K. 2012, \apj, 752,
  106, \dodoi{10.1088/0004-637X/752/2/106}

\bibitem[{{Okuzumi} {et~al.}(2009){Okuzumi}, {Tanaka}, \&
  {Sakagami}}]{2009ApJ...707.1247O}
{Okuzumi}, S., {Tanaka}, H., \& {Sakagami}, M.-a. 2009, \apj, 707, 1247,
  \dodoi{10.1088/0004-637X/707/2/1247}

\bibitem[{{Omura} {et~al.}(2024){Omura}, {Tokuda}, \&
  {Machida}}]{2024arXiv240103086O}
{Omura}, M., {Tokuda}, K., \& {Machida}, M.~N. 2024, arXiv e-prints,
  arXiv:2401.03086, \dodoi{10.48550/arXiv.2401.03086}

\bibitem[{{Ormel} \& {Cuzzi}(2007)}]{2007A&A...466..413O}
{Ormel}, C.~W., \& {Cuzzi}, J.~N. 2007, \aap, 466, 413,
  \dodoi{10.1051/0004-6361:20066899}

\bibitem[{{Ormel} {et~al.}(2007){Ormel}, {Spaans}, \&
  {Tielens}}]{2007A&A...461..215O}
{Ormel}, C.~W., {Spaans}, M., \& {Tielens}, A.~G.~G.~M. 2007, \aap, 461, 215,
  \dodoi{10.1051/0004-6361:20065949}

\bibitem[{{Owen} {et~al.}(2010){Owen}, {Ercolano}, {Clarke}, \&
  {Alexander}}]{2010MNRAS.401.1415O}
{Owen}, J.~E., {Ercolano}, B., {Clarke}, C.~J., \& {Alexander}, R.~D. 2010,
  \mnras, 401, 1415, \dodoi{10.1111/j.1365-2966.2009.15771.x}

\bibitem[{{Pinilla} {et~al.}(2016){Pinilla}, {Flock}, {Ovelar}, \&
  {Birnstiel}}]{2016A&A...596A..81P}
{Pinilla}, P., {Flock}, M., {Ovelar}, M. d.~J., \& {Birnstiel}, T. 2016, \aap,
  596, A81, \dodoi{10.1051/0004-6361/201628441}

\bibitem[{{Pudritz} \& {Ray}(2019)}]{2019FrASS...6...54P}
{Pudritz}, R.~E., \& {Ray}, T.~P. 2019, Frontiers in Astronomy and Space
  Sciences, 6, 54, \dodoi{10.3389/fspas.2019.00054}

\bibitem[{{Ruden} \& {Pollack}(1991)}]{1991ApJ...375..740R}
{Ruden}, S.~P., \& {Pollack}, J.~B. 1991, \apj, 375, 740,
  \dodoi{10.1086/170239}

\bibitem[{{Sano} {et~al.}(2000){Sano}, {Miyama}, {Umebayashi}, \&
  {Nakano}}]{2000ApJ...543..486S}
{Sano}, T., {Miyama}, S.~M., {Umebayashi}, T., \& {Nakano}, T. 2000, \apj, 543,
  486, \dodoi{10.1086/317075}

\bibitem[{{Sato} {et~al.}(2016){Sato}, {Okuzumi}, \&
  {Ida}}]{2016A&A...589A..15S}
{Sato}, T., {Okuzumi}, S., \& {Ida}, S. 2016, \aap, 589, A15,
  \dodoi{10.1051/0004-6361/201527069}

\bibitem[{{Shakura} \& {Sunyaev}(1973)}]{1973A&A....24..337S}
{Shakura}, N.~I., \& {Sunyaev}, R.~A. 1973, \aap, 24, 337

\bibitem[{{Sheehan} {et~al.}(2020){Sheehan}, {Tobin}, {Federman}, {Megeath}, \&
  {Looney}}]{2020ApJ...902..141S}
{Sheehan}, P.~D., {Tobin}, J.~J., {Federman}, S., {Megeath}, S.~T., \&
  {Looney}, L.~W. 2020, \apj, 902, 141, \dodoi{10.3847/1538-4357/abbad5}

\bibitem[{{Shoshi} {et~al.}(2023){Shoshi}, {Harada}, {Tokuda}, {Kawasaki},
  {Yamasaki}, {Sato}, {Omura}, {Yamaguchi}, {Tachihara}, \&
  {Machida}}]{2023arXiv231202504S}
{Shoshi}, A., {Harada}, N., {Tokuda}, K., {et~al.} 2023, arXiv e-prints,
  arXiv:2312.02504, \dodoi{10.48550/arXiv.2312.02504}

\bibitem[{{Smoluchowski}(1916)}]{1916ZPhy...17..557S}
{Smoluchowski}, M.~V. 1916, Zeitschrift fur Physik, 17, 557

\bibitem[{{Stammler} \& {Birnstiel}(2022)}]{2022ApJ...935...35S}
{Stammler}, S.~M., \& {Birnstiel}, T. 2022, \apj, 935, 35,
  \dodoi{10.3847/1538-4357/ac7d58}

\bibitem[{{Suzuki} {et~al.}(2010){Suzuki}, {Muto}, \&
  {Inutsuka}}]{2010ApJ...718.1289S}
{Suzuki}, T.~K., {Muto}, T., \& {Inutsuka}, S.-i. 2010, \apj, 718, 1289,
  \dodoi{10.1088/0004-637X/718/2/1289}

\bibitem[{{Suzuki} {et~al.}(2016){Suzuki}, {Ogihara}, {Morbidelli}, {Crida}, \&
  {Guillot}}]{2016A&A...596A..74S}
{Suzuki}, T.~K., {Ogihara}, M., {Morbidelli}, A., {Crida}, A., \& {Guillot}, T.
  2016, \aap, 596, A74, \dodoi{10.1051/0004-6361/201628955}

\bibitem[{{Tabone} {et~al.}(2022){Tabone}, {Rosotti}, {Cridland}, {Armitage},
  \& {Lodato}}]{2022MNRAS.512.2290T}
{Tabone}, B., {Rosotti}, G.~P., {Cridland}, A.~J., {Armitage}, P.~J., \&
  {Lodato}, G. 2022, \mnras, 512, 2290, \dodoi{10.1093/mnras/stab3442}

\bibitem[{{Takahashi} {et~al.}(2013){Takahashi}, {Inutsuka}, \&
  {Machida}}]{2013ApJ...770...71T}
{Takahashi}, S.~Z., {Inutsuka}, S.-i., \& {Machida}, M.~N. 2013, \apj, 770, 71,
  \dodoi{10.1088/0004-637X/770/1/71}

\bibitem[{{Takahashi} \& {Muto}(2018)}]{2018ApJ...865..102T}
{Takahashi}, S.~Z., \& {Muto}, T. 2018, \apj, 865, 102,
  \dodoi{10.3847/1538-4357/aadda0}

\bibitem[{{Tanaka} {et~al.}(2013){Tanaka}, {Nakamoto}, \&
  {Omukai}}]{2013ApJ...773..155T}
{Tanaka}, K. E.~I., {Nakamoto}, T., \& {Omukai}, K. 2013, \apj, 773, 155,
  \dodoi{10.1088/0004-637X/773/2/155}

\bibitem[{{Testi} {et~al.}(2014){Testi}, {Birnstiel}, {Ricci}, {Andrews},
  {Blum}, {Carpenter}, {Dominik}, {Isella}, {Natta}, {Williams}, \&
  {Wilner}}]{2014prpl.conf..339T}
{Testi}, L., {Birnstiel}, T., {Ricci}, L., {et~al.} 2014, in Protostars and
  Planets VI, ed. H.~{Beuther}, R.~S. {Klessen}, C.~P. {Dullemond}, \&
  T.~{Henning}, 339--361, \dodoi{10.2458/azu_uapress_9780816531240-ch015}

\bibitem[{{Tokuda} {et~al.}(2020){Tokuda}, {Fujishiro}, {Tachihara},
  {Takashima}, {Fukui}, {Zahorecz}, {Saigo}, {Matsumoto}, {Tomida}, {Machida},
  {Inutsuka}, {Andr{\'e}}, {Kawamura}, \& {Onishi}}]{2020ApJ...899...10T}
{Tokuda}, K., {Fujishiro}, K., {Tachihara}, K., {et~al.} 2020, \apj, 899, 10,
  \dodoi{10.3847/1538-4357/ab9ca7}

\bibitem[{{Tomida} {et~al.}(2017){Tomida}, {Machida}, {Hosokawa}, {Sakurai}, \&
  {Lin}}]{2017ApJ...835L..11T}
{Tomida}, K., {Machida}, M.~N., {Hosokawa}, T., {Sakurai}, Y., \& {Lin}, C.~H.
  2017, \apjl, 835, L11, \dodoi{10.3847/2041-8213/835/1/L11}

\bibitem[{{Tomida} {et~al.}(2013){Tomida}, {Tomisaka}, {Matsumoto}, {Hori},
  {Okuzumi}, {Machida}, \& {Saigo}}]{2013ApJ...763....6T}
{Tomida}, K., {Tomisaka}, K., {Matsumoto}, T., {et~al.} 2013, \apj, 763, 6,
  \dodoi{10.1088/0004-637X/763/1/6}

\bibitem[{{Tominaga} \& {Tanaka}(2023)}]{2023ApJ...958..168T}
{Tominaga}, R.~T., \& {Tanaka}, H. 2023, \apj, 958, 168,
  \dodoi{10.3847/1538-4357/ad002e}

\bibitem[{{Tsukamoto} {et~al.}(2021{\natexlab{a}}){Tsukamoto}, {Machida}, \&
  {Inutsuka}}]{2021ApJ...913..148T}
{Tsukamoto}, Y., {Machida}, M.~N., \& {Inutsuka}, S. 2021{\natexlab{a}}, \apj,
  913, 148, \dodoi{10.3847/1538-4357/abf5db}

\bibitem[{{Tsukamoto} {et~al.}(2021{\natexlab{b}}){Tsukamoto}, {Machida}, \&
  {Inutsuka}}]{2021ApJ...920L..35T}
{Tsukamoto}, Y., {Machida}, M.~N., \& {Inutsuka}, S.-i. 2021{\natexlab{b}},
  \apjl, 920, L35, \dodoi{10.3847/2041-8213/ac2b2f}

\bibitem[{{Tsukamoto} {et~al.}(2023){Tsukamoto}, {Machida}, \&
  {Inutsuka}}]{2023PASJ...75..835T}
---. 2023, \pasj, 75, 835, \dodoi{10.1093/pasj/psad040}

\bibitem[{{Tsukamoto} {et~al.}(2020){Tsukamoto}, {Machida}, {Susa}, {Nomura},
  \& {Inutsuka}}]{2020ApJ...896..158T}
{Tsukamoto}, Y., {Machida}, M.~N., {Susa}, H., {Nomura}, H., \& {Inutsuka}, S.
  2020, \apj, 896, 158, \dodoi{10.3847/1538-4357/ab93d0}

\bibitem[{{Tsukamoto} {et~al.}(2017){Tsukamoto}, {Okuzumi}, \&
  {Kataoka}}]{2017ApJ...838..151T}
{Tsukamoto}, Y., {Okuzumi}, S., \& {Kataoka}, A. 2017, \apj, 838, 151,
  \dodoi{10.3847/1538-4357/aa6081}

\bibitem[{{Ueda} {et~al.}(2019){Ueda}, {Flock}, \&
  {Okuzumi}}]{2019ApJ...871...10U}
{Ueda}, T., {Flock}, M., \& {Okuzumi}, S. 2019, \apj, 871, 10,
  \dodoi{10.3847/1538-4357/aaf3a1}

\bibitem[{{Umebayashi} \& {Nakano}(1990)}]{1990MNRAS.243..103U}
{Umebayashi}, T., \& {Nakano}, T. 1990, \mnras, 243, 103,
  \dodoi{10.1093/mnras/243.1.103}

\bibitem[{{Vorobyov} \& {Basu}(2006)}]{2006ApJ...650..956V}
{Vorobyov}, E.~I., \& {Basu}, S. 2006, \apj, 650, 956, \dodoi{10.1086/507320}

\bibitem[{{Wada} {et~al.}(2013){Wada}, {Tanaka}, {Okuzumi}, {Kobayashi},
  {Suyama}, {Kimura}, \& {Yamamoto}}]{2013A&A...559A..62W}
{Wada}, K., {Tanaka}, H., {Okuzumi}, S., {et~al.} 2013, \aap, 559, A62,
  \dodoi{10.1051/0004-6361/201322259}

\bibitem[{{Wada} {et~al.}(2009){Wada}, {Tanaka}, {Suyama}, {Kimura}, \&
  {Yamamoto}}]{2009ApJ...702.1490W}
{Wada}, K., {Tanaka}, H., {Suyama}, T., {Kimura}, H., \& {Yamamoto}, T. 2009,
  \apj, 702, 1490, \dodoi{10.1088/0004-637X/702/2/1490}

\bibitem[{{Wardle} \& {Salmeron}(2012)}]{2012MNRAS.422.2737W}
{Wardle}, M., \& {Salmeron}, R. 2012, \mnras, 422, 2737,
  \dodoi{10.1111/j.1365-2966.2011.20022.x}

\bibitem[{{Weidenschilling}(1977)}]{1977MNRAS.180...57W}
{Weidenschilling}, S.~J. 1977, \mnras, 180, 57, \dodoi{10.1093/mnras/180.2.57}

\bibitem[{{Whipple}(1972)}]{1972fpp..conf..211W}
{Whipple}, F.~L. 1972, in From Plasma to Planet, ed. A.~{Elvius}, 211

\bibitem[{{Windmark} {et~al.}(2012{\natexlab{a}}){Windmark}, {Birnstiel},
  {G{\"u}ttler}, {Blum}, {Dullemond}, \& {Henning}}]{2012A&A...540A..73W}
{Windmark}, F., {Birnstiel}, T., {G{\"u}ttler}, C., {et~al.}
  2012{\natexlab{a}}, \aap, 540, A73, \dodoi{10.1051/0004-6361/201118475}

\bibitem[{{Windmark} {et~al.}(2012{\natexlab{b}}){Windmark}, {Birnstiel},
  {Ormel}, \& {Dullemond}}]{2012A&A...544L..16W}
{Windmark}, F., {Birnstiel}, T., {Ormel}, C.~W., \& {Dullemond}, C.~P.
  2012{\natexlab{b}}, \aap, 544, L16, \dodoi{10.1051/0004-6361/201220004}

\bibitem[{{Wurm} {et~al.}(2005){Wurm}, {Paraskov}, \&
  {Krauss}}]{2005Icar..178..253W}
{Wurm}, G., {Paraskov}, G., \& {Krauss}, O. 2005, \icarus, 178, 253,
  \dodoi{10.1016/j.icarus.2005.04.002}

\bibitem[{{Xu} \& {Armitage}(2023)}]{2023ApJ...946...94X}
{Xu}, W., \& {Armitage}, P.~J. 2023, \apj, 946, 94,
  \dodoi{10.3847/1538-4357/acb7e5}

\bibitem[{{Xu} \& {Kunz}(2021{\natexlab{a}})}]{2021MNRAS.502.4911X}
{Xu}, W., \& {Kunz}, M.~W. 2021{\natexlab{a}}, \mnras, 502, 4911,
  \dodoi{10.1093/mnras/stab314}

\bibitem[{{Xu} \& {Kunz}(2021{\natexlab{b}})}]{2021MNRAS.508.2142X}
---. 2021{\natexlab{b}}, \mnras, 508, 2142, \dodoi{10.1093/mnras/stab2715}

\bibitem[{{Yang} {et~al.}(2017){Yang}, {Johansen}, \&
  {Carrera}}]{2017A&A...606A..80Y}
{Yang}, C.-C., {Johansen}, A., \& {Carrera}, D. 2017, \aap, 606, A80,
  \dodoi{10.1051/0004-6361/201630106}

\bibitem[{{Youdin} \& {Goodman}(2005)}]{2005ApJ...620..459Y}
{Youdin}, A.~N., \& {Goodman}, J. 2005, \apj, 620, 459, \dodoi{10.1086/426895}

\bibitem[{{Youdin} \& {Lithwick}(2007)}]{2007Icar..192..588Y}
{Youdin}, A.~N., \& {Lithwick}, Y. 2007, \icarus, 192, 588,
  \dodoi{10.1016/j.icarus.2007.07.012}

\bibitem[{{Zhu} {et~al.}(2010){Zhu}, {Hartmann}, \&
  {Gammie}}]{2010ApJ...713.1143Z}
{Zhu}, Z., {Hartmann}, L., \& {Gammie}, C. 2010, \apj, 713, 1143,
  \dodoi{10.1088/0004-637X/713/2/1143}

\bibitem[{{Zhu} {et~al.}(2012){Zhu}, {Hartmann}, {Nelson}, \&
  {Gammie}}]{2012ApJ...746..110Z}
{Zhu}, Z., {Hartmann}, L., {Nelson}, R.~P., \& {Gammie}, C.~F. 2012, \apj, 746,
  110, \dodoi{10.1088/0004-637X/746/1/110}

\end{thebibliography}


\begin{thebibliography}{}
\expandafter\ifx\csname natexlab\endcsname\relax\def\natexlab#1{#1}\fi
\providecommand{\url}[1]{\href{#1}{#1}}
\providecommand{\dodoi}[1]{doi:~\href{http://doi.org/#1}{\nolinkurl{#1}}}
\providecommand{\doeprint}[1]{\href{http://ascl.net/#1}{\nolinkurl{http://ascl.net/#1}}}
\providecommand{\doarXiv}[1]{\href{https://arxiv.org/abs/#1}{\nolinkurl{https://arxiv.org/abs/#1}}}

\bibitem[{{Alexander} {et~al.}(2023){Alexander}, {Rosotti}, {Armitage},
  {Herczeg}, {Manara}, \& {Tabone}}]{2023MNRAS.524.3948A}
{Alexander}, R., {Rosotti}, G., {Armitage}, P.~J., {et~al.} 2023, \mnras, 524,
  3948, \dodoi{10.1093/mnras/stad1983}

\bibitem[{{Andrews} {et~al.}(2018){Andrews}, {Huang}, {P{\'e}rez}, {Isella},
  {Dullemond}, {Kurtovic}, {Guzm{\'a}n}, {Carpenter}, {Wilner}, {Zhang}, {Zhu},
  {Birnstiel}, {Bai}, {Benisty}, {Hughes}, {{\"O}berg}, \&
  {Ricci}}]{2018ApJ...869L..41A}
{Andrews}, S.~M., {Huang}, J., {P{\'e}rez}, L.~M., {et~al.} 2018, \apjl, 869,
  L41, \dodoi{10.3847/2041-8213/aaf741}

\bibitem[{{Bae} {et~al.}(2013){Bae}, {Hartmann}, {Zhu}, \&
  {Gammie}}]{2013ApJ...764..141B}
{Bae}, J., {Hartmann}, L., {Zhu}, Z., \& {Gammie}, C. 2013, \apj, 764, 141,
  \dodoi{10.1088/0004-637X/764/2/141}

\bibitem[{{Bae} {et~al.}(2014){Bae}, {Hartmann}, {Zhu}, \&
  {Nelson}}]{2014ApJ...795...61B}
{Bae}, J., {Hartmann}, L., {Zhu}, Z., \& {Nelson}, R.~P. 2014, \apj, 795, 61,
  \dodoi{10.1088/0004-637X/795/1/61}

\bibitem[{{Bai}(2016)}]{2016ApJ...821...80B}
{Bai}, X.-N. 2016, \apj, 821, 80, \dodoi{10.3847/0004-637X/821/2/80}

\bibitem[{{Bai} {et~al.}(2016){Bai}, {Ye}, {Goodman}, \&
  {Yuan}}]{2016ApJ...818..152B}
{Bai}, X.-N., {Ye}, J., {Goodman}, J., \& {Yuan}, F. 2016, \apj, 818, 152,
  \dodoi{10.3847/0004-637X/818/2/152}

\bibitem[{{Birnstiel} {et~al.}(2010){Birnstiel}, {Dullemond}, \&
  {Brauer}}]{2010A&A...513A..79B}
{Birnstiel}, T., {Dullemond}, C.~P., \& {Brauer}, F. 2010, \aap, 513, A79,
  \dodoi{10.1051/0004-6361/200913731}

\bibitem[{{Birnstiel} {et~al.}(2012){Birnstiel}, {Klahr}, \&
  {Ercolano}}]{2012A&A...539A.148B}
{Birnstiel}, T., {Klahr}, H., \& {Ercolano}, B. 2012, \aap, 539, A148,
  \dodoi{10.1051/0004-6361/201118136}

\bibitem[{{Bonnor}(1956)}]{1956MNRAS.116..351B}
{Bonnor}, W.~B. 1956, \mnras, 116, 351, \dodoi{10.1093/mnras/116.3.351}

\bibitem[{{Brauer} {et~al.}(2008){Brauer}, {Dullemond}, \&
  {Henning}}]{2008A&A...480..859B}
{Brauer}, F., {Dullemond}, C.~P., \& {Henning}, T. 2008, \aap, 480, 859,
  \dodoi{10.1051/0004-6361:20077759}

\bibitem[{{Carrera} {et~al.}(2015){Carrera}, {Johansen}, \&
  {Davies}}]{2015A&A...579A..43C}
{Carrera}, D., {Johansen}, A., \& {Davies}, M.~B. 2015, \aap, 579, A43,
  \dodoi{10.1051/0004-6361/201425120}

\bibitem[{{Cassen} \& {Moosman}(1981)}]{1981Icar...48..353C}
{Cassen}, P., \& {Moosman}, A. 1981, \icarus, 48, 353,
  \dodoi{10.1016/0019-1035(81)90051-8}

\bibitem[{{Chen} \& {Lin}(2020)}]{2020ApJ...891..132C}
{Chen}, K., \& {Lin}, M.-K. 2020, \apj, 891, 132,
  \dodoi{10.3847/1538-4357/ab76ca}

\bibitem[{{Desch} \& {Turner}(2015)}]{2015ApJ...811..156D}
{Desch}, S.~J., \& {Turner}, N.~J. 2015, \apj, 811, 156,
  \dodoi{10.1088/0004-637X/811/2/156}

\bibitem[{{Ebert}(1955)}]{1955ZA.....37..217E}
{Ebert}, R. 1955, \zap, 37, 217

\bibitem[{{Flock} {et~al.}(2016){Flock}, {Fromang}, {Turner}, \&
  {Benisty}}]{2016ApJ...827..144F}
{Flock}, M., {Fromang}, S., {Turner}, N.~J., \& {Benisty}, M. 2016, \apj, 827,
  144, \dodoi{10.3847/0004-637X/827/2/144}

\bibitem[{{Homma} \& {Nakamoto}(2018)}]{2018ApJ...868..118H}
{Homma}, K., \& {Nakamoto}, T. 2018, \apj, 868, 118,
  \dodoi{10.3847/1538-4357/aae0fb}

\bibitem[{{Hueso} \& {Guillot}(2005)}]{2005A&A...442..703H}
{Hueso}, R., \& {Guillot}, T. 2005, \aap, 442, 703,
  \dodoi{10.1051/0004-6361:20041905}

\bibitem[{{Johansen} \& {Youdin}(2007)}]{2007ApJ...662..627J}
{Johansen}, A., \& {Youdin}, A. 2007, \apj, 662, 627, \dodoi{10.1086/516730}

\bibitem[{{Kataoka} {et~al.}(2013){Kataoka}, {Tanaka}, {Okuzumi}, \&
  {Wada}}]{2013A&A...557L...4K}
{Kataoka}, A., {Tanaka}, H., {Okuzumi}, S., \& {Wada}, K. 2013, \aap, 557, L4,
  \dodoi{10.1051/0004-6361/201322151}

\bibitem[{{Kawasaki} {et~al.}(2021){Kawasaki}, {Koga}, \&
  {Machida}}]{2021MNRAS.504.5588K}
{Kawasaki}, Y., {Koga}, S., \& {Machida}, M.~N. 2021, \mnras, 504, 5588,
  \dodoi{10.1093/mnras/stab1224}

\bibitem[{{Kawasaki} {et~al.}(2022){Kawasaki}, {Koga}, \&
  {Machida}}]{2022MNRAS.515.2072K}
---. 2022, \mnras, 515, 2072, \dodoi{10.1093/mnras/stac1919}

\bibitem[{{Kobayashi} \& {Tanaka}(2021)}]{2021ApJ...922...16K}
{Kobayashi}, H., \& {Tanaka}, H. 2021, \apj, 922, 16,
  \dodoi{10.3847/1538-4357/ac289c}

\bibitem[{{Koga} {et~al.}(2022){Koga}, {Kawasaki}, \&
  {Machida}}]{2022MNRAS.515.6073K}
{Koga}, S., {Kawasaki}, Y., \& {Machida}, M.~N. 2022, \mnras, 515, 6073,
  \dodoi{10.1093/mnras/stac2115}

\bibitem[{{Koga} \& {Machida}(2023)}]{2023MNRAS.519.3595K}
{Koga}, S., \& {Machida}, M.~N. 2023, \mnras, 519, 3595,
  \dodoi{10.1093/mnras/stac3503}

\bibitem[{{Kudoh} \& {Shibata}(1997)}]{1997ApJ...474..362K}
{Kudoh}, T., \& {Shibata}, K. 1997, \apj, 474, 362, \dodoi{10.1086/303437}

\bibitem[{{Lebreuilly} {et~al.}(2019){Lebreuilly}, {Commer{\c{c}}on}, \&
  {Laibe}}]{2019A&A...626A..96L}
{Lebreuilly}, U., {Commer{\c{c}}on}, B., \& {Laibe}, G. 2019, \aap, 626, A96,
  \dodoi{10.1051/0004-6361/201834147}

\bibitem[{{Lebreuilly} {et~al.}(2020){Lebreuilly}, {Commer{\c{c}}on}, \&
  {Laibe}}]{2020A&A...641A.112L}
---. 2020, \aap, 641, A112, \dodoi{10.1051/0004-6361/202038174}

\bibitem[{{Lesur} {et~al.}(2023){Lesur}, {Flock}, {Ercolano}, {Lin}, {Yang},
  {Barranco}, {Benitez-Llambay}, {Goodman}, {Johansen}, {Klahr}, {Laibe},
  {Lyra}, {Marcus}, {Nelson}, {Squire}, {Simon}, {Turner}, {Umurhan}, \&
  {Youdin}}]{2023ASPC..534..465L}
{Lesur}, G., {Flock}, M., {Ercolano}, B., {et~al.} 2023, in Astronomical
  Society of the Pacific Conference Series, Vol. 534, Astronomical Society of
  the Pacific Conference Series, ed. S.~{Inutsuka}, Y.~{Aikawa}, T.~{Muto},
  K.~{Tomida}, \& M.~{Tamura}, 465

\bibitem[{{Li} \& {Youdin}(2021)}]{2021ApJ...919..107L}
{Li}, R., \& {Youdin}, A.~N. 2021, \apj, 919, 107,
  \dodoi{10.3847/1538-4357/ac0e9f}

\bibitem[{{Lin} \& {Hsu}(2022)}]{2022ApJ...926...14L}
{Lin}, M.-K., \& {Hsu}, C.-Y. 2022, \apj, 926, 14,
  \dodoi{10.3847/1538-4357/ac3bb9}

\bibitem[{{Liu} {et~al.}(2023){Liu}, {Takahashi}, {Machida}, {Tomisaka},
  {Girart}, {Ho}, {Nakanishi}, \& {Sato}}]{2023arXiv231213573L}
{Liu}, Y., {Takahashi}, S., {Machida}, M., {et~al.} 2023, arXiv e-prints,
  arXiv:2312.13573, \dodoi{10.48550/arXiv.2312.13573}

\bibitem[{{Machida} \& {Basu}(2019)}]{2019ApJ...876..149M}
{Machida}, M.~N., \& {Basu}, S. 2019, \apj, 876, 149,
  \dodoi{10.3847/1538-4357/ab18a7}

\bibitem[{{Machida} \& {Basu}(2020)}]{2020MNRAS.494..827M}
---. 2020, \mnras, 494, 827, \dodoi{10.1093/mnras/staa672}

\bibitem[{{Machida} \& {Hosokawa}(2020)}]{2020MNRAS.499.4490M}
{Machida}, M.~N., \& {Hosokawa}, T. 2020, \mnras, 499, 4490,
  \dodoi{10.1093/mnras/staa3139}

\bibitem[{{Machida} {et~al.}(2010){Machida}, {Inutsuka}, \&
  {Matsumoto}}]{2010ApJ...724.1006M}
{Machida}, M.~N., {Inutsuka}, S.-i., \& {Matsumoto}, T. 2010, \apj, 724, 1006,
  \dodoi{10.1088/0004-637X/724/2/1006}

\bibitem[{{Machida} \& {Matsumoto}(2011)}]{2011MNRAS.413.2767M}
{Machida}, M.~N., \& {Matsumoto}, T. 2011, \mnras, 413, 2767,
  \dodoi{10.1111/j.1365-2966.2011.18349.x}

\bibitem[{{Machida} {et~al.}(2006){Machida}, {Matsumoto}, {Hanawa}, \&
  {Tomisaka}}]{2006ApJ...645.1227M}
{Machida}, M.~N., {Matsumoto}, T., {Hanawa}, T., \& {Tomisaka}, K. 2006, \apj,
  645, 1227, \dodoi{10.1086/504423}

\bibitem[{{Marchand} {et~al.}(2023){Marchand}, {Lebreuilly}, {Mac Low}, \&
  {Guillet}}]{2023A&A...670A..61M}
{Marchand}, P., {Lebreuilly}, U., {Mac Low}, M.~M., \& {Guillet}, V. 2023,
  \aap, 670, A61, \dodoi{10.1051/0004-6361/202244291}

\bibitem[{{Masson} {et~al.}(2016){Masson}, {Chabrier}, {Hennebelle}, {Vaytet},
  \& {Commer{\c{c}}on}}]{2016A&A...587A..32M}
{Masson}, J., {Chabrier}, G., {Hennebelle}, P., {Vaytet}, N., \&
  {Commer{\c{c}}on}, B. 2016, \aap, 587, A32,
  \dodoi{10.1051/0004-6361/201526371}

\bibitem[{{Miyake} {et~al.}(2016){Miyake}, {Suzuki}, \&
  {Inutsuka}}]{2016ApJ...821....3M}
{Miyake}, T., {Suzuki}, T.~K., \& {Inutsuka}, S.-i. 2016, \apj, 821, 3,
  \dodoi{10.3847/0004-637X/821/1/3}

\bibitem[{{Nakamoto} \& {Nakagawa}(1994)}]{1994ApJ...421..640N}
{Nakamoto}, T., \& {Nakagawa}, Y. 1994, \apj, 421, 640, \dodoi{10.1086/173678}

\bibitem[{{Ohashi} {et~al.}(2023{\natexlab{a}}){Ohashi}, {Tobin},
  {J{\o}rgensen}, {Takakuwa}, {Sheehan}, {Aikawa}, {Li}, {Looney}, {Williams},
  {Aso}, {Sharma}, {Sai}, {Yamato}, {Lee}, {Tomida}, {Yen}, {Encalada},
  {Flores}, {Gavino}, {Kido}, {Han}, {Lin}, {Narayanan}, {Phuong},
  {Santamar{\'\i}a-Miranda}, {Thieme}, {van't Hoff}, {de Gregorio-Monsalvo},
  {Koch}, {Kwon}, {Lai}, {Lee}, {Plunkett}, {Saigo}, {Hirano}, {Lam}, \&
  {Mori}}]{2023ApJ...951....8O}
{Ohashi}, N., {Tobin}, J.~J., {J{\o}rgensen}, J.~K., {et~al.}
  2023{\natexlab{a}}, \apj, 951, 8, \dodoi{10.3847/1538-4357/acd384}

\bibitem[{{Ohashi} {et~al.}(2021){Ohashi}, {Kobayashi}, {Nakatani}, {Okuzumi},
  {Tanaka}, {Murakawa}, {Zhang}, {Liu}, \& {Sakai}}]{2021ApJ...907...80O}
{Ohashi}, S., {Kobayashi}, H., {Nakatani}, R., {et~al.} 2021, \apj, 907, 80,
  \dodoi{10.3847/1538-4357/abd0fa}

\bibitem[{{Ohashi} {et~al.}(2023{\natexlab{b}}){Ohashi}, {Momose}, {Kataoka},
  {Higuchi}, {Tsukagoshi}, {Ueda}, {Codella}, {Podio}, {Hanawa}, {Sakai},
  {Kobayashi}, {Okuzumi}, \& {Tanaka}}]{2023ApJ...954..110O}
{Ohashi}, S., {Momose}, M., {Kataoka}, A., {et~al.} 2023{\natexlab{b}}, \apj,
  954, 110, \dodoi{10.3847/1538-4357/ace9b9}

\bibitem[{{Okuzumi} {et~al.}(2016){Okuzumi}, {Momose}, {Sirono}, {Kobayashi},
  \& {Tanaka}}]{2016ApJ...821...82O}
{Okuzumi}, S., {Momose}, M., {Sirono}, S.-i., {Kobayashi}, H., \& {Tanaka}, H.
  2016, \apj, 821, 82, \dodoi{10.3847/0004-637X/821/2/82}

\bibitem[{{Okuzumi} {et~al.}(2012){Okuzumi}, {Tanaka}, {Kobayashi}, \&
  {Wada}}]{2012ApJ...752..106O}
{Okuzumi}, S., {Tanaka}, H., {Kobayashi}, H., \& {Wada}, K. 2012, \apj, 752,
  106, \dodoi{10.1088/0004-637X/752/2/106}

\bibitem[{{Ormel} {et~al.}(2007){Ormel}, {Spaans}, \&
  {Tielens}}]{2007A&A...461..215O}
{Ormel}, C.~W., {Spaans}, M., \& {Tielens}, A.~G.~G.~M. 2007, \aap, 461, 215,
  \dodoi{10.1051/0004-6361:20065949}

\bibitem[{{Pinilla} {et~al.}(2016){Pinilla}, {Flock}, {Ovelar}, \&
  {Birnstiel}}]{2016A&A...596A..81P}
{Pinilla}, P., {Flock}, M., {Ovelar}, M. d.~J., \& {Birnstiel}, T. 2016, \aap,
  596, A81, \dodoi{10.1051/0004-6361/201628441}

\bibitem[{{Ruden} \& {Pollack}(1991)}]{1991ApJ...375..740R}
{Ruden}, S.~P., \& {Pollack}, J.~B. 1991, \apj, 375, 740,
  \dodoi{10.1086/170239}

\bibitem[{{Sato} {et~al.}(2016){Sato}, {Okuzumi}, \&
  {Ida}}]{2016A&A...589A..15S}
{Sato}, T., {Okuzumi}, S., \& {Ida}, S. 2016, \aap, 589, A15,
  \dodoi{10.1051/0004-6361/201527069}

\bibitem[{{Shakura} \& {Sunyaev}(1973)}]{1973A&A....24..337S}
{Shakura}, N.~I., \& {Sunyaev}, R.~A. 1973, \aap, 24, 337

\bibitem[{{Sheehan} {et~al.}(2020){Sheehan}, {Tobin}, {Federman}, {Megeath}, \&
  {Looney}}]{2020ApJ...902..141S}
{Sheehan}, P.~D., {Tobin}, J.~J., {Federman}, S., {Megeath}, S.~T., \&
  {Looney}, L.~W. 2020, \apj, 902, 141, \dodoi{10.3847/1538-4357/abbad5}

\bibitem[{{Shoshi} {et~al.}(2023){Shoshi}, {Harada}, {Tokuda}, {Kawasaki},
  {Yamasaki}, {Sato}, {Omura}, {Yamaguchi}, {Tachihara}, \&
  {Machida}}]{2023arXiv231202504S}
{Shoshi}, A., {Harada}, N., {Tokuda}, K., {et~al.} 2023, arXiv e-prints,
  arXiv:2312.02504, \dodoi{10.48550/arXiv.2312.02504}

\bibitem[{{Smoluchowski}(1916)}]{1916ZPhy...17..557S}
{Smoluchowski}, M.~V. 1916, Zeitschrift fur Physik, 17, 557

\bibitem[{{Stammler} \& {Birnstiel}(2022)}]{2022ApJ...935...35S}
{Stammler}, S.~M., \& {Birnstiel}, T. 2022, \apj, 935, 35,
  \dodoi{10.3847/1538-4357/ac7d58}

\bibitem[{{Suzuki} {et~al.}(2010){Suzuki}, {Muto}, \&
  {Inutsuka}}]{2010ApJ...718.1289S}
{Suzuki}, T.~K., {Muto}, T., \& {Inutsuka}, S.-i. 2010, \apj, 718, 1289,
  \dodoi{10.1088/0004-637X/718/2/1289}

\bibitem[{{Suzuki} {et~al.}(2016){Suzuki}, {Ogihara}, {Morbidelli}, {Crida}, \&
  {Guillot}}]{2016A&A...596A..74S}
{Suzuki}, T.~K., {Ogihara}, M., {Morbidelli}, A., {Crida}, A., \& {Guillot}, T.
  2016, \aap, 596, A74, \dodoi{10.1051/0004-6361/201628955}

\bibitem[{{Tabone} {et~al.}(2022){Tabone}, {Rosotti}, {Cridland}, {Armitage},
  \& {Lodato}}]{2022MNRAS.512.2290T}
{Tabone}, B., {Rosotti}, G.~P., {Cridland}, A.~J., {Armitage}, P.~J., \&
  {Lodato}, G. 2022, \mnras, 512, 2290, \dodoi{10.1093/mnras/stab3442}

\bibitem[{{Takahashi} {et~al.}(2013){Takahashi}, {Inutsuka}, \&
  {Machida}}]{2013ApJ...770...71T}
{Takahashi}, S.~Z., {Inutsuka}, S.-i., \& {Machida}, M.~N. 2013, \apj, 770, 71,
  \dodoi{10.1088/0004-637X/770/1/71}

\bibitem[{{Takahashi} \& {Muto}(2018)}]{2018ApJ...865..102T}
{Takahashi}, S.~Z., \& {Muto}, T. 2018, \apj, 865, 102,
  \dodoi{10.3847/1538-4357/aadda0}

\bibitem[{{Testi} {et~al.}(2014){Testi}, {Birnstiel}, {Ricci}, {Andrews},
  {Blum}, {Carpenter}, {Dominik}, {Isella}, {Natta}, {Williams}, \&
  {Wilner}}]{2014prpl.conf..339T}
{Testi}, L., {Birnstiel}, T., {Ricci}, L., {et~al.} 2014, in Protostars and
  Planets VI, ed. H.~{Beuther}, R.~S. {Klessen}, C.~P. {Dullemond}, \&
  T.~{Henning}, 339--361, \dodoi{10.2458/azu_uapress_9780816531240-ch015}

\bibitem[{{Tomida} {et~al.}(2017){Tomida}, {Machida}, {Hosokawa}, {Sakurai}, \&
  {Lin}}]{2017ApJ...835L..11T}
{Tomida}, K., {Machida}, M.~N., {Hosokawa}, T., {Sakurai}, Y., \& {Lin}, C.~H.
  2017, \apjl, 835, L11, \dodoi{10.3847/2041-8213/835/1/L11}

\bibitem[{{Tomida} {et~al.}(2013){Tomida}, {Tomisaka}, {Matsumoto}, {Hori},
  {Okuzumi}, {Machida}, \& {Saigo}}]{2013ApJ...763....6T}
{Tomida}, K., {Tomisaka}, K., {Matsumoto}, T., {et~al.} 2013, \apj, 763, 6,
  \dodoi{10.1088/0004-637X/763/1/6}

\bibitem[{{Tominaga} \& {Tanaka}(2023)}]{2023ApJ...958..168T}
{Tominaga}, R.~T., \& {Tanaka}, H. 2023, \apj, 958, 168,
  \dodoi{10.3847/1538-4357/ad002e}

\bibitem[{{Tsukamoto} {et~al.}(2021{\natexlab{a}}){Tsukamoto}, {Machida}, \&
  {Inutsuka}}]{2021ApJ...913..148T}
{Tsukamoto}, Y., {Machida}, M.~N., \& {Inutsuka}, S. 2021{\natexlab{a}}, \apj,
  913, 148, \dodoi{10.3847/1538-4357/abf5db}

\bibitem[{{Tsukamoto} {et~al.}(2021{\natexlab{b}}){Tsukamoto}, {Machida}, \&
  {Inutsuka}}]{2021ApJ...920L..35T}
{Tsukamoto}, Y., {Machida}, M.~N., \& {Inutsuka}, S.-i. 2021{\natexlab{b}},
  \apjl, 920, L35, \dodoi{10.3847/2041-8213/ac2b2f}

\bibitem[{{Tsukamoto} {et~al.}(2023){Tsukamoto}, {Machida}, \&
  {Inutsuka}}]{2023PASJ...75..835T}
---. 2023, \pasj, 75, 835, \dodoi{10.1093/pasj/psad040}

\bibitem[{{Tsukamoto} {et~al.}(2020){Tsukamoto}, {Machida}, {Susa}, {Nomura},
  \& {Inutsuka}}]{2020ApJ...896..158T}
{Tsukamoto}, Y., {Machida}, M.~N., {Susa}, H., {Nomura}, H., \& {Inutsuka}, S.
  2020, \apj, 896, 158, \dodoi{10.3847/1538-4357/ab93d0}

\bibitem[{{Tsukamoto} {et~al.}(2017){Tsukamoto}, {Okuzumi}, \&
  {Kataoka}}]{2017ApJ...838..151T}
{Tsukamoto}, Y., {Okuzumi}, S., \& {Kataoka}, A. 2017, \apj, 838, 151,
  \dodoi{10.3847/1538-4357/aa6081}

\bibitem[{{Ueda} {et~al.}(2019){Ueda}, {Flock}, \&
  {Okuzumi}}]{2019ApJ...871...10U}
{Ueda}, T., {Flock}, M., \& {Okuzumi}, S. 2019, \apj, 871, 10,
  \dodoi{10.3847/1538-4357/aaf3a1}

\bibitem[{{Wada} {et~al.}(2013){Wada}, {Tanaka}, {Okuzumi}, {Kobayashi},
  {Suyama}, {Kimura}, \& {Yamamoto}}]{2013A&A...559A..62W}
{Wada}, K., {Tanaka}, H., {Okuzumi}, S., {et~al.} 2013, \aap, 559, A62,
  \dodoi{10.1051/0004-6361/201322259}

\bibitem[{{Wada} {et~al.}(2009){Wada}, {Tanaka}, {Suyama}, {Kimura}, \&
  {Yamamoto}}]{2009ApJ...702.1490W}
{Wada}, K., {Tanaka}, H., {Suyama}, T., {Kimura}, H., \& {Yamamoto}, T. 2009,
  \apj, 702, 1490, \dodoi{10.1088/0004-637X/702/2/1490}

\bibitem[{{Windmark} {et~al.}(2012){Windmark}, {Birnstiel}, {G{\"u}ttler},
  {Blum}, {Dullemond}, \& {Henning}}]{2012A&A...540A..73W}
{Windmark}, F., {Birnstiel}, T., {G{\"u}ttler}, C., {et~al.} 2012, \aap, 540,
  A73, \dodoi{10.1051/0004-6361/201118475}

\bibitem[{{Wu} \& {Chen}(2025)}]{2025MNRAS.536L..13W}
{Wu}, Y., \& {Chen}, Y.-X. 2025, \mnras, 536, L13,
  \dodoi{10.1093/mnrasl/slae102}

\bibitem[{{Wu} {et~al.}(2023){Wu}, {Chen}, {Jiang}, {Dong}, {Mac{\'\i}as},
  {Lin}, {Rosotti}, \& {Elbakyan}}]{2023MNRAS.523.2630W}
{Wu}, Y., {Chen}, Y.-X., {Jiang}, H., {et~al.} 2023, \mnras, 523, 2630,
  \dodoi{10.1093/mnras/stad1553}

\bibitem[{{Wu} {et~al.}(2024){Wu}, {Lin}, {Cui}, {Krapp}, {Lee}, \&
  {Youdin}}]{2024ApJ...962..173W}
{Wu}, Y., {Lin}, M.-K., {Cui}, C., {et~al.} 2024, \apj, 962, 173,
  \dodoi{10.3847/1538-4357/ad15fe}

\bibitem[{{Xu} \& {Armitage}(2023)}]{2023ApJ...946...94X}
{Xu}, W., \& {Armitage}, P.~J. 2023, \apj, 946, 94,
  \dodoi{10.3847/1538-4357/acb7e5}

\bibitem[{{Xu} {et~al.}(2024){Xu}, {Jiang}, {Kunz}, \&
  {Stone}}]{2024arXiv241012042X}
{Xu}, W., {Jiang}, Y.-F., {Kunz}, M.~W., \& {Stone}, J.~M. 2024, arXiv
  e-prints, arXiv:2410.12042, \dodoi{10.48550/arXiv.2410.12042}

\bibitem[{{Xu} \& {Kunz}(2021{\natexlab{a}})}]{2021MNRAS.502.4911X}
{Xu}, W., \& {Kunz}, M.~W. 2021{\natexlab{a}}, \mnras, 502, 4911,
  \dodoi{10.1093/mnras/stab314}

\bibitem[{{Xu} \& {Kunz}(2021{\natexlab{b}})}]{2021MNRAS.508.2142X}
---. 2021{\natexlab{b}}, \mnras, 508, 2142, \dodoi{10.1093/mnras/stab2715}

\bibitem[{{Yang} {et~al.}(2017){Yang}, {Johansen}, \&
  {Carrera}}]{2017A&A...606A..80Y}
{Yang}, C.-C., {Johansen}, A., \& {Carrera}, D. 2017, \aap, 606, A80,
  \dodoi{10.1051/0004-6361/201630106}

\bibitem[{{Youdin} \& {Goodman}(2005)}]{2005ApJ...620..459Y}
{Youdin}, A.~N., \& {Goodman}, J. 2005, \apj, 620, 459, \dodoi{10.1086/426895}

\bibitem[{{Youdin} \& {Lithwick}(2007)}]{2007Icar..192..588Y}
{Youdin}, A.~N., \& {Lithwick}, Y. 2007, \icarus, 192, 588,
  \dodoi{10.1016/j.icarus.2007.07.012}

\bibitem[{{Zhu} {et~al.}(2010){Zhu}, {Hartmann}, \&
  {Gammie}}]{2010ApJ...713.1143Z}
{Zhu}, Z., {Hartmann}, L., \& {Gammie}, C. 2010, \apj, 713, 1143,
  \dodoi{10.1088/0004-637X/713/2/1143}

\bibitem[{{Zhu} {et~al.}(2012){Zhu}, {Hartmann}, {Nelson}, \&
  {Gammie}}]{2012ApJ...746..110Z}
{Zhu}, Z., {Hartmann}, L., {Nelson}, R.~P., \& {Gammie}, C.~F. 2012, \apj, 746,
  110, \dodoi{10.1088/0004-637X/746/1/110}

\bibitem[{{Zhu} {et~al.}(2015){Zhu}, {Stone}, \& {Bai}}]{2015ApJ...801...81Z}
{Zhu}, Z., {Stone}, J.~M., \& {Bai}, X.-N. 2015, \apj, 801, 81,
  \dodoi{10.1088/0004-637X/801/2/81}

\end{thebibliography}
\bibliographystyle{aasjournal}

\end{document}